\documentclass[aps,prc,showpacs,amssymb,amsmath,amsfonts,nofootinbib,floatfix]{revtex4-2}
\usepackage{graphicx}

\makeatletter
\def\p@subsection{}

\def\p@subsubsection{}
\makeatother
\usepackage{natbib}
\usepackage[usenames]{color}
\usepackage{epsfig,graphicx}
\usepackage{epstopdf}
\usepackage{amssymb,amsmath,eqparbox}
\usepackage{subfig}
\usepackage{epstopdf}
\usepackage{enumitem}
\usepackage{comment}
\usepackage[bookmarks={false}]{hyperref}
\usepackage{caption}
\definecolor{grey}{rgb}{0.9,0.9,0.9}

\definecolor{black}{rgb}{0,0,0}

\hypersetup{pdfstartview={XYZ null null 1.}}

\def \LP{Svarc2013,Svarc2016}
\def \AA/PWA{Svarc2020,Svarc2022}
\def \BGED{Anisovich2017,Anisovich2017a}

\def  \Yannick/Lothar{Tiator2012,YannickPhD,Wunderlich2014}
\def  \LPapplication{Svarc2014,Svarc2014a,Svarc2015,Svarc2016a}

\newcommand{\be}{\begin{eqnarray}}
\newcommand{\ee}{\end{eqnarray}}
\newcommand{\bc}{\begin{center}}
\newcommand{\ec}{\end{center}}

\newcommand{\RudjerBoskovic}{Rudjer Bo\v{s}kovi\'{c} Institute, Bijeni\v{c}ka cesta 54, P.O. Box 180, 10002 Zagreb, Croatia}
\newcommand{\Tesla}{Tesla Biotech d.o.o., Mandlova 7, 10000 Zagreb, Croatia}

\begin{document}

\allowdisplaybreaks

\title{Laurent+Pietarinen Partial-Wave Analysis}
\author{A.~{\v{S}}varc}\email[Corresponding author: ]{svarc@irb.hr}
\affiliation{\RudjerBoskovic,\Tesla}

\author{R.L. Workman}
\affiliation{Institute for Nuclear Studies, Department of Physics, \
The George Washington University, Washington D.C. 20052, USA}

\date{\today}
\begin{abstract}
\vspace{0.5cm}
A new energy-dependent fit strategy, independent of any specific microscopic theory, is applied to kaon photoproduction data with center-of-mass energies ranging from  1625 MeV to 2296 MeV. Experimental data are fitted in terms of a modified Laurent expansion (Laurent+Pietarinen expansion)  which previously has been successfully applied to multipoles. The present aim is to extract resonance pole parameters directly from the data, rather than from sets of multipoles. A constrained single-energy fit is then used to search for missing structures. In this proof-of-principle study, the data are well-described by the initial L+P fit, and it is shown that only a moderate amount of structure, mostly in higher multipoles, is missing from the original fit. Problems due to an unmeasurable overall phase, plaguing single-channel multipole analyses, are mitigated by implementing a form of phase limitation, fixing the initial values of fit parameters using a multi-channel analysis.

\end{abstract}

\maketitle

\newpage

\section{Introduction} \label{sec:Intro}

 Measurements of meson-nucleon scattering and photoproduction, with polarized beams, targets and recoil particles, have a long history, largely motivated by the search for baryon resonances and their properties.
The resulting data have been analyzed and re-analyzed as new measurements and analysis techniques have become available. Most current studies employ elaborate multi-channel formalisms that cannot easily be reproduced by experimental groups providing the data.
Analysis groups can obtain partial-wave amplitudes and continue these to poles in the complex energy plane but experimental groups, seeking to determine the influence of their measurements, must often rely on less-elaborate methods or fits taken from another group. Differences in these models also introduce systematic errors, complicating the confirmation or comparison of different results. If would be useful to have a single-channel technique available to perform this task. In practice, some minimal phase information from a multi-channel analysis is required to avoid the continuum ambiguity in a single-energy analysis. Some ways to implement this multi-channel constraint have been explored in Refs.\cite{Svarc2020,Svarc2022}.

 In the present study, a more direct connection between pole parameters and data is developed. This method avoids the construction of an explicit model for the process, instead employing analyticity properties of complex functions in the vicinity of poles and cuts. This approach, the Laurent+Pietarinen (L+P) expansion, provides an approximation to the analytic function under consideration using the Laurent theorem, and represents the regular background term as a fast converging series in a conformal variable with chosen branch-points~\cite{Ciulli,CiulliFisher,Pietarinen,Pietarinen1}. This method was at first applied to the extraction of poles from partial waves or multipoles~\cite{\LP} with notable success~\cite{\LPapplication}.

 However, the main problem with this approach is that it requires having partial waves (multipoles) already at ones disposal. The source may be some theoretical model, or the result of a single-energy partial-wave analysis (SE PWA) which is constrained to be reasonably continuous\footnote{It is well known that any unconstrained single-channel, single-energy partial wave analysis (SC SE PWA) has discontinuities due to the continuum ambiguity~\cite{SE-PWA-uniqueness,Svarc2018}.}.
 Here, the L+P analysis technique is extended to analyze experimental data.
 Instead of analyzing single multipoles individually, all multipoles are fitted simultaneously. An L+P decomposition of
 a finite set of multipoles was made and used to reconstruct all available observables. The L+P parameters were then fitted directly to measured data. In this way, complications of a theoretical model were replaced by the selection of terms in the L+P decomposition, the relevant singularities (poles and branch-points) of each partial wave.
While this is, in principle, a single-step procedure, fitting available data sets with a sufficient number of parameters, the goodness of fit will depend on where one cuts off the L+P expansion. To address this issue, the single-energy
fit method is used to search for structure missing in the L+P expansion.
As a first proof-of-principle study, a minimal L+P expansion is fitted to
kaon photoproduction data, followed by
a single-energy fit.

In the next section, the formalism used in the L+P and
single-energy fits is outlined. Particular attention is paid to the cutoff in L+P
expansion terms and the starting point and constraints used in the fit. Results
are then presented to show the fit quality for observables and consistency of the
L+P and single-energy amplitudes. Prospects for an enhanced analysis, utilizing a more complete set of L+P parameters, is considered. Finally, some conclusions from this study are listed.

 \section{Formalism}
\subsection{Laurent+Pietarinen Partial-Wave Analysis}
In references~\cite{\LP}, a method to analyze the analytic structure of any complex function  was formulated. The function, in particular a partial-wave or multipole amplitude, was locally represented in terms of a Laurent decomposition where the regular (non-pole) part was expanded in a sum of finite rapidly-converging power-series in a conveniently defined conformal variable (one series per branch-point). The most general form used to analyze an analytic function of interest was given by~\cite{PDG}:
\be  \label{L+P-general}
T(W)  & = & \sum_{i=1}^{N}\frac{u_i+  \imath \, v_i}{W-W_i}+ \sum_{j=1}^{M} \sum_{n=0}^{n_{max}^j} c_{n}^{j}\left( \frac{\alpha_j-\sqrt{x_j-W}}{\alpha_j+\sqrt{x_j-W}} \right)^n,
\ee
where $W$ is center-of-mass energy, $N$ is the number of poles, $W_i$, $u_i, v_i$ are pole positions and residues, $M$ is the number of Pietarinen expansions, $n_{max}^j$ is the number of coefficients in the j-th Pietarinen expansion, and $c_n^j$ is the real expansion coefficient. For analyses of Refs.~\cite{\LP}, we have used three Pietarinen expansions ($M=3$) with maximum number of Pietarinen coefficients up to six ($n_ {max}^j=6$), and up to three resonances per partial wave. With these choices, there was very good agreement with the input function, and reliable pole positions and residues could be obtained.

For each partial wave, exactly this form of the expansion could be used to fit experimental data from kaon photoproduction measurements. However, all multipoles should be minimized at the same time, as observables are given as functions of all multipoles. In Appendix~\ref{AppendixA}, the relevant formulae for pseudoscalar meson photoproduction are given.
Replacing $E_{\ell \pm}$ and $M_{\ell \pm}$ of Eqs.~(\ref{eq:MultExpF1})-(\ref{eq:MultExpF4})  with the L+P parameterization given in Eq~(\ref{L+P-general}), for each multipole, there exists a well-defined system of equations ready to be fitted to data, using formulas given in Table.~\ref{tab:PhotoproductionObservables}.

In principle, the formalism is well defined with a clear advantage that the pole fitting parameters are physical quantities for which, from other processes and/or other analyses, one has approximate values. In addition, one knows which branch-points could be important for a particular partial wave, and what the threshold behaviour should be.
However, in reality, fitting the full database with the number of free parameters listed above is not straightforward. For a typical fit of a multipole in pseudoscalar photoproduction in Refs.\cite{\LP},  three Pietarinen expansions were used with up to 6 terms each, and up to three resonances. That amounts to as many as  27+12=39 parameters per multipole.  For a realistic fit, one has to include waves up to least to $L=5$. Knowing that one has to fit $J=L+1/2$, and $J=L-1/2$ electric and magnetic multipoles, this requires 20 multipoles (recalling that $M_{0+}$ and $E_{1-}$ multipoles are unphysical) to get a good fit. Taking into account that certain multipoles couple to the same $J^P$ ($E_{L \pm}$ and $M_{L \pm}$), and hence have the same pole position, one is still left with typically 800 free parameters. There does exist a sufficient number of measurements in the data base to obtain a reliable fit (approximately 8000 data points, including many single- and double-polarization quantities), but the complicated non-linear structure of the fitted formulas results~\cite{CPU} in Mathematica run times measured in CPU days. Recalling that one goal is a useful tool for experimentalists, a less CPU-intensive problem was solved for the present proof-of-principle study.

In the simplified L+P PWA, one Pietarinen expansion per multipole was used, instead of three, retaining the dominant background contribution associated with the $K \Lambda$ threshold. The inclusion of resonances was restricted to 4* states quoted in the baryon summary tables of the PDG~\cite{PDG}. This simplifies the background, a quantity which is generally the hardest to fully calculate, and which is usually simplified in a model.  The pole complexity remains unchanged. The number of poles contributing to a multipole is generally less than three; often
only a single pole is included in Eq.(\ref{L+P-general}). This simplified form of Eq.(\ref{L+P-general}) is used for
$T \in \{E_{\ell+},E_{\ell-},M_{\ell+},M_{\ell-} \}$.

With these approximations, the number of multipole background parameters was reduced from 27 to 9 and, with the reduction of pole terms, the number of utilized parameters was about 300. To further accelerate testing, an interpolation method was used on the database.
  The fit at interpolated values of all observables was performed, using 36 interpolated instead of 121 measured energy points. The interpolation was done in both variables (energy and angle) simultaneously, as described in Refs.~\cite{\AA/PWA}. This reduced the number of minimization points from 8000 to about 3000, reducing fit times to a manageable 12 CPU hours.
\\ \\ \indent
\subsection{Combining Simplified Energy-Dependent and
Single-Energy Partial-Wave Analysis}
  Experience gained in Refs.\cite{\AA/PWA}  was used to estimate how much the simplified L+P PWA actually differed from the full solution. In these previous studies, the best single-channel single-energy multipoles were obtained using constrained PWA in a two-step procedure. In the first step, we obtained reaction amplitudes which  fit the data at all energies where measurements were done, and these amplitudes were taken  as constraining functions for a constrained SE PWA in the second step. The essential equations governing our two-step amplitude analysis (AA) method from Ref.~\cite{Svarc2020,Svarc2022} are given below.

  A standard approach of any constrained partial wave analysis is to penalize partial waves that stray too far away from each other, while simultaneously fitting the set of measured observables:

\be
\label{Eq1}
\chi^2(W) & = & \sum_{i=1}^{N_{data}}w^i \left[ {\cal O}^{exp}_i (W,\Theta_i) - {\cal O}^{th}_i ({\cal M}^{fit}(W),\Theta_i) \right]^2 + \lambda_{pen} \sum_{j=1}^{N_{mult}} | {\cal M}_j^{fit}(W)- {\cal M}_j^{th}(W) | ^2
\ee
where
\be
{\cal M} & \stackrel{def}{=} & \left\{ {\cal M}_0, {\cal M}_1, {\cal M}_2, ..., {\cal M}_{N_{mult}} \right\} \nonumber
\ee
$w_i$   is the statistical weight and $N_{mult}$ is the number of  partial waves (multipoles).
Here  ${\cal M}^{fit}$  are fitting parameters and ${\cal M}^{th}$ are continuous functions taken from a particular theoretical model; ${\cal O}^{exp}$ and ${\cal O}^{th}$ are respectively the measured and calculated observables.

The possibility to make the penalty function
independent of a particular model was first formulated in the Karlsruhe-Helsinki
elastic pion-nucleon scattering analysis, by G.~H\"ohler in the 1980s. There,
partial waves, which are inherently model-dependent, are replaced by a penalty
function constructed from reaction amplitudes which can be more directly linked to
experimental data without any model in the amplitude reconstruction procedure, a point revisited in the next section. This leads to a change from
Eq.(\ref{Eq1} to:
\be
\label{Eq2}
\chi^2(W) & = & \sum_{i=1}^{N_{data}}w^i \left[ {\cal O}^{exp}_i (W,\Theta_i) - {\cal O}^{fit}_i ({\cal M}^{fit}(W),\Theta_i) \right]^2
+ \cal{P}  \\ \nonumber
 \cal{P}   &=& \lambda_{pen} \sum_{i=1}^{N_{data}} \sum_{k=1}^{N_{amp}}\left| {\cal A}_k({\cal M}^{fit}(W),\Theta_i)- {\cal A}_k({\cal M}^{pen}(W),\Theta_i) \right|^2
\ee
where ${\cal A}_k$  is the generic name for a reaction amplitude (invariant, helicity, or transversity) which is a function of multipoles $\cal{M}$ and angles $\Theta_i$.  In the present study, we take the penalty functions  ${\cal A}_k({\cal M}^{pen}(W),\Theta_i)$ to be transversity amplitudes, obtained using multipoles from the simplified L+P PWA.

  The constraining function in the second step serves
  to avoid the continuum ambiguity, which makes unconstrained SE PWA  discontinuous, by stabilizing the phase. It need not be exact; it is enough that it is smooth, and close to the true value. To determine how the simplified L+P PWA differs from the full solution, the formalism developed in  Refs.~\cite{\AA/PWA} was modified  by replacing SE amplitude-analysis constraining functions with the simplified L+P PWA amplitudes, which are energy-dependent and smooth, but do not reproduce the experimental data exactly.  Then, using the proposed formalism, we obtain the best SE result. Differences between the energy-dependent (ED) constraining partial waves, obtained by a simplified L+P PWA, and the final SE result, obtained in a constrained PWA, reveal where the analytic simplifications used in simplified L+P PWA are too crude. This process could be repeated iteratively to produce an ED solution which is closer to the SE values.
\\ \\ \indent
\subsection{Remarks on Phase Ambiguity and Initial Parameters} \label{PhaseandInitial}

Observe that the overall energy- and angle-dependent phase remains undetermined in this method of finding a single-energy solution (see Ref.~\cite{Phase-ambiguity}).
No explicit constraint of the final fit phase to the phase of a chosen theoretical model, in this case the Bonn-Gatchina analysis~\cite{Anisovich2017,Anisovich2017a}, was used. However, the initial phase enters through the choice of initial parameters for the simplified L+P fit. In spite of the fact that one has prior knowledge of the pole parameters entering the L+P fit (it is assumed that they are not far from the  PDG~\cite{PDG} values), the background parameters (thresholds and Pietarinen coefficients) are numerous and less-well determined.
Therefore, one has to determine the initial values of remaining Pietarinen coefficients in some way to constrain the phase, otherwise the fit would have the tendency to produce a final phase which disagrees with the phase coming from  multichannel unitarity\footnote{Multichannel unitarity determines all phases in all channels.}. This initial-phase information is taken from the transversity amplitudes corresponding to the Bonn-Gatchina multi-channel analysis~\cite{\BGED}.  These obtained coefficients (Pietarinen and pole parameters) were used as initial values in the ED L+P fit of experimental data. In this way one starts with a solution whose overall phase is close to the multi-channel phase of the BG model~\cite{Anisovich2017,Anisovich2017a}.
This phase will change in the final solution, but only as much as is needed to improve the fit to polarization data, which depend on the combination of individual reaction amplitude phases. As an aside, this is very similar to the phase treatment in the Karlsruhe-Helsinki fixed-t PWA~\cite{Hoehler,Osmanovic2018,Osmanovic2019,Osmanovic2021}. In that fit to elastic pion-nucleon scattering, the overall phase is not mentioned, but is implicitly introduced as the phase of the solution whose Pietarinen parameters are taken as starting values in a fixed-t analysis of data.

\subsection{The $\gamma p \to K^+ \Lambda$ Data Base} \label{Data:base}
The $\gamma p\to K^+ \Lambda$ data base, used in this study, is identical to one fitted in Ref.~\cite{Svarc2022}.
In Table~\ref{tab:expdata} our data base is summarized. It has been taken, in numeric form, from the Bonn-Gatchina and George-Washington-University (SAID) web pages~\cite{BG-web,GWU-web}:
\begin{table*}[htb]
\begin{center}

\caption{\label{tab:expdata} Experimental data from CLAS, and GRAAL used in our PWA. Note that the observables $C_x$ and $C_z$
are measured in a rotated coordinate frame~\cite{Bradford}. They are related to the standard observables $C_{x'}$ and $C_{z'}$ in
the $c.m.$ frame by an angular rotation: $C_x= C_{z'} \sin(\theta)+ C_{x'} \cos(\theta)$, and $C_z= C_{z'} \cos(\theta)- C_{x'}
\sin(\theta)$, see Ref.~\cite{Anisovich2017a}. }
\bigskip
\begin{ruledtabular}
\begin{tabular}{ccccccc}
 Obs.        & $N$ & $E_{c.m.}$~[MeV] & $N_E$  & $\theta_{cm}$~[deg] & $N_\theta$ & Reference    \\
\hline
 $d\sigma/d\Omega \equiv \sigma_0$ & $3615$ & $1625-2295$ & $268$  & $28 - 152$ & $5-19$ & CLAS(2007)~\cite{Bradford}, CLAS(2010)~\cite{McCracken} \\
 $\Sigma$   & $ 400$ & $1649 - 2179$ & $ 34$  & $35 - 143$ & $6-16$ & GRAAL(2007)~\cite{Lleres}, CLAS(2016)~\cite{Paterson} \\
 $T$        & $ 408$ & $1645 - 2179$ & $ 34$  & $31 - 142$ & $6-16$ &  GRAAL(2007)~\cite{Lleres},CLAS(2016)~\cite{Paterson}  \\
 $P$        & $ 1597$ & $1625 - 2295$ & $ 78$  & $28 - 143$ & $6-18$ & CLAS(2010)~\cite{McCracken},  GRAAL(2007)~\cite{Lleres} \\
 $O_{x'}$   & $ 415$   & $1645 - 2179$ & $ 34 $  & $31 - 143$ & $6-16$ & GRAAL(2007)~\cite{Lleres}, CLAS(2016)~\cite{Paterson} \\
 $O_{z'}$   & $ 415$  & $1645 - 2179 $ & $ 34 $  & $31 - 143$ & $6-16$ &  GRAAL(2007)~\cite{Lleres}, CLAS(2016)~\cite{Paterson} \\
 $C_x$     & $ 138$  & $1678 - 2296$ & $ 14 $  & $31 - 139$ & $9 $ &  CLAS(2007)~\cite{Bradford} \\
 $C_z$      & $ 138 $ & $1678 - 2296 $ & $ 14 $  & $31 - 139$ & $9$ &  CLAS(2007)~\cite{Bradford}
\end{tabular}
\end{ruledtabular}
\end{center}
\end{table*}
For general details related to the 2-dimensional interpolation and its implementation, see Ref.~\cite{\AA/PWA}. However, the interpolating/extrapolating stability in the present study is significantly improved with respect to Ref.~\cite{\AA/PWA}. Observe that, in angular range, not all observables overlap, and for some data groups extrapolations are needed. However, this extrapolation at extreme forward and backward angles can become rather ambiguous if it is completely determined by the fitting software. Therefore, we have introduced additional kinematical constraints to the measured data at the angular limits:
\be
\Sigma = P = T = O_{x'} = O _{z'}  =  C_{x} = 0 \hspace{0.7cm}  \& \hspace{0.7cm} C_{z} = 1 \hspace{0.7cm} {\rm at} \hspace{0.7cm}  \cos \theta  = \pm 1
\ee
For the differential cross section $d\sigma/d\Omega$, the Bonn-Gatchina theoretical values were used as a constraint at these angles.
\\ \noindent
This stabilizes the extrapolations at low and high angles significantly, and enables us to increase the angular range from experimentally measured -0.7 $< \cos \theta <$ 0.8 to a broader -0.9 $ < \cos \theta <$ 0.9, and this notably increases the reliability of partial wave reconstruction.

\subsection{The Fit Procedure} \label{Procedure}
As discussed above, the proposed fit procedure constitutes a proof-of-principle study, utilizing fewer parameters than
have been used in previous pole extractions involving single multipoles.
However,  this simplification still retains a realistic complexity of the analytic structure, but reduces the number of free parameters. As background is concerned, the procedure takes only a single L+P expansion point, instead of three, and its branch-point is fixed to the pion production threshold.
Also, the intention was to keep the same number of poles as established in the multichannel fits of the Bonn-Gatchina group, and collected by the PDG~\cite{PDG}. As a large number of parameters is used, we have to carefully select initial parameters for poles and Pietarinen coefficients, essential for the success of the fit. To accomplish this,  Bonn-Gatchina transversity amplitudes were fitted with the simplified analytic structure described above. To simplify the problem further, all poles which showed signs of having little influence, or even being redundant\footnote{The poles which Bonn-Gatchina reports are the result of a multi-channel fit, so it is a realistic possibility that some of these poles couple weakly to $K \Lambda$ channel, so their influence in this channel could be is low.} (negative width, small residue...) were disregarded. Very good and stable results in this preliminary fitting of Bonn-Gatchina results were obtained. The obtained parameters, which almost perfectly reproduce Bonn-Gatchina multipoles, were used as initial values in the simplified L+P PWA. The fit to the described experimental data base, in the energy range \mbox{1625 MeV $< W <$  2296 MeV} and angular range  \mbox{-0.9 $ < \cos \theta <$ 0.9}, required acceptable CPU times of about 12-14 hours.  The result of this simplified L+P minimization is a set of multipoles which fit the data better than the ED theoretical model, and are by definition smooth. In order to check the influence of simplifications to the L+P analysis, the procedure of single-channel and single-energy (SE) AA/PWA, defined in Ref.~\cite{\AA/PWA}, was applied and followed directly. Transversity amplitudes were obtained from the smooth multipoles of the simplified L+P PWA, as given by Eqs.~(\ref{eq:MultExpF1})-(\ref{eq:b4BasicForm}), and used as a constraining functions in the constrained SE PWA. The penalty factor $\lambda$ introduced in Eq.~\ref{Eq2} (for a full explanation see Refs.~\cite{\AA/PWA}\footnote{Choosing the value of the penalty factor is a delicate procedure. On one hand we require that it is large enough to ensure that our SE solution does not jump between two neighbouring solutions, but it must be small enough to avoid a larger influence which would bias the final result.}) is {optimized in a way to ensure the penalty function contribution to the total $\chi^2$ is no larger than 10 \% }, and is set to $\lambda=150$.  The final result, which is in quality identical to the result given in Ref.~\cite{\AA/PWA}, was obtained. A slightly different phase was found, as the constraining functions do not have the exact Bonn-Gatchina phase as they did in Ref.~\cite{\AA/PWA}. In spite of being close to the smooth result of the simplified L+P minimization, the obtained discrete set of multipoles is an improvement, in terms of the agreement with data.  In comparing discrete multipoles of AA/PWA with the smooth values obtained in simplified L+P PWA (see Figs.~\ref{Multipoles:a} and \ref{Multipoles:b}) one sees that some discrete sets have more structure than the present simplified L+P version, so the simplified L+P PWA can be improved by performing the full L+P PWA of obtained SE multipoles with the formalism described in Refs.~\cite{\LP,\LPapplication} adding more structure exactly where needed. This result can then be taken as a complete L+P ED PWA solution of the problem.  In future studies, the development can be imagined where faster minimization software is used (MINUIT in FORTRAN 90 for example instead of presently used Mathematica 11.0 which is known not to be ideal for minimizations), and applied with improved hardware. This would reduce CPU time, and enable the introduction of more complex analytic forms (more poles, more Pietarinen expansions having more terms than were used in the simplified L+P expansion). In principle, the present iterative three-step
procedure could then be replaced by a more elaborate single-step L+P fit.

\section{Results from the Simplified LEP ED and SE AA/PWA Analyses}
\begin{figure}[h!]
\bc
\includegraphics[width=0.37\textwidth]{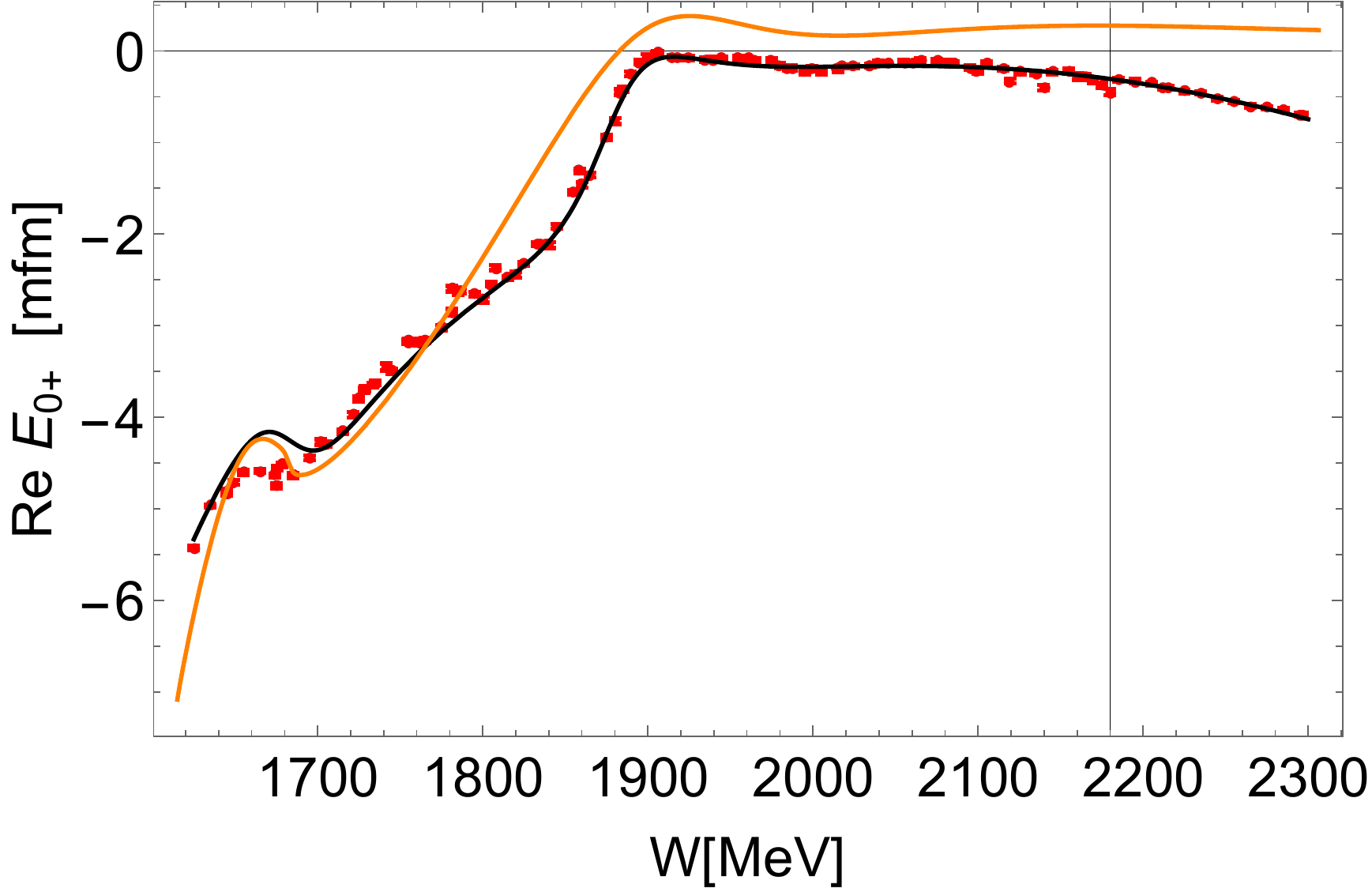} \hspace{0.5cm}
\includegraphics[width=0.37\textwidth]{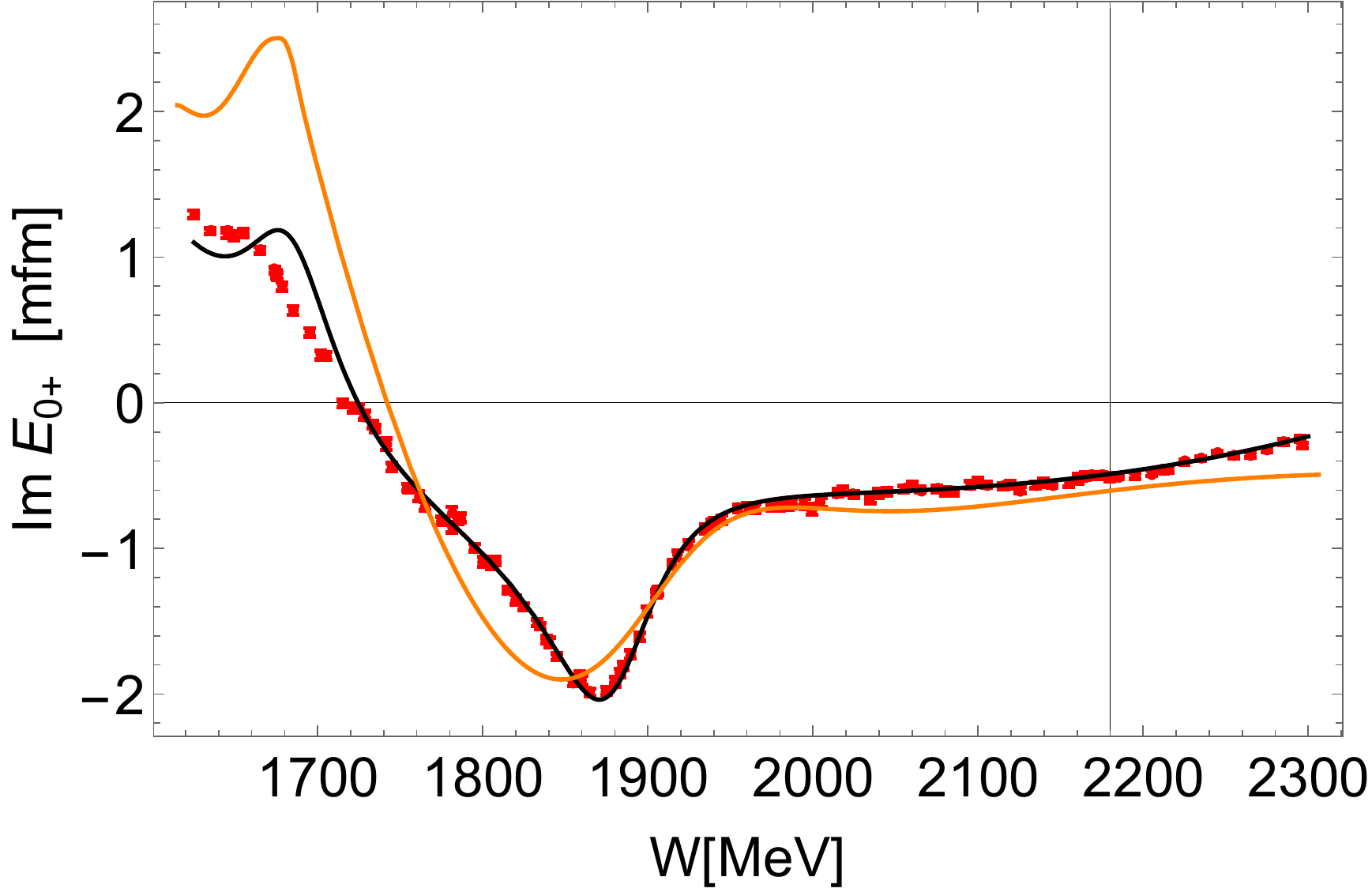}  \\
\includegraphics[width=0.37\textwidth]{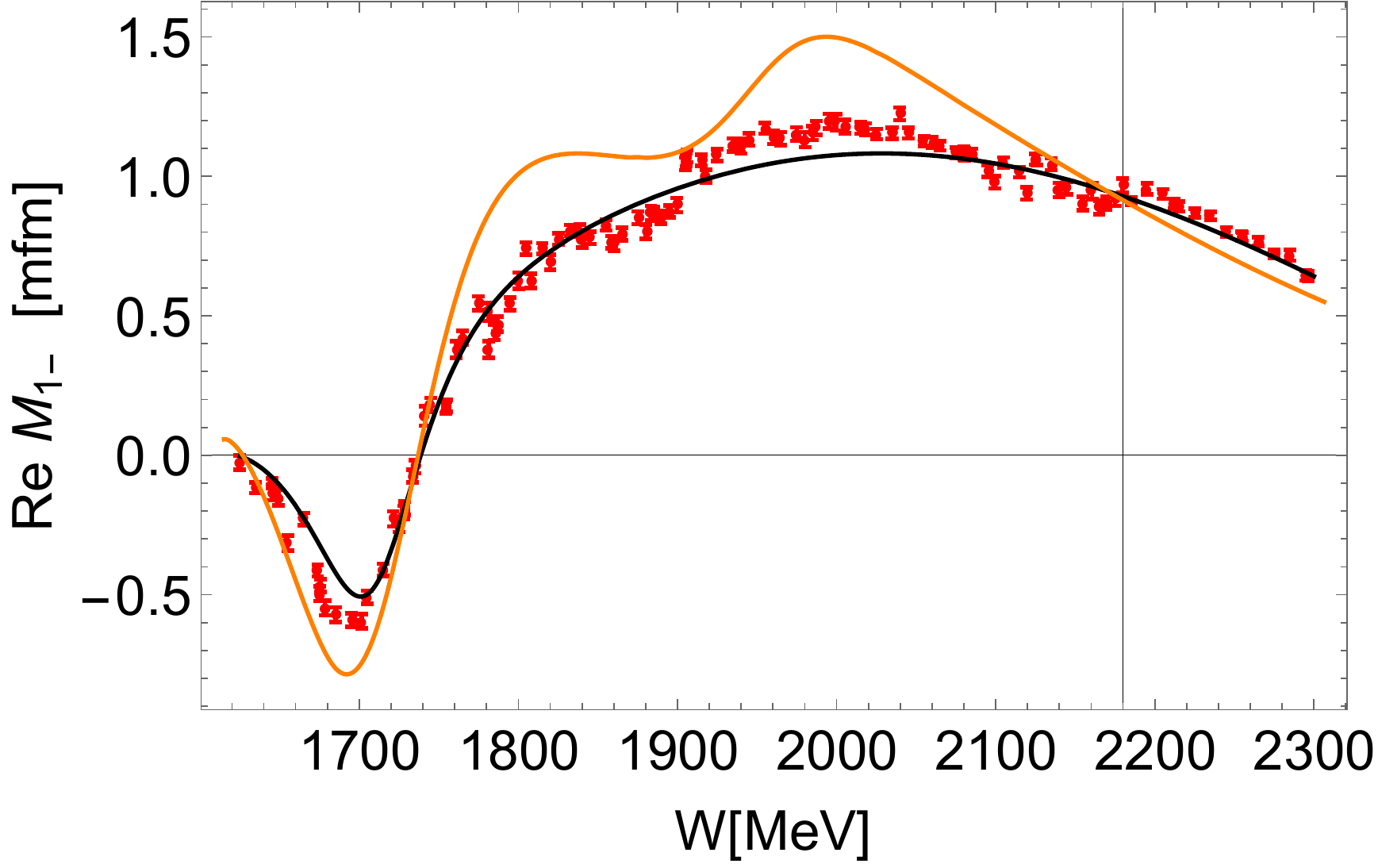} \hspace{0.5cm}
\includegraphics[width=0.37\textwidth]{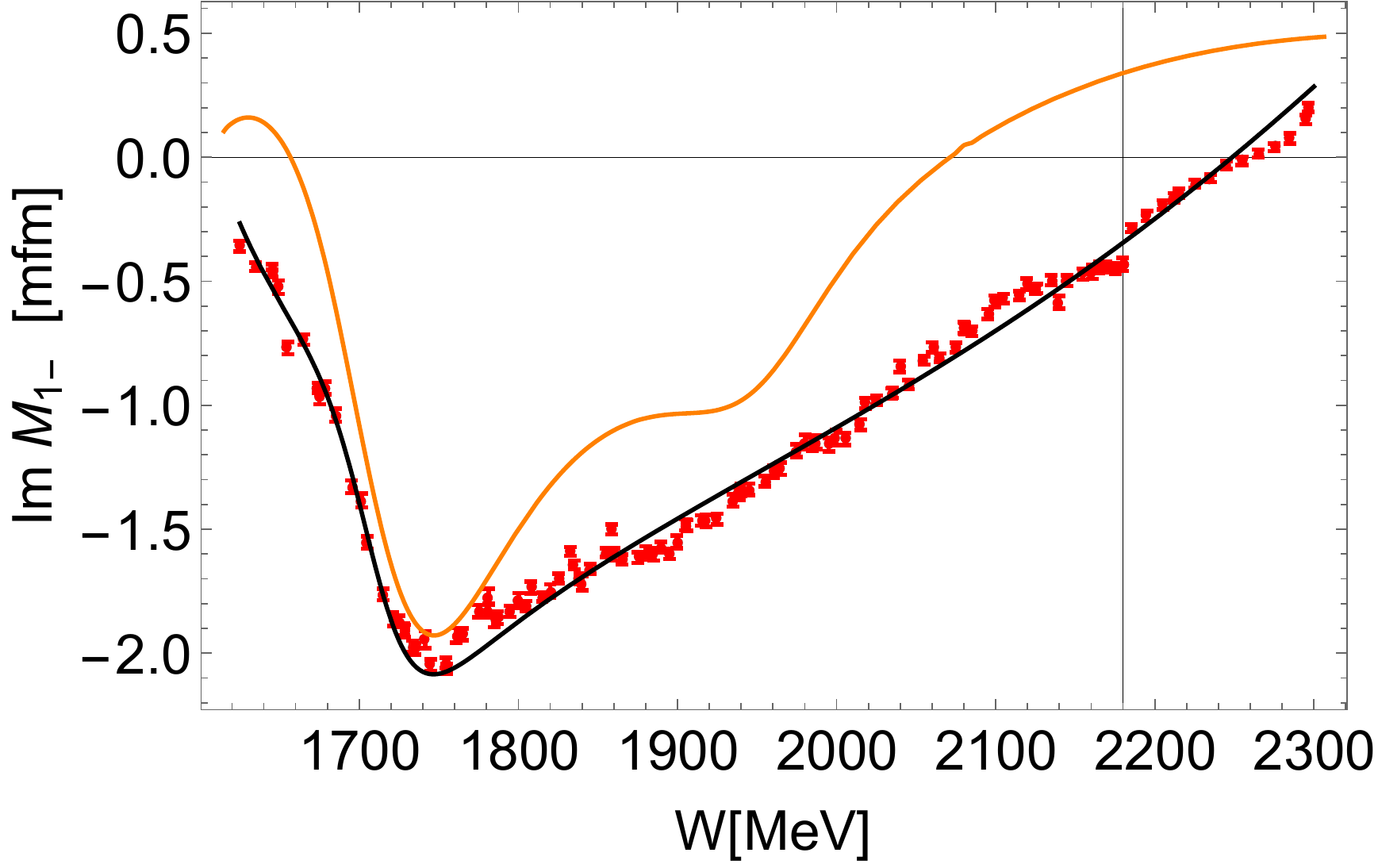}  \\
\includegraphics[width=0.37\textwidth]{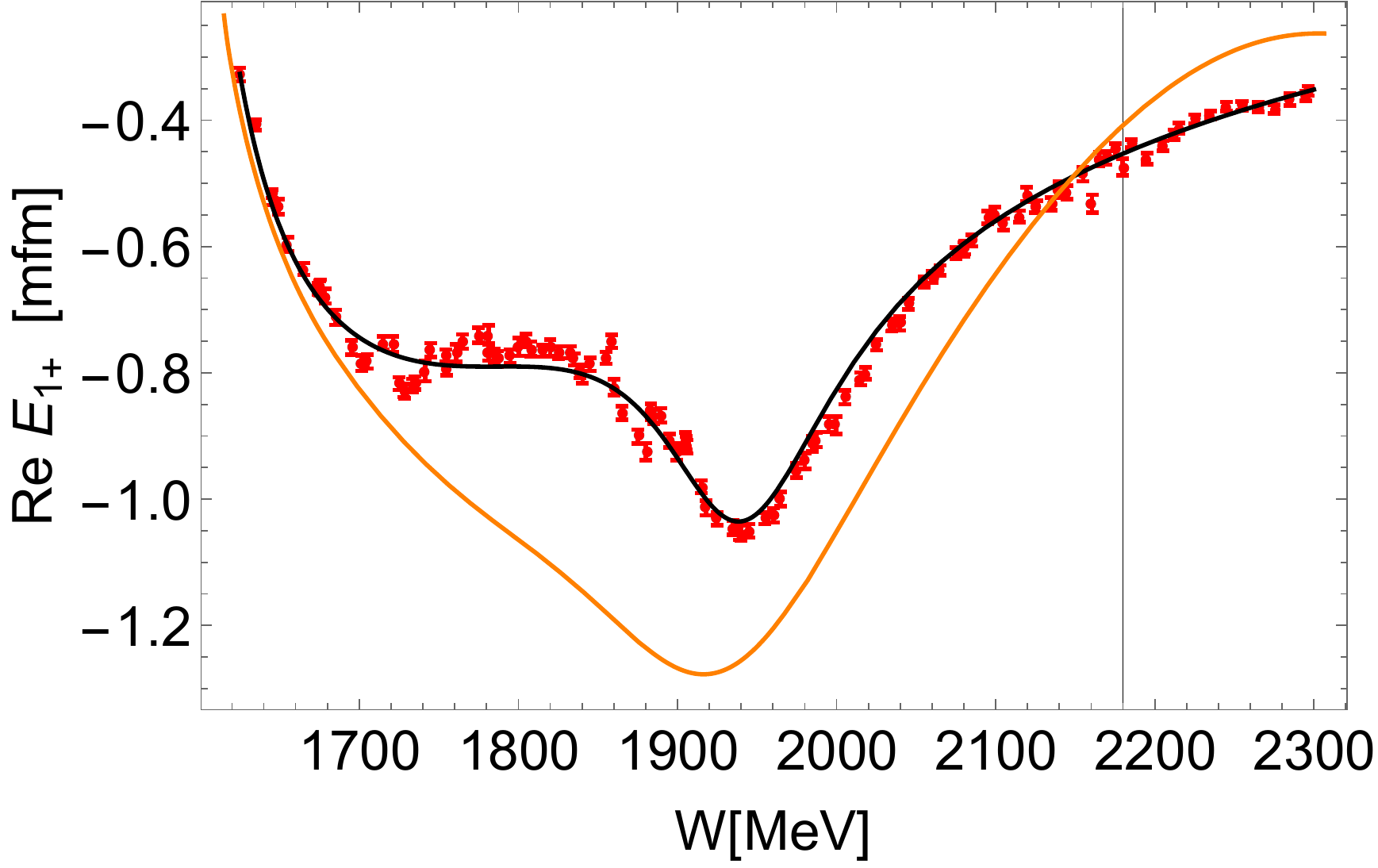} \hspace{0.5cm}
\includegraphics[width=0.37\textwidth]{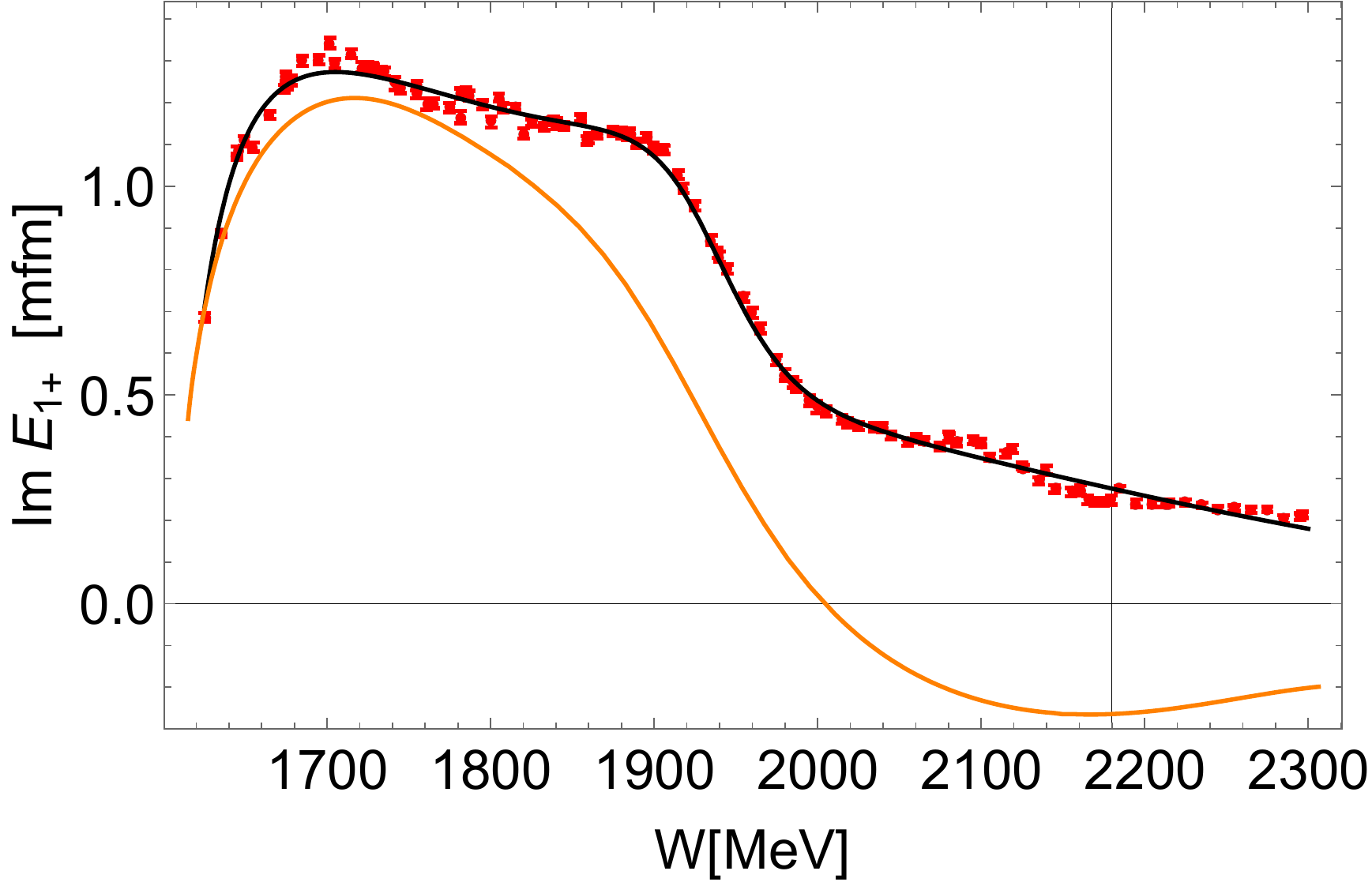}  \\
\includegraphics[width=0.37\textwidth]{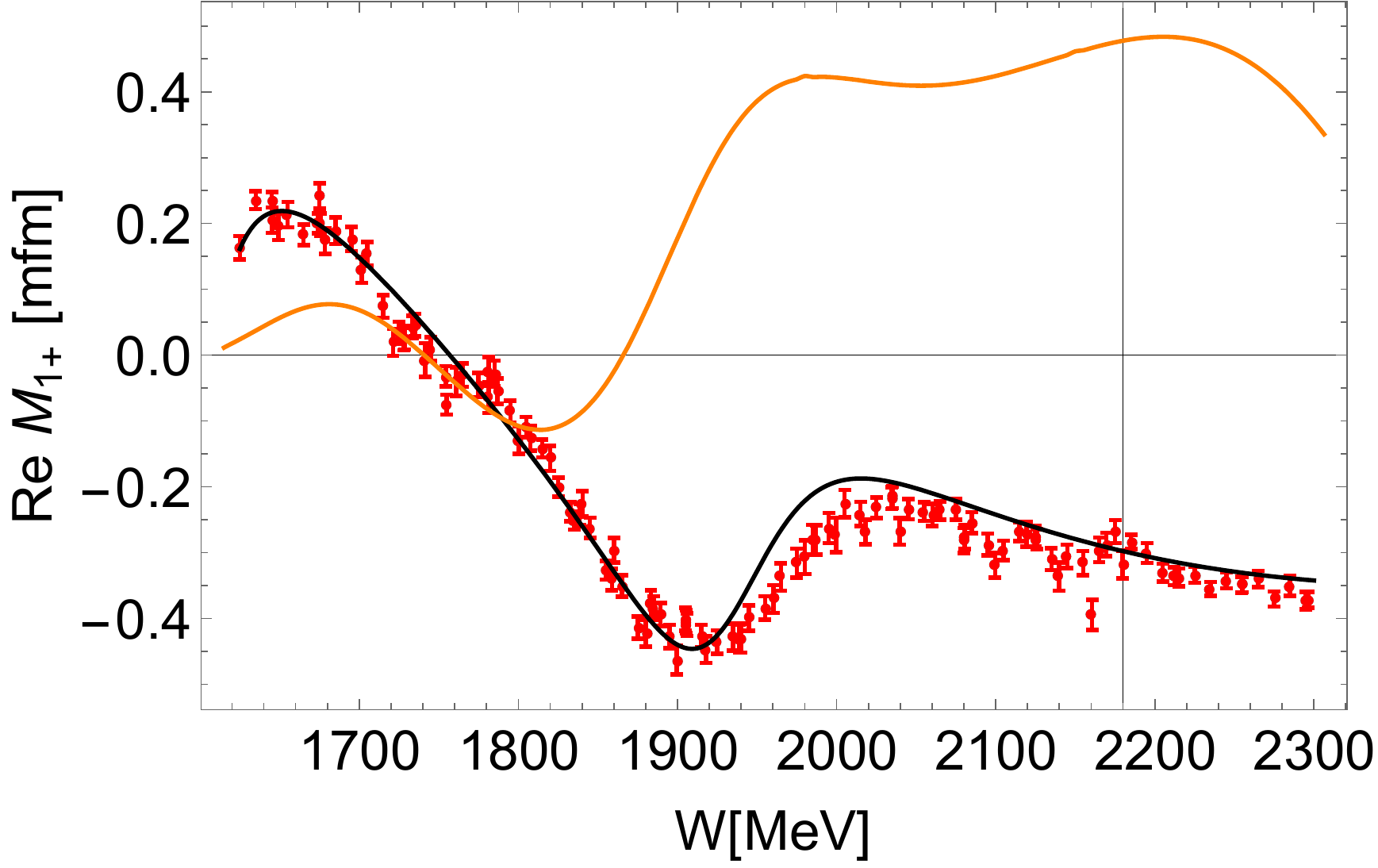} \hspace{0.5cm}
\includegraphics[width=0.37\textwidth]{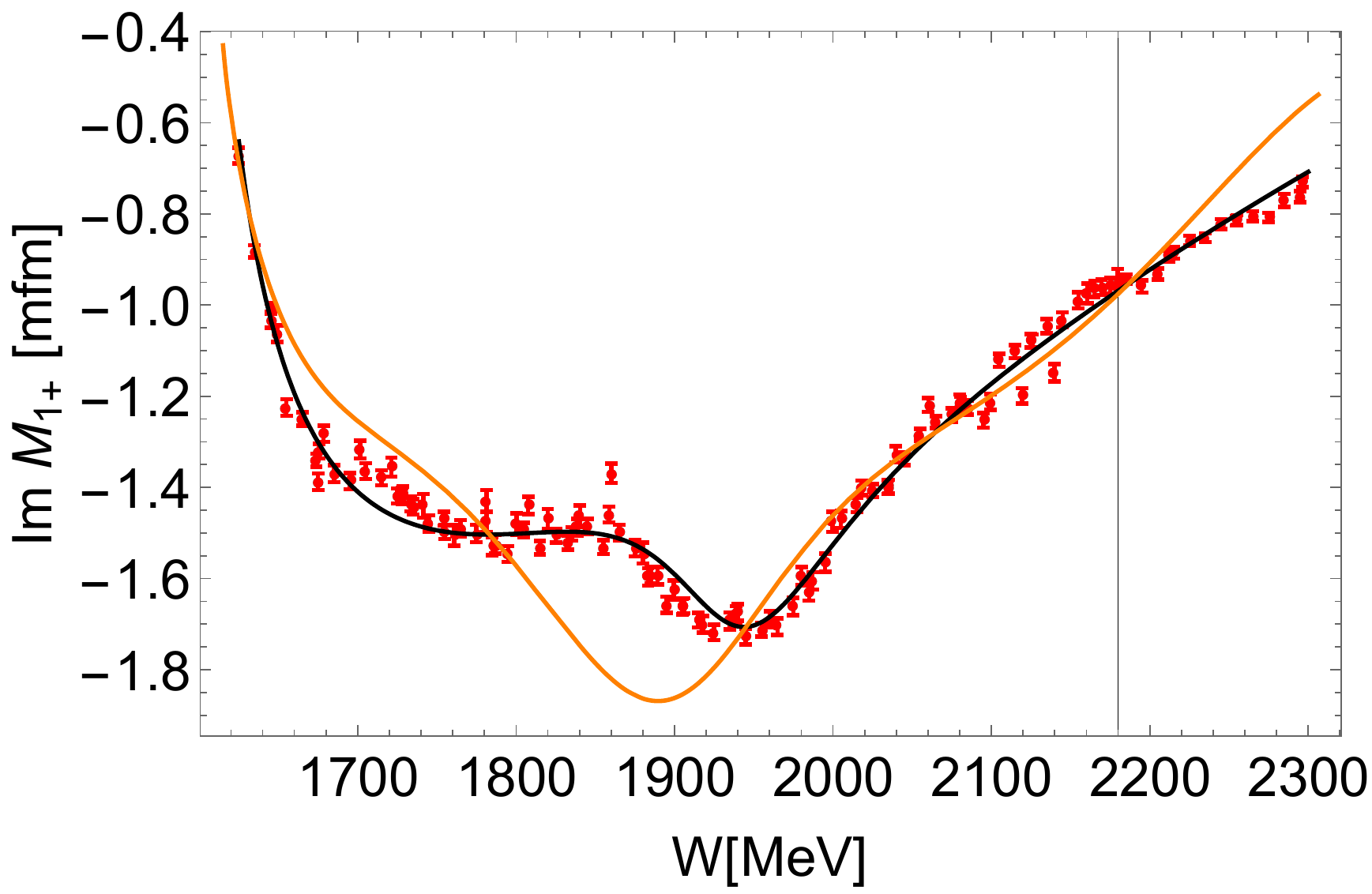}  \\
\includegraphics[width=0.37\textwidth]{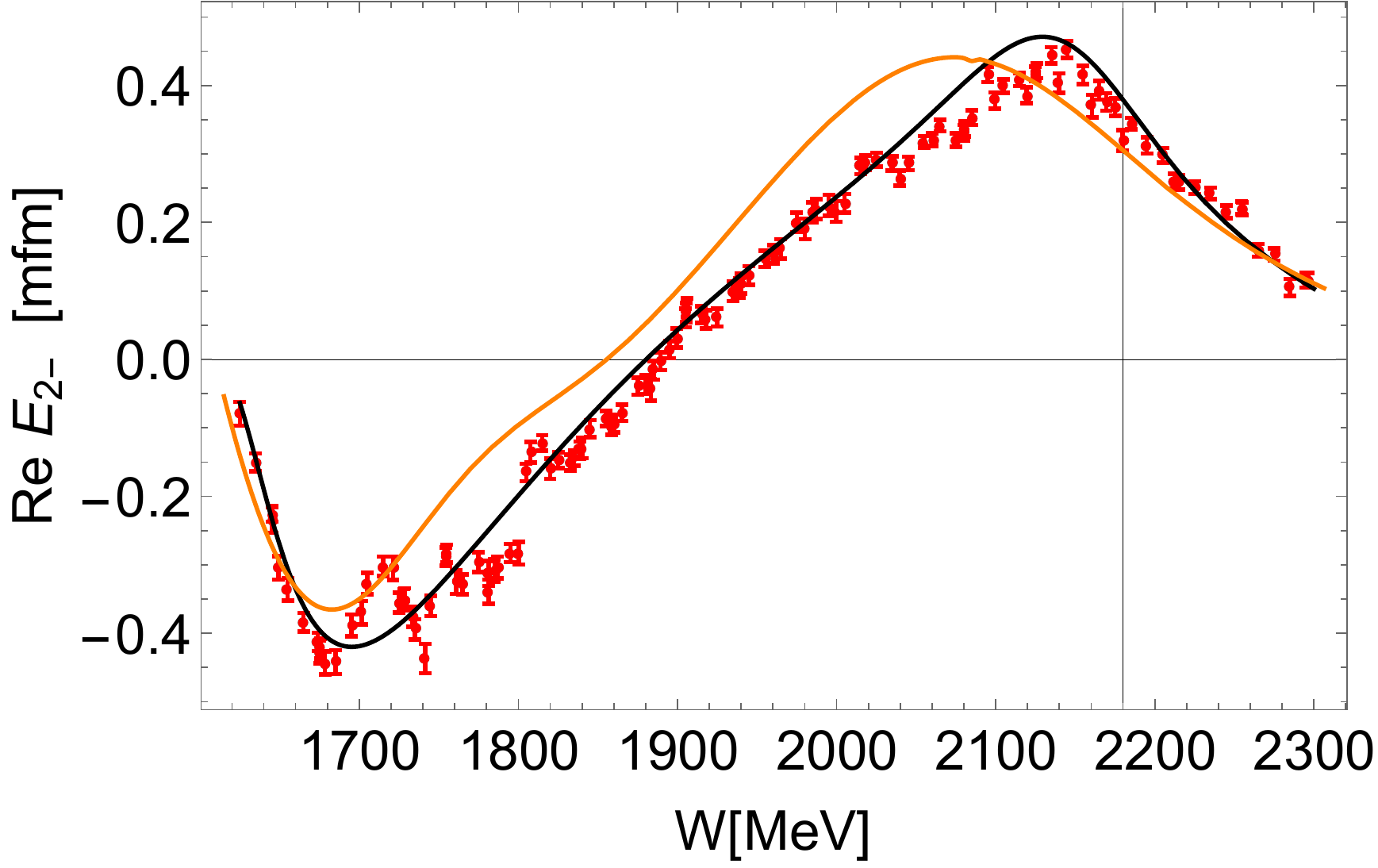} \hspace{0.5cm}
\includegraphics[width=0.37\textwidth]{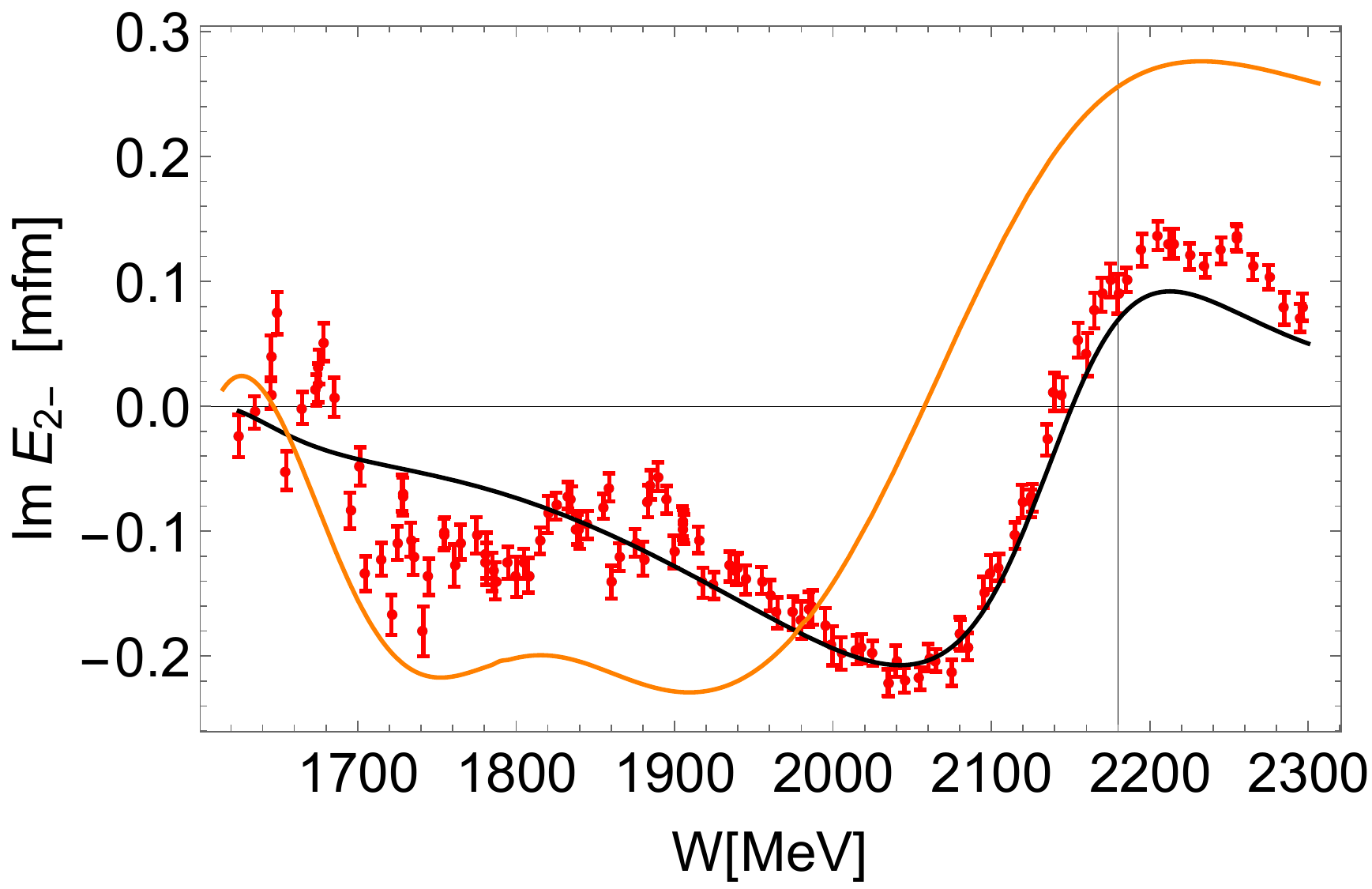}  \\
\caption{\label{Multipoles:a}(Color online) The multipoles for the $L=0$, $1$ and $2$ partial waves. Red discrete symbols correspond to our Step 2 SE AA/PWA solution, the black full line gives the result of the Step 1 L+P ED PWA, and the full orange line gives the BG2017 ED solution for comparison. The
thin vertical black line marks the energy beyond which only 4 observables are measured instead of 8 (cf. Table~\ref{tab:expdata}). } \ec
\end{figure}

\begin{figure}[h!]
\bc
\includegraphics[width=0.37\textwidth]{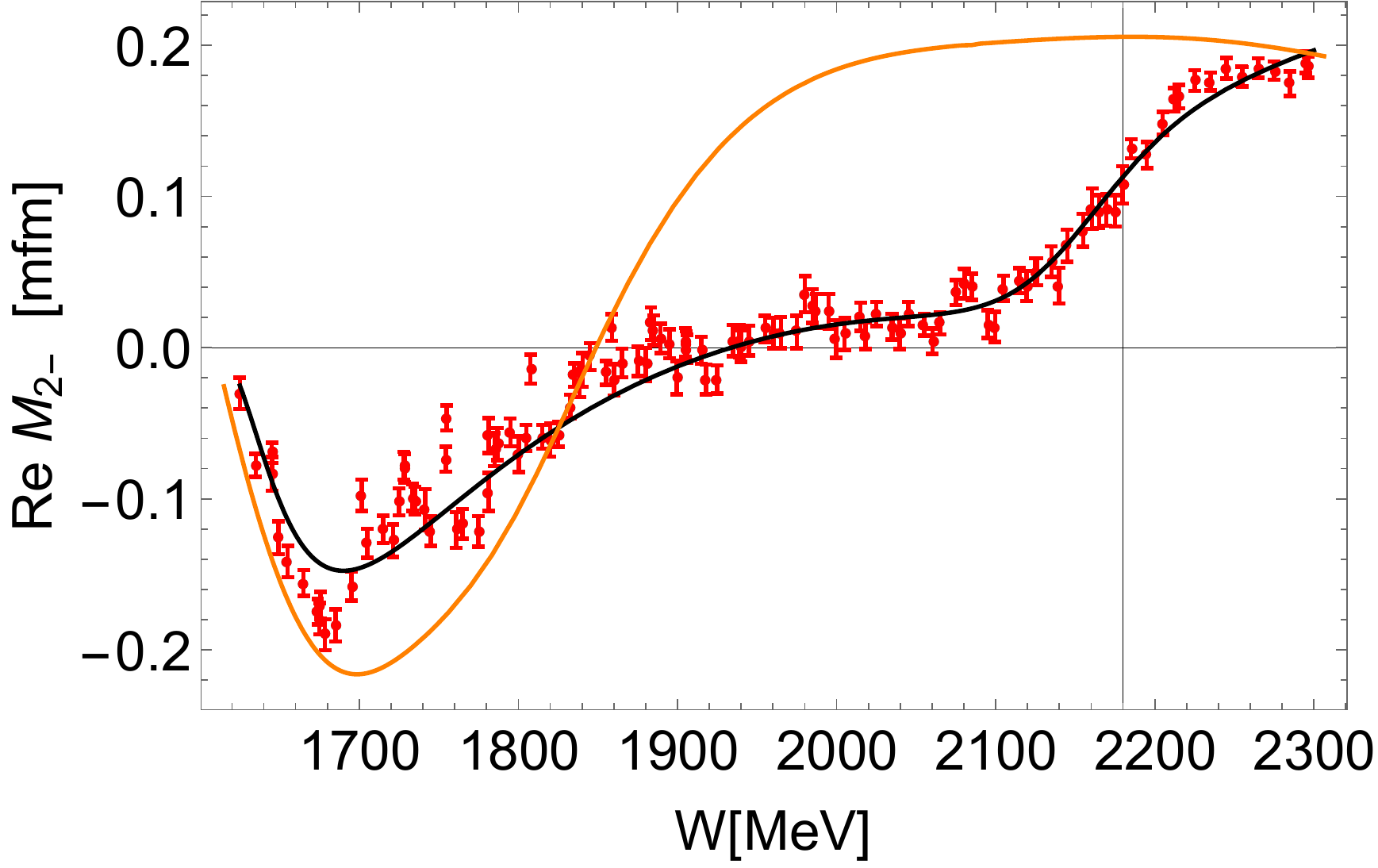} \hspace{0.5cm}
\includegraphics[width=0.37\textwidth]{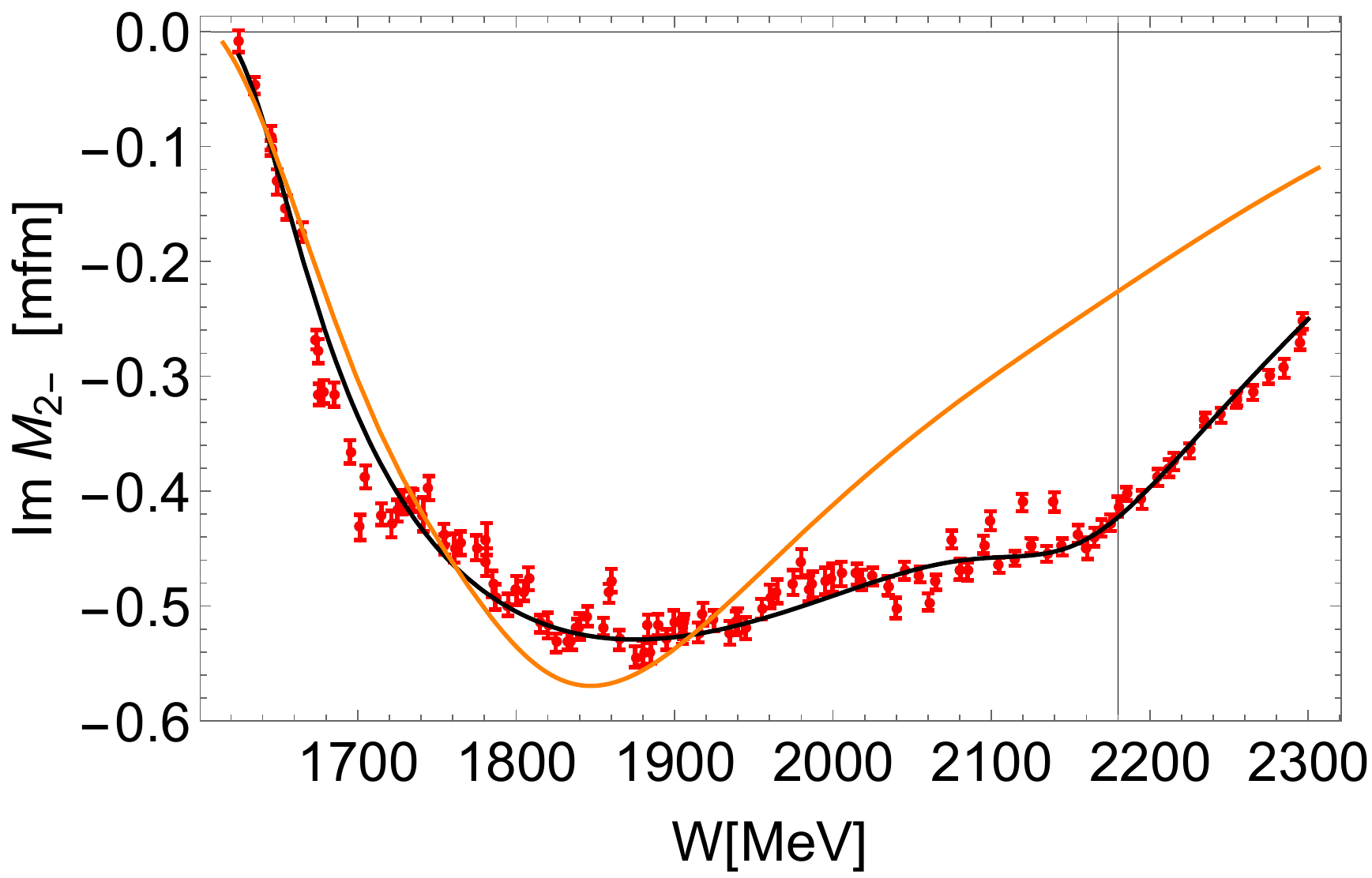}  \\
\includegraphics[width=0.37\textwidth]{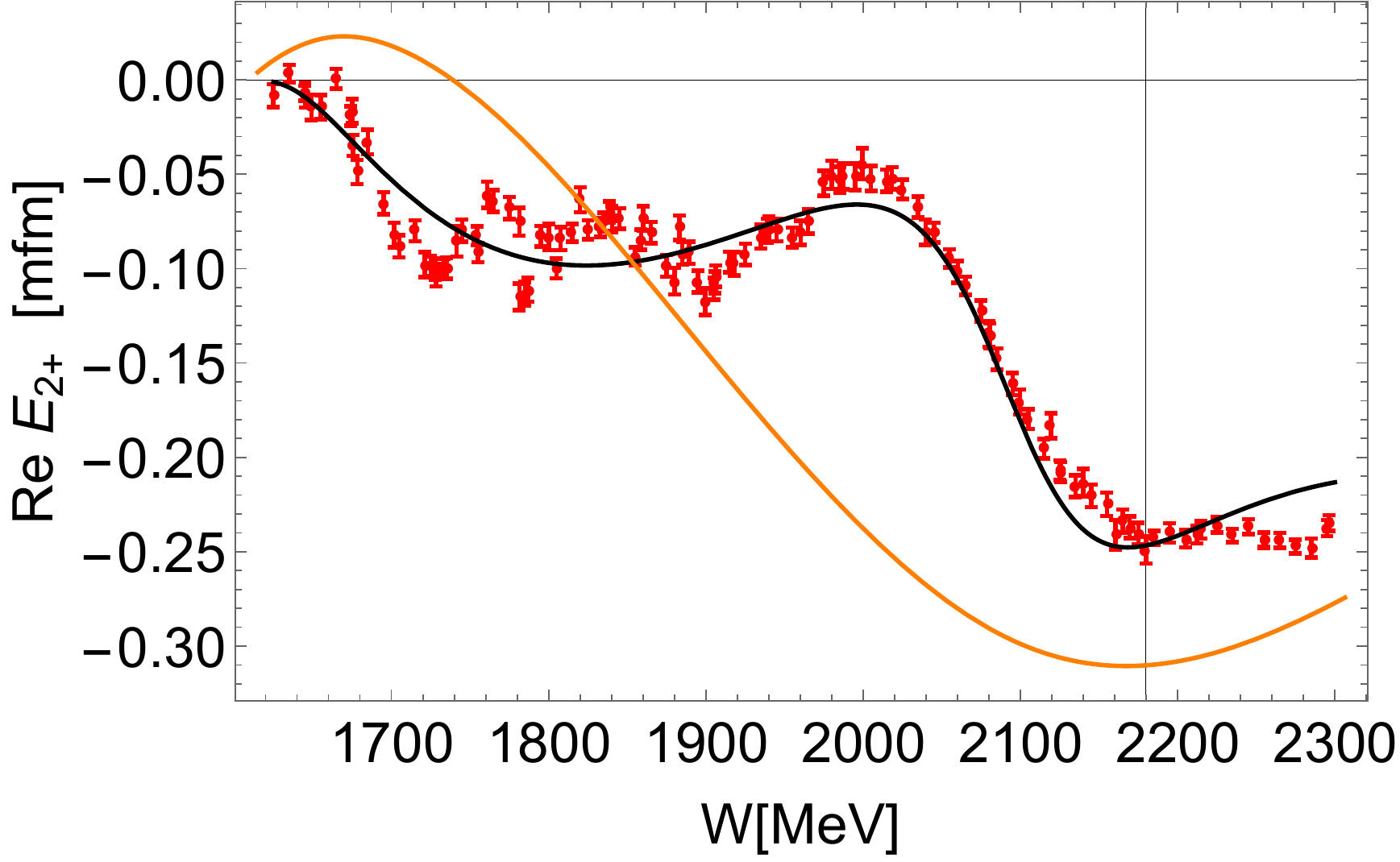} \hspace{0.5cm}
\includegraphics[width=0.37\textwidth]{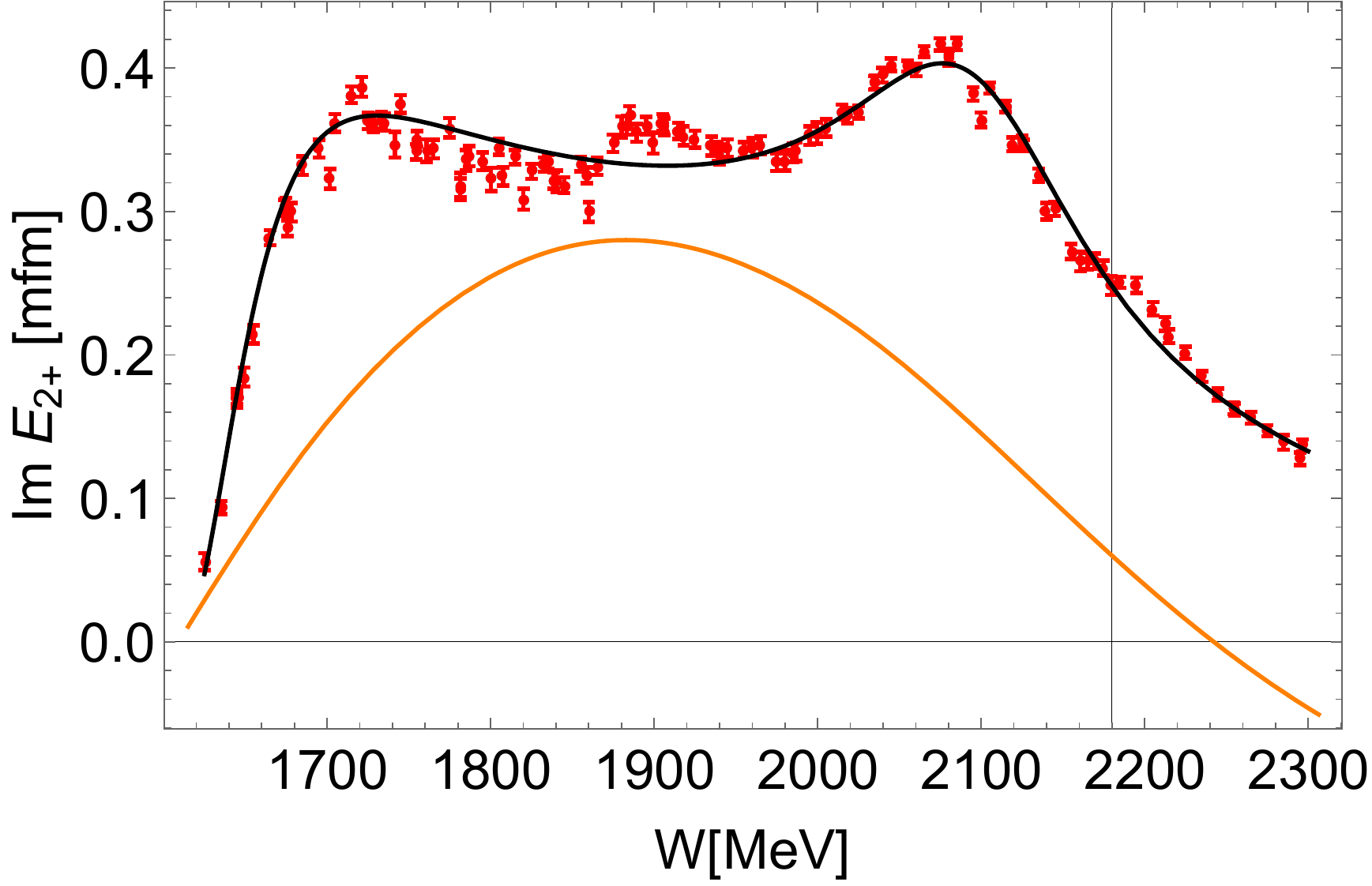}  \\
\includegraphics[width=0.37\textwidth]{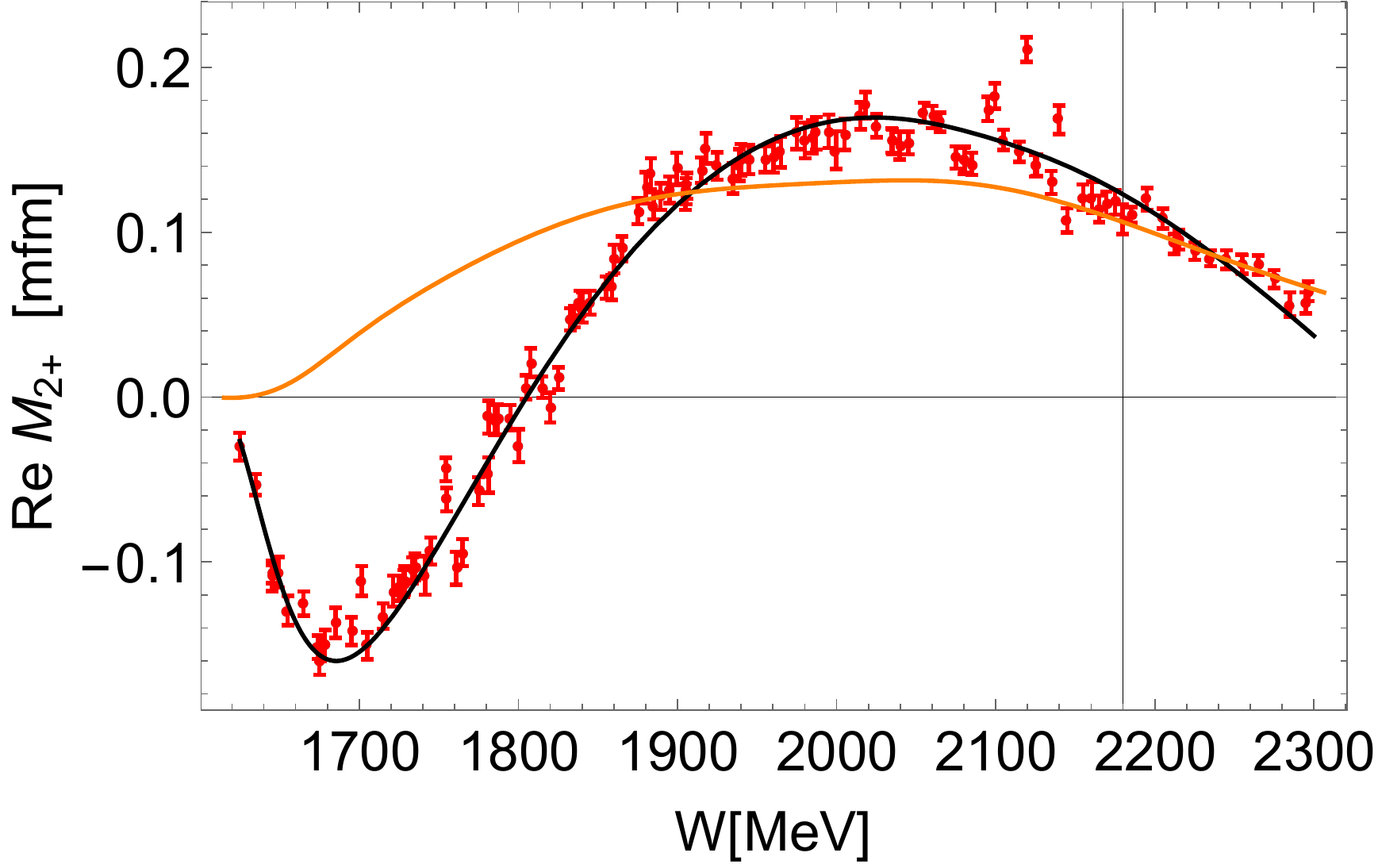} \hspace{0.5cm}
\includegraphics[width=0.37\textwidth]{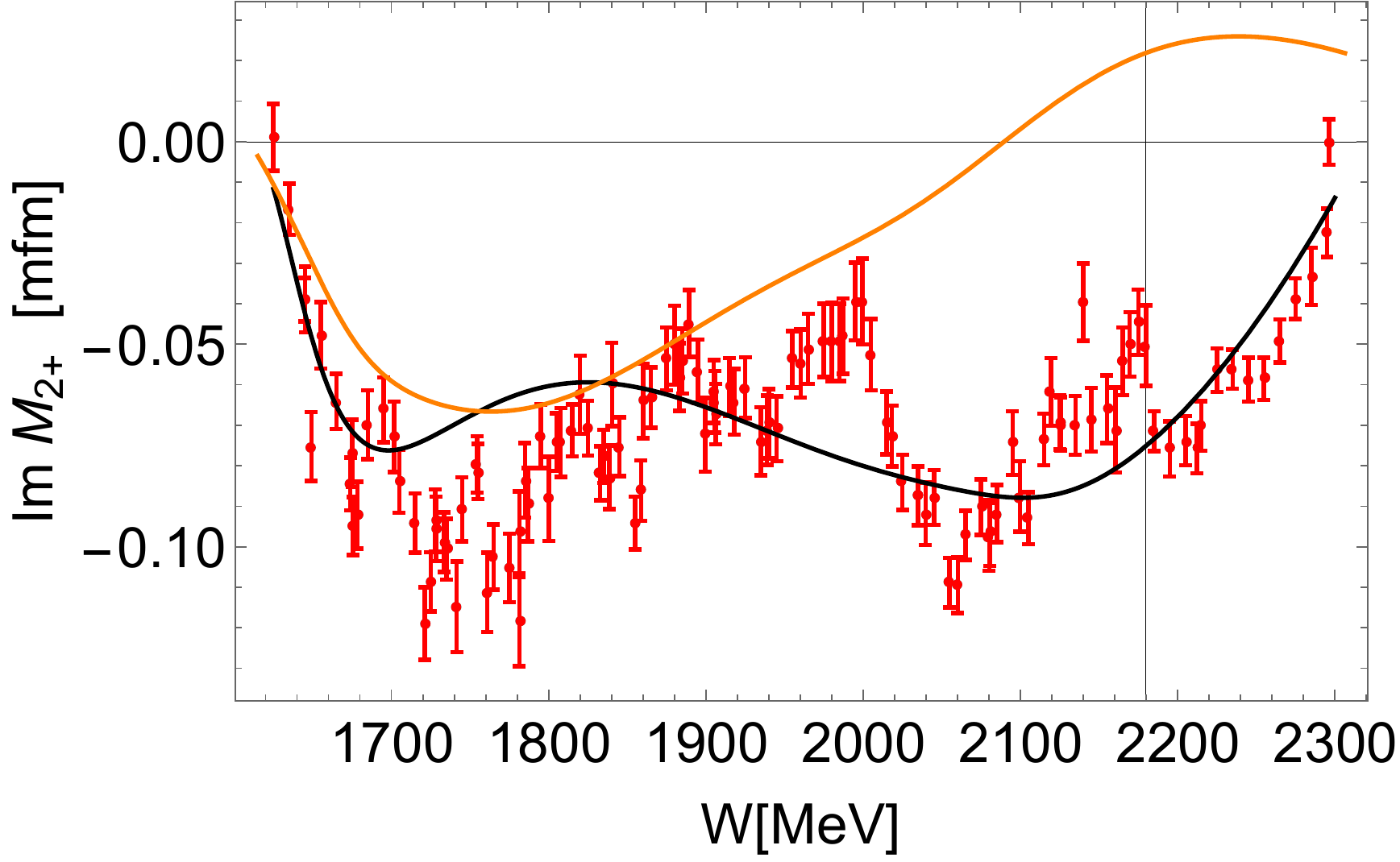}  \\
\includegraphics[width=0.37\textwidth]{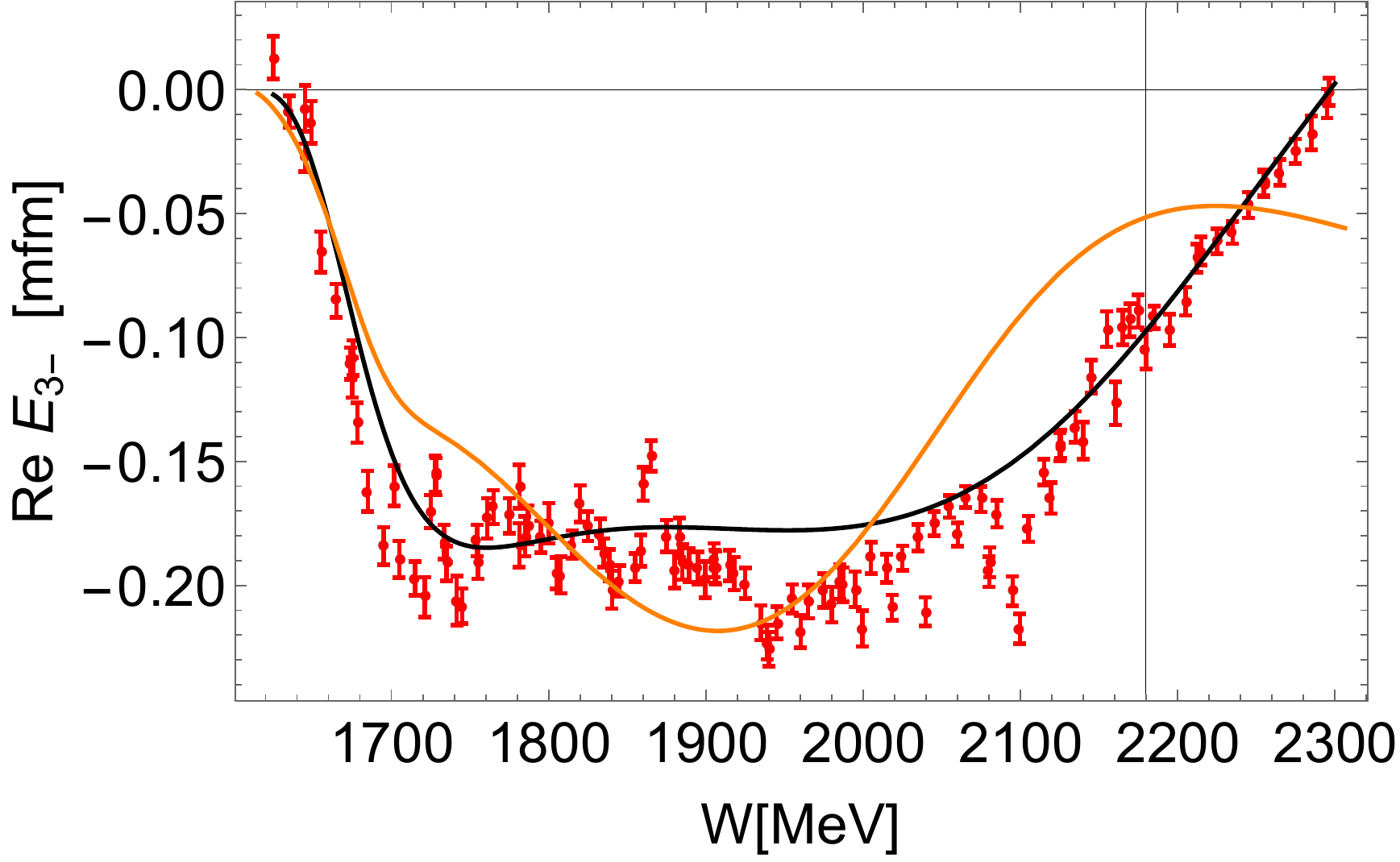} \hspace{0.5cm}
\includegraphics[width=0.37\textwidth]{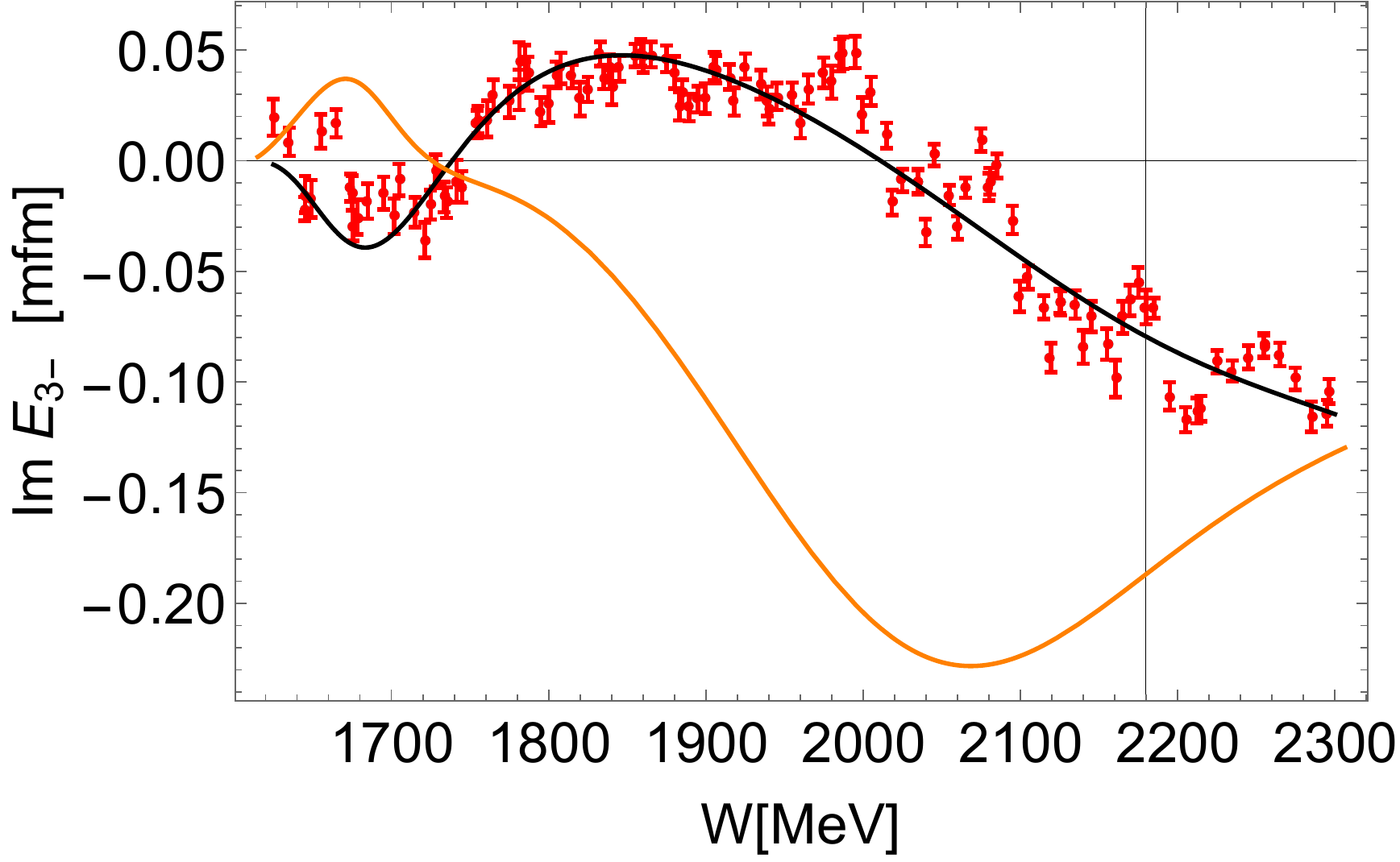}  \\
\includegraphics[width=0.37\textwidth]{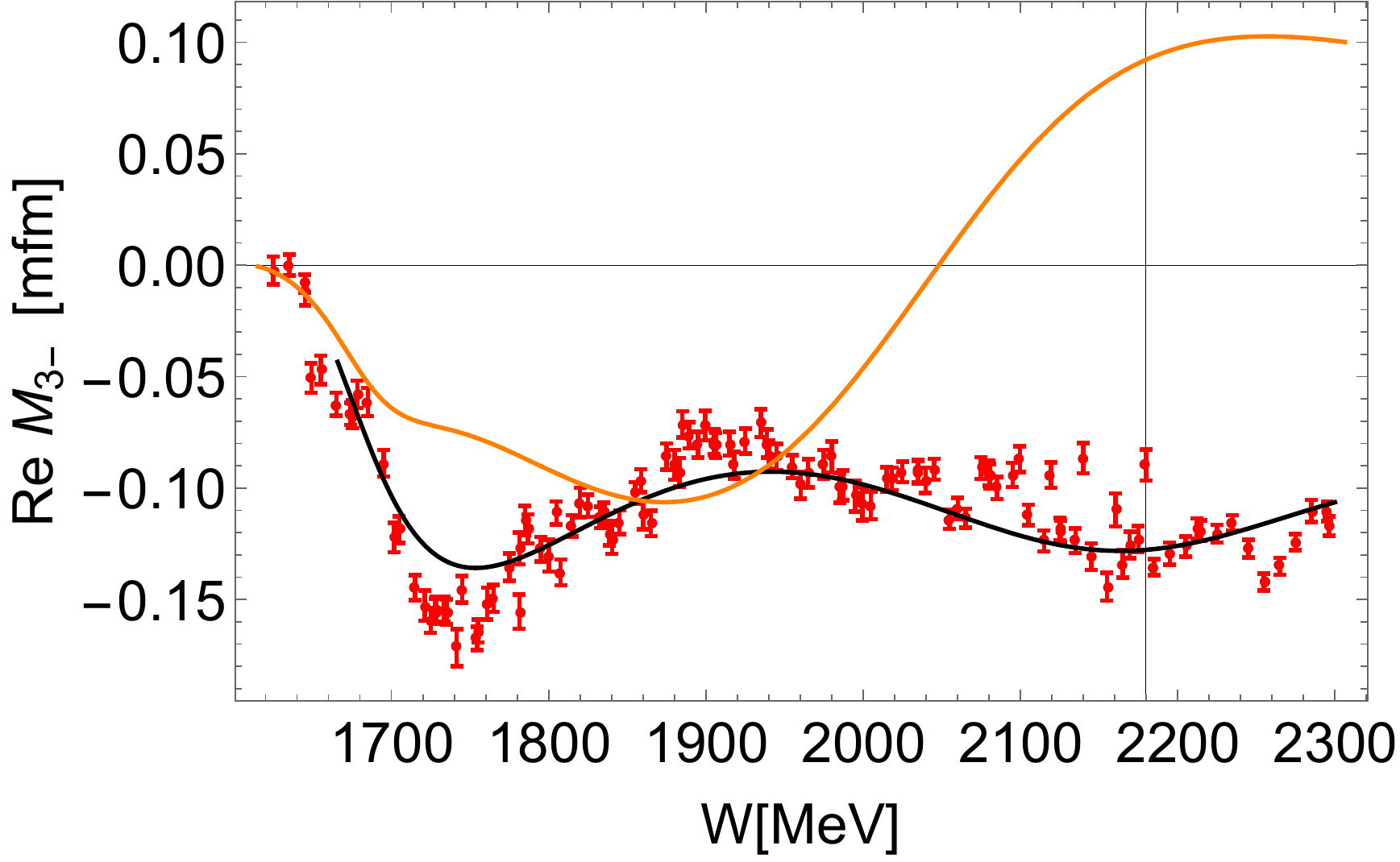} \hspace{0.5cm}
\includegraphics[width=0.37\textwidth]{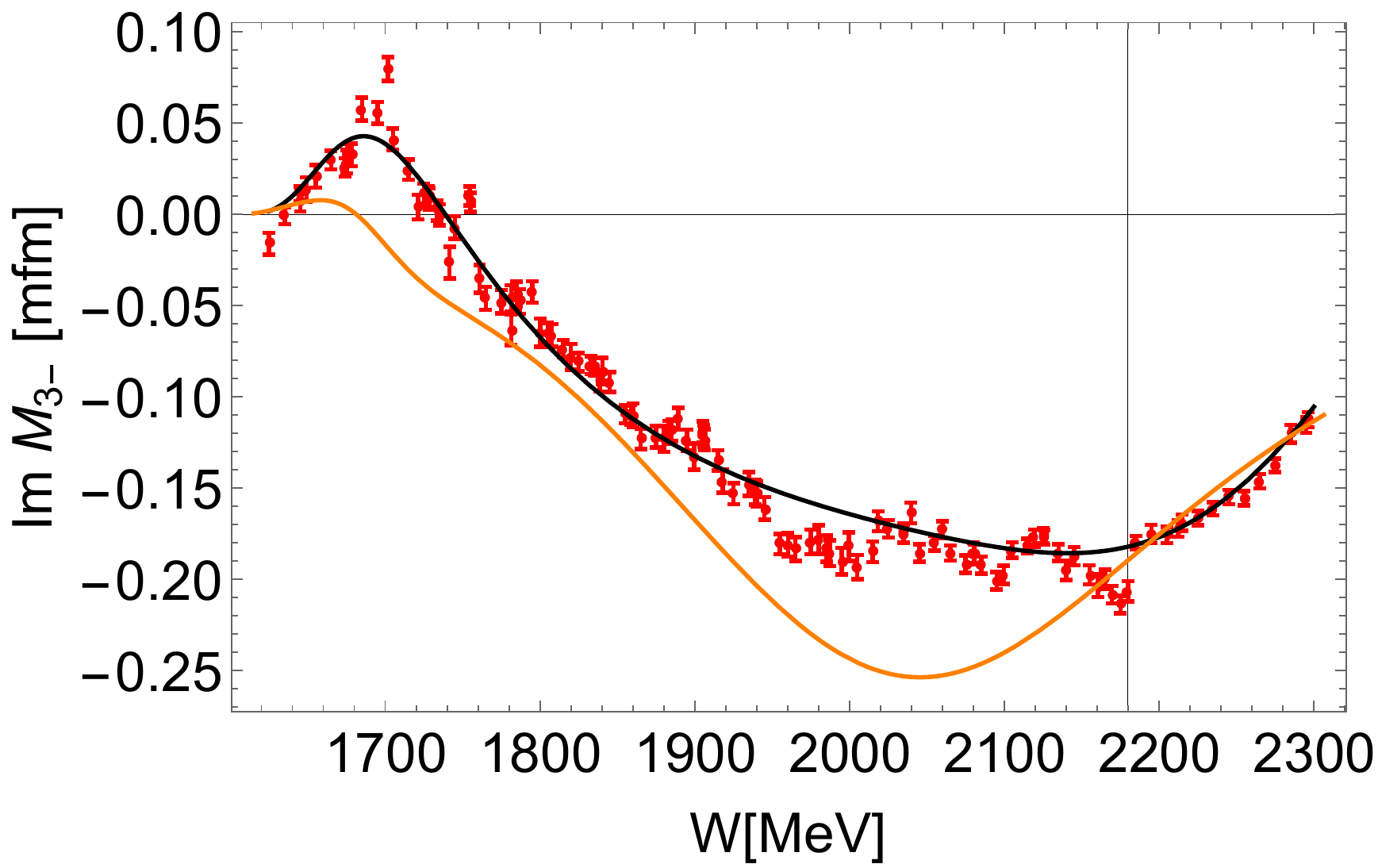}  \\
\caption{\label{Multipoles:b}(Color online) The multipoles for the $L=2$ and $3$ partial waves.
Red discrete symbols correspond to our Step 2 SE AA/PWA solution, the black full line gives the result of the Step 1 ED L+P ED PWA,  and the orange full line gives the BG2017 ED solution for comparison. The thin vertical black line marks the energy beyond which only 4 observables are measured instead of 8 (cf. Table~\ref{tab:expdata}).} \ec
\end{figure}

In Figs.~\ref{Multipoles:a} and \ref{Multipoles:b}, we show the final results for multipoles over the full set of energies. Red symbols give the values of multipoles for our step-two SE AA/PWA
solution, which scatter around the black full line produced using the simplified ED L+P fit. Note that this scatter of the SE points is limited
by the penalty function and results in a better fit to data. The SE and ED multipoles show better agreement in Fig.~\ref{Multipoles:a} for the
lower and dominant multipoles; more variation is seen in the higher multipoles of Fig.~\ref{Multipoles:b}. For comparison, the full orange
line gives the BG2017 solution~\cite{BG-web} used to initialize the L+P fit parameters. For lower multipoles, the orange and black lines
show qualitative agreement, apart from the real part of $M_{1+}$. Closer agreement should not be expected as the Bonn-Gatchina fits simultaneously account for channels beyond kaon photoproduction. Note that the vertical black line in Fig~\ref{Multipoles:a}, indicating
the energy where the number of measured observables drops from 8 to 4, is not reflected in jumps to equivalent solutions for the SE AA/PWA
points, ostensibly due to the penalty function constraint. Some discontinuities do begin to appear in the higher multipoles of
Ref.~\ref{Multipoles:b}.

\begin{figure*}[h!]
\bc \hspace*{0.5cm} \includegraphics[width=1.\textwidth]{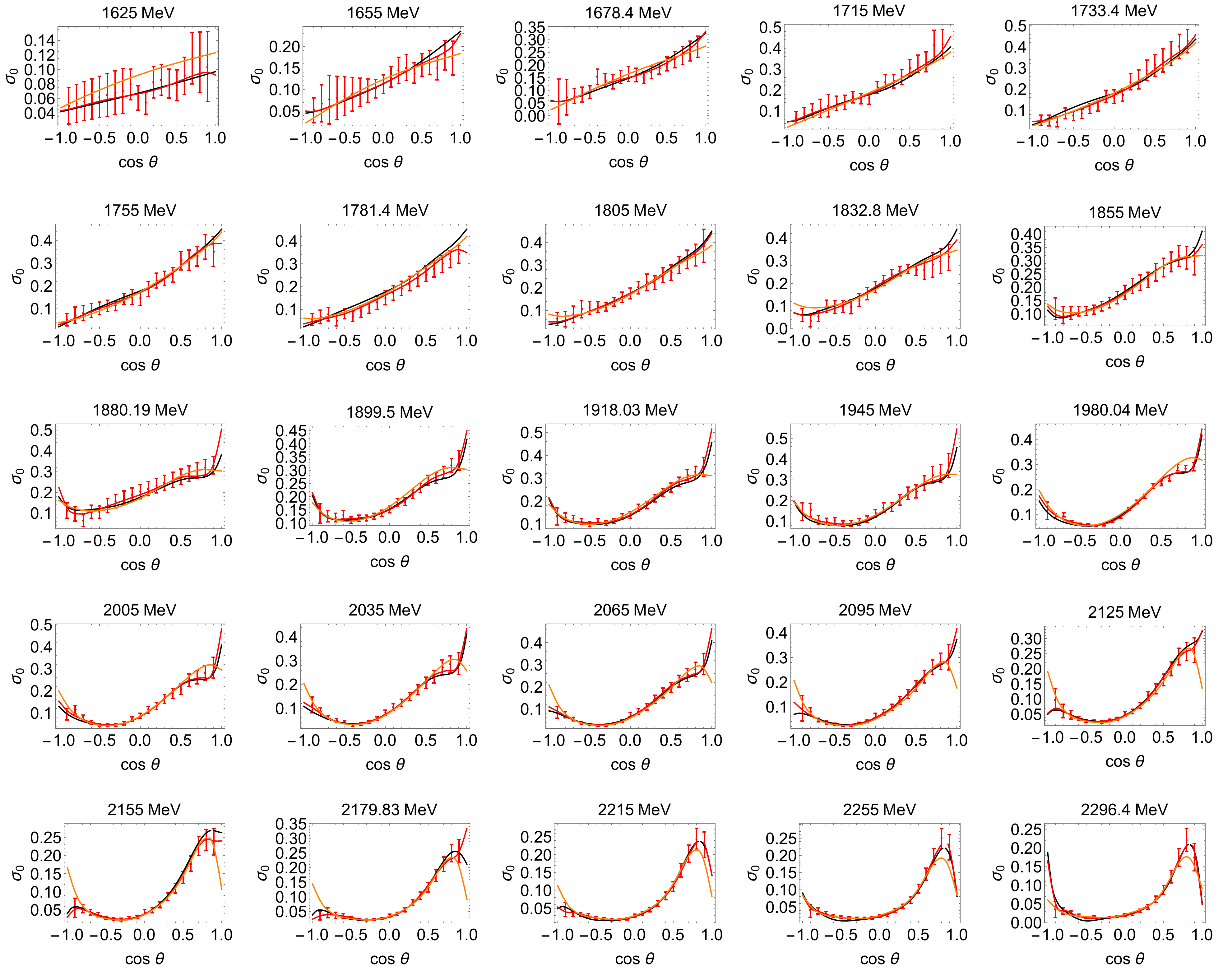} \caption{\label{Sol1:DCS}(Color online) Comparison of experimental data for $\sigma_0$ (red discrete symbols) with results from our SE AA/PWA (red full line), with our L+P ED PWA (black full line),  and with the BG2017 fit (orange
line) at representative energies.} \ec
\end{figure*}

\begin{figure*}[h!]
\bc \hspace*{0.5cm} \includegraphics[width=1.\textwidth]{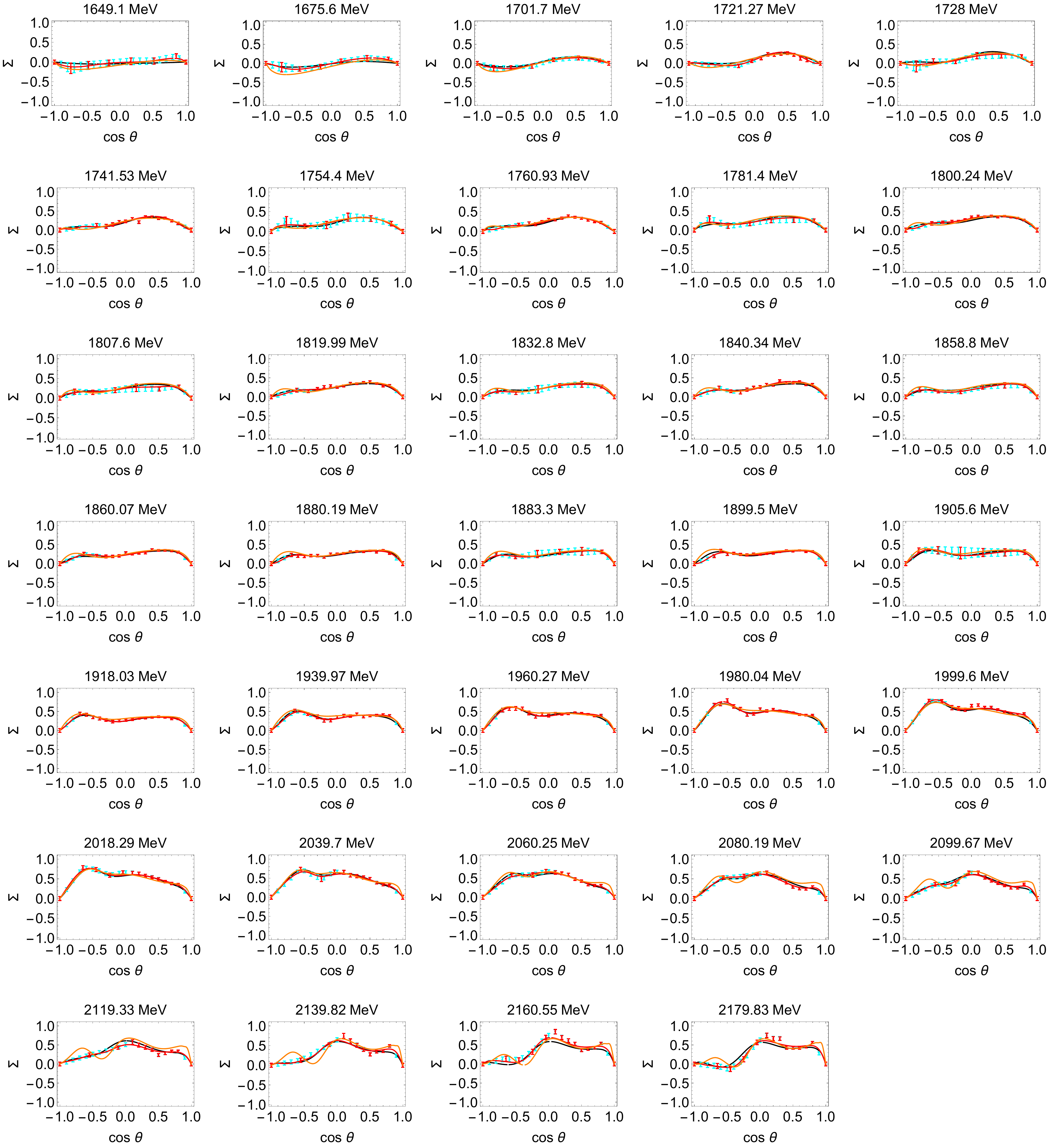} \caption{\label{Sol1:Sigma}(Color online) Comparison of experimental data for $\Sigma$ (red discrete symbols are measured values and cyan symbols are interpolated values) with results from our SE AA/PWA (red full
line), with our  L+P ED PWA (black full line), and with the BG2017 fit (orange line) at representative energies.} \ec
\end{figure*}

\begin{figure*}[h!]
\bc \hspace*{0.5cm} \includegraphics[width=1.\textwidth]{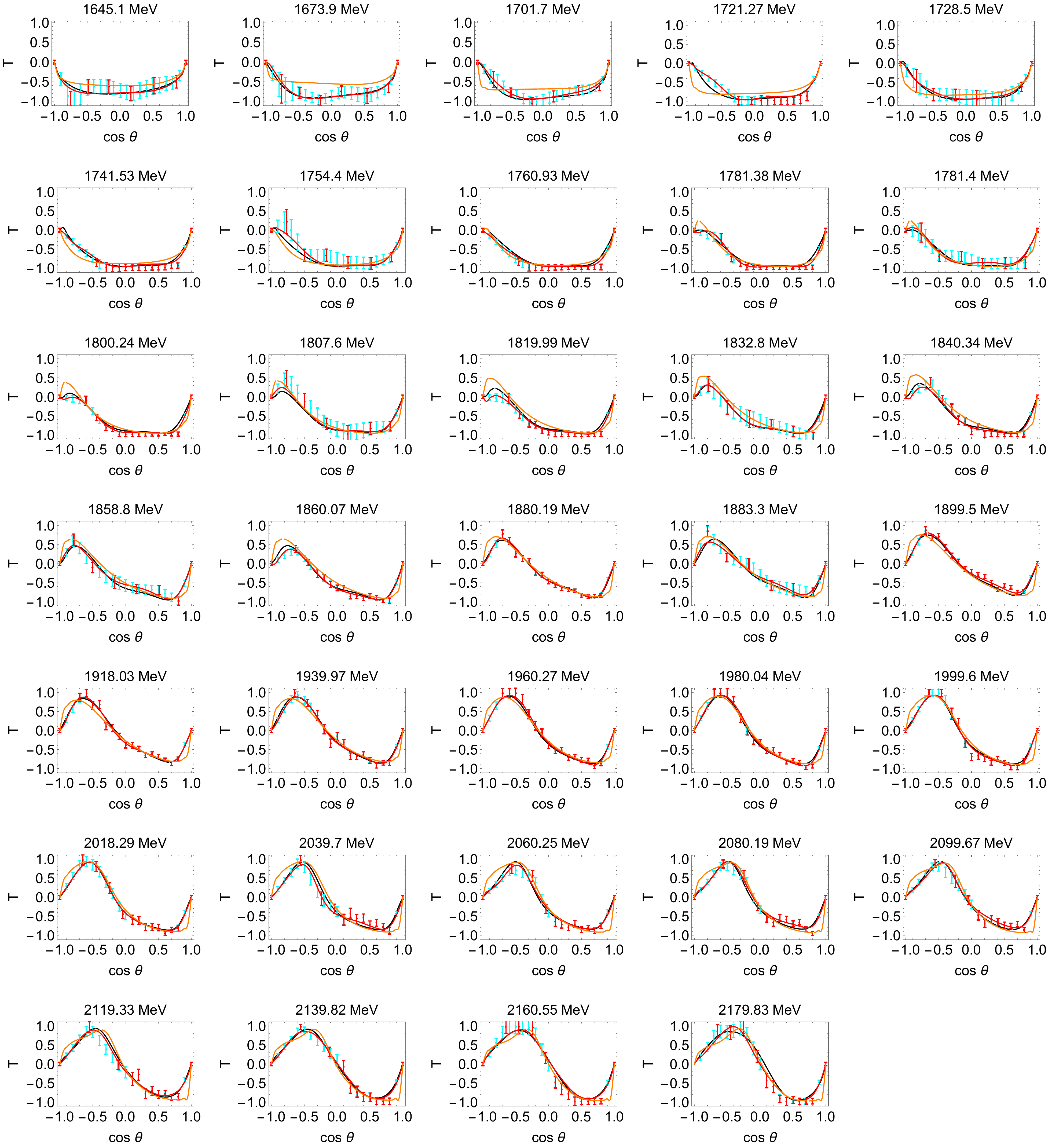} \caption{\label{Sol1:T}(Color online) Comparison of experimental data for $T$ (red discrete symbols are measured values and cyan symbols are interpolated values) with results from our SE AA/PWA (red full
line), with our L+P ED PWA (black full line), and with the BG2017 fit (orange line) at representative energies.} \ec
\end{figure*}

\begin{figure*}[h!]
\bc \hspace*{0.5cm} \includegraphics[width=1.\textwidth]{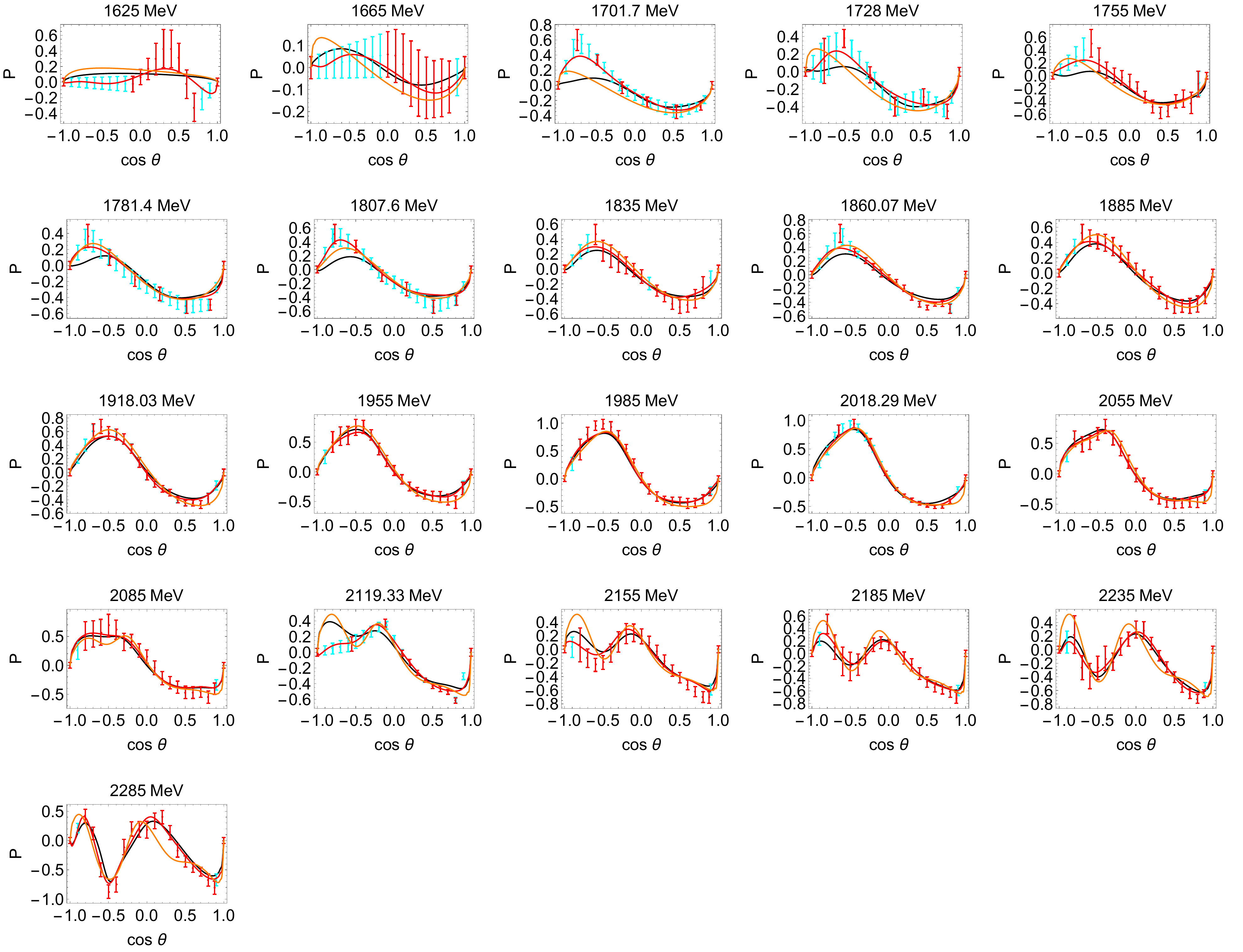} \caption{\label{Sol1:P}(Color online) Comparison of experimental data for $P$ (red discrete symbols are measured values and cyan symbols are interpolated values) with results from our SE AA/PWA (red full
line), with our  L+P ED PWA (black full line), and with the BG2017 fit (orange line) at representative energies.} \ec
\end{figure*}

\begin{figure*}[h!]
\bc \hspace*{0.5cm} \includegraphics[width=1.\textwidth]{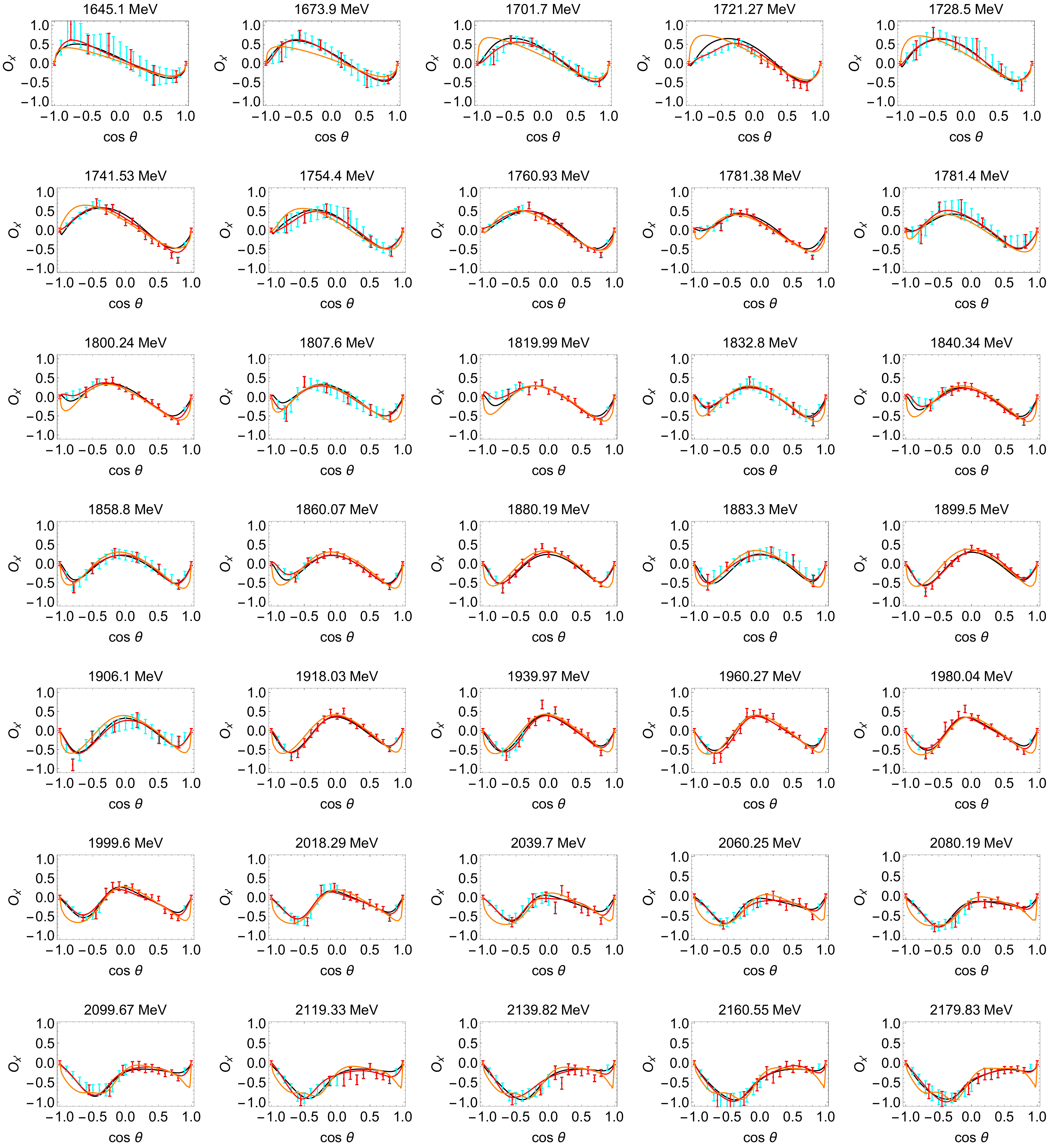} \caption{\label{Sol1:Oxprime}(Color online)Comparison of experimental data for $O_{x'}$ (red discrete symbols are measured values and cyan symbols are interpolated values) with results from our SE AA/PWA (red full line), with our  L+P ED PWA (black full line), and with the BG2017 fit (orange line) at representative energies.} \ec
\end{figure*}

\begin{figure*}[h!]
\bc \hspace*{0.5cm} \includegraphics[width=1.\textwidth]{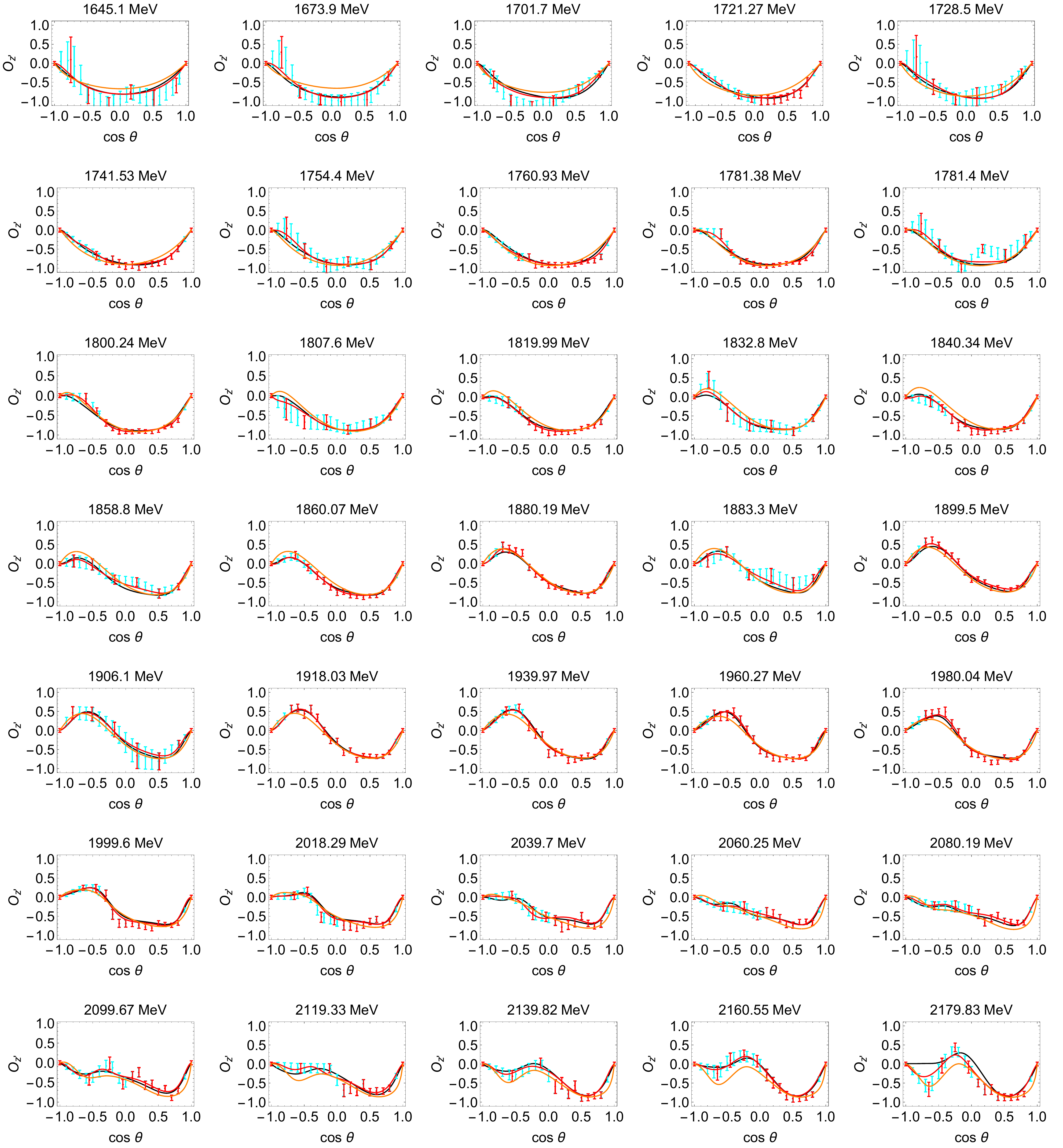} \caption{\label{Sol1:Ozprime}(Color online))Comparison of experimental data for $O_{z'}$ (red discrete symbols are measured values and cyan symbols are interpolated values) with results from our SE AA/PWA (red full line), with our  L+P ED PWA (black full line), and with the BG2017 fit (orange line) at representative energies.} \ec
\end{figure*}

\begin{figure*}[h!]
\bc \hspace*{0.5cm} \includegraphics[width=1.\textwidth]{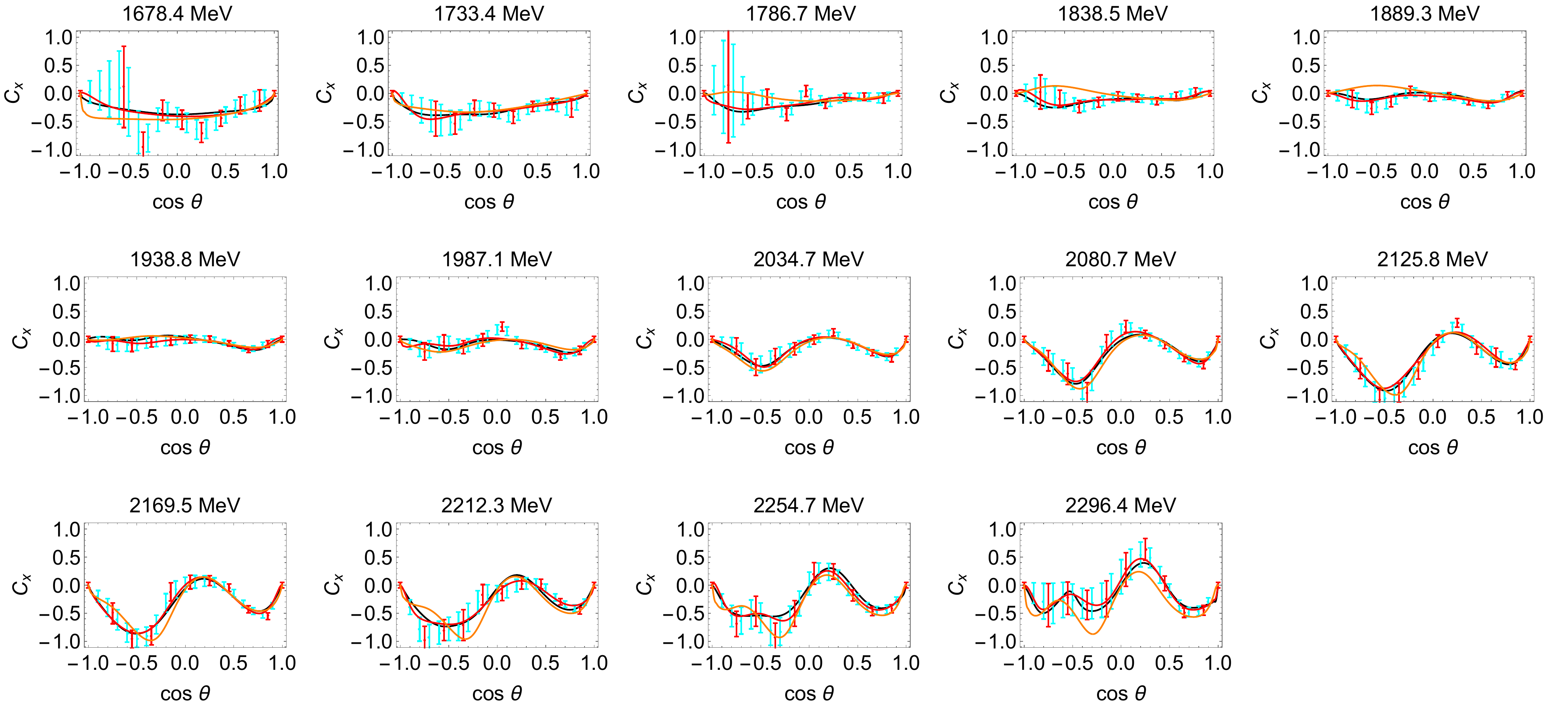} \caption{\label{Sol1:Cx}(Color online) )Comparison of experimental data for $C_{x}$ (red discrete symbols are measured values and cyan symbols are interpolated values) with results from our SE AA/PWA (red full line), with our  L+P ED PWA (black full line), and with the BG2017 fit (orange line) at representative energies.} \ec
\end{figure*}

\begin{figure*}[h!]
\bc \hspace*{0.5cm} \includegraphics[width=1.\textwidth]{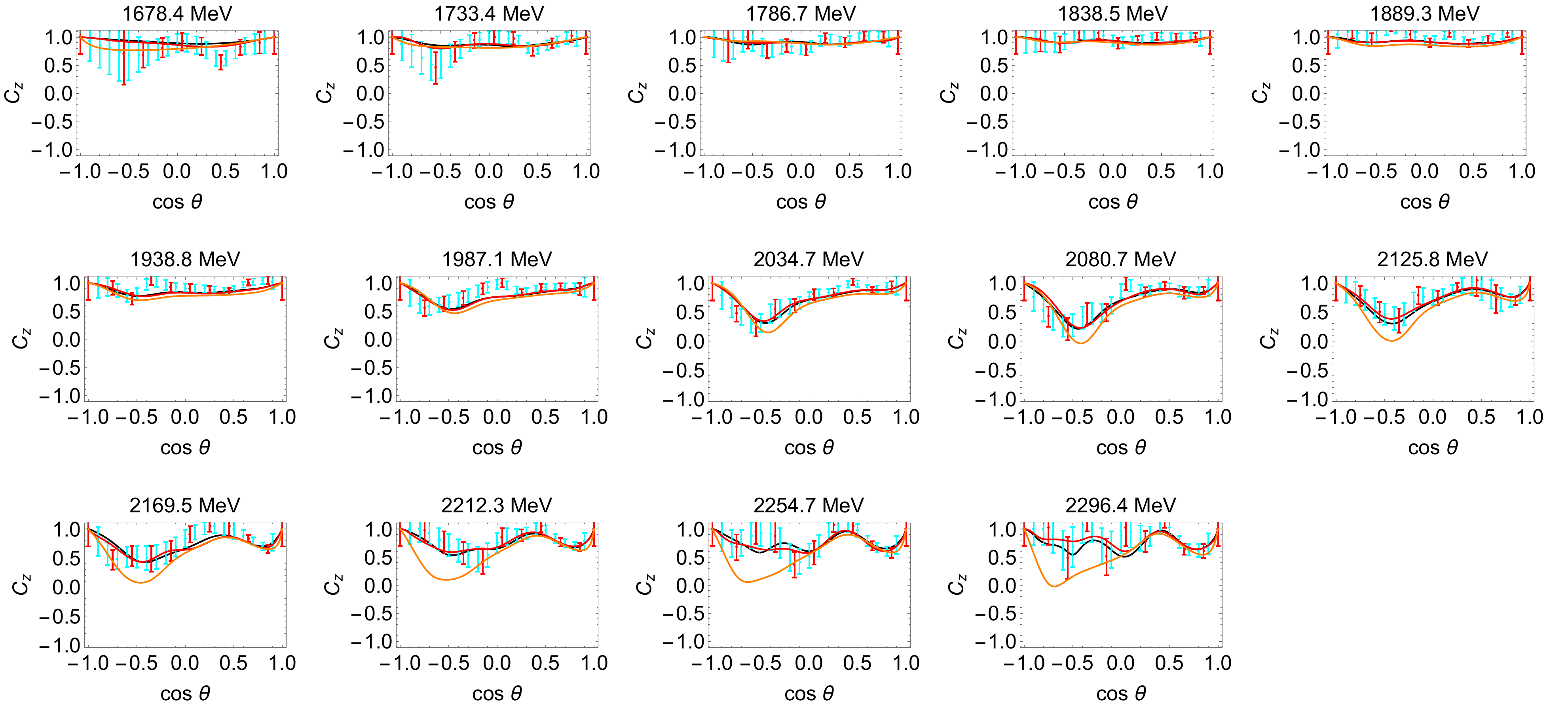} \caption{\label{Sol1:Cz}(Color online) )Comparison of experimental data for $C_{z}$ (red discrete symbols are measured values and cyan symbols are interpolated values) with results from our SE AA/PWA (red full line), with our  L+P ED PWA (black full line), and with the BG2017 fit (orange line) at representative energies.} \ec
\end{figure*}

In Figs.~\ref{Sol1:DCS} - \ref{Sol1:Cz},  the fits to
measured observables resulting from our simplified L+P PWA (black line) and SE AA/PWA method (red full line) are given, as well as predictions from the ED BG2017 solution~\cite{BG-web} (full orange line) at representative energies only. All further energies are available upon demand. Red symbols represent the actually measured data points as given in references collected in Table~\ref{tab:expdata}, and cyan symbols denote the interpolated values obtained with the procedure described in Section~\ref{Data:base}
Here we see that all curves give a good representation of the data, where it has been measured. Some small deviations exist mainly at extreme
forward and backward angles where experimental coverage is incomplete. Sharp structures in these regions are influenced by higher partial waves
and may be linked to some deviations seen in the multipoles of Fig.~\ref{Multipoles:b}.

\newpage

\newpage

\clearpage

\section{Full L+P analysis of the (Step 2) SE AA/PWA multipoles}

\begin{figure}[h!]
\bc
\includegraphics[width=0.37\textwidth]{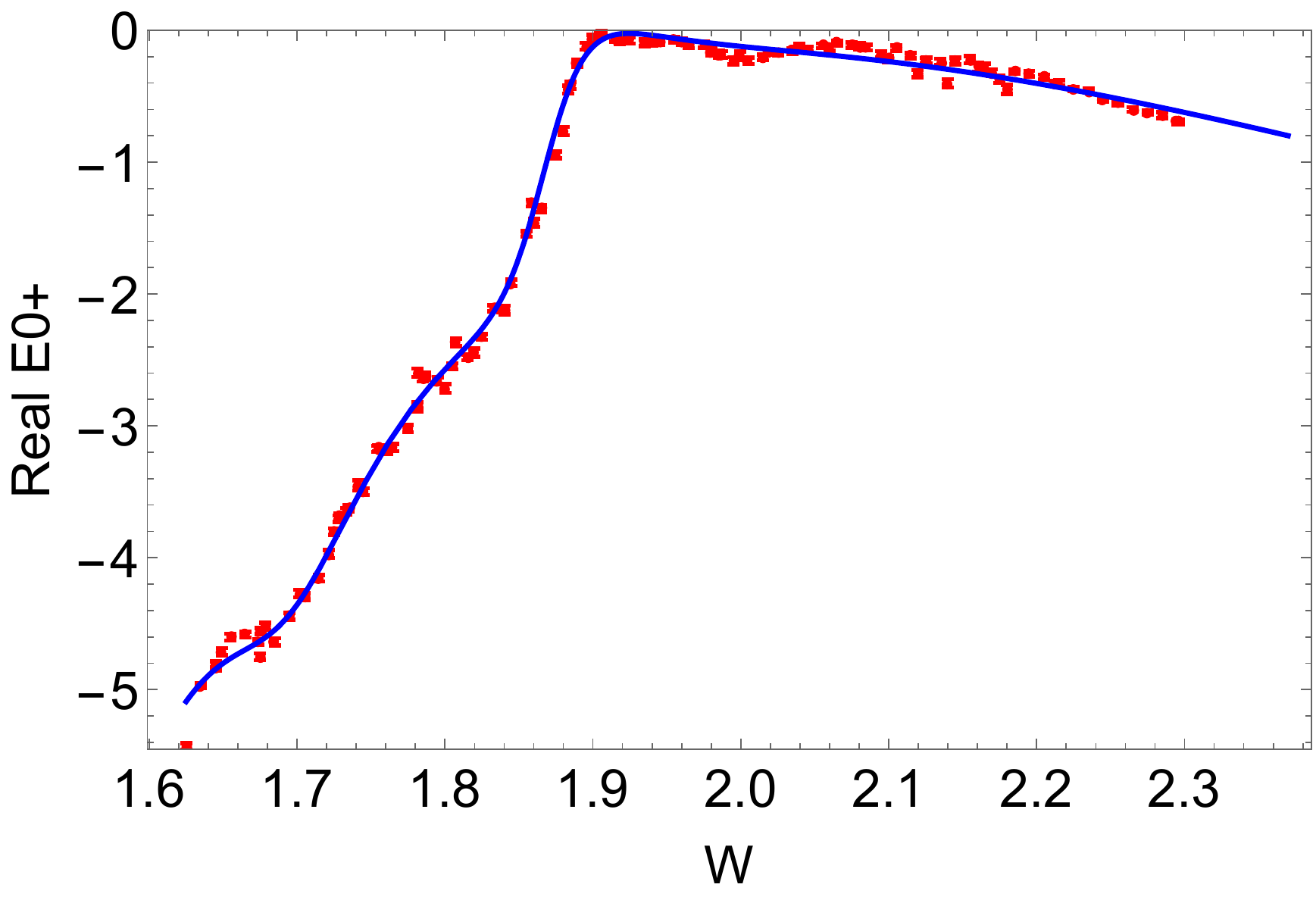} \hspace{0.5cm}
\includegraphics[width=0.37\textwidth]{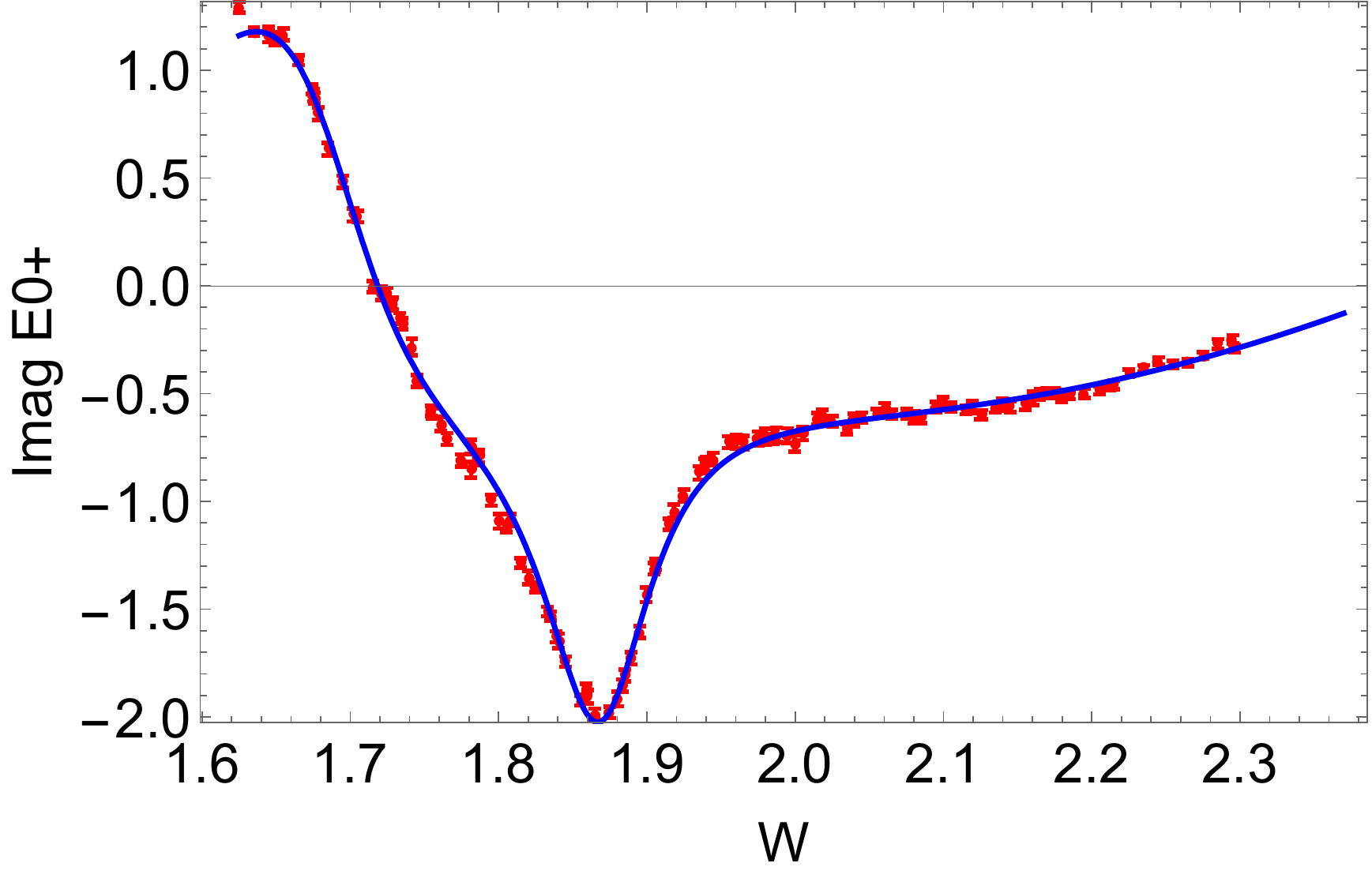} \hspace{0.5cm}
\includegraphics[width=0.37\textwidth]{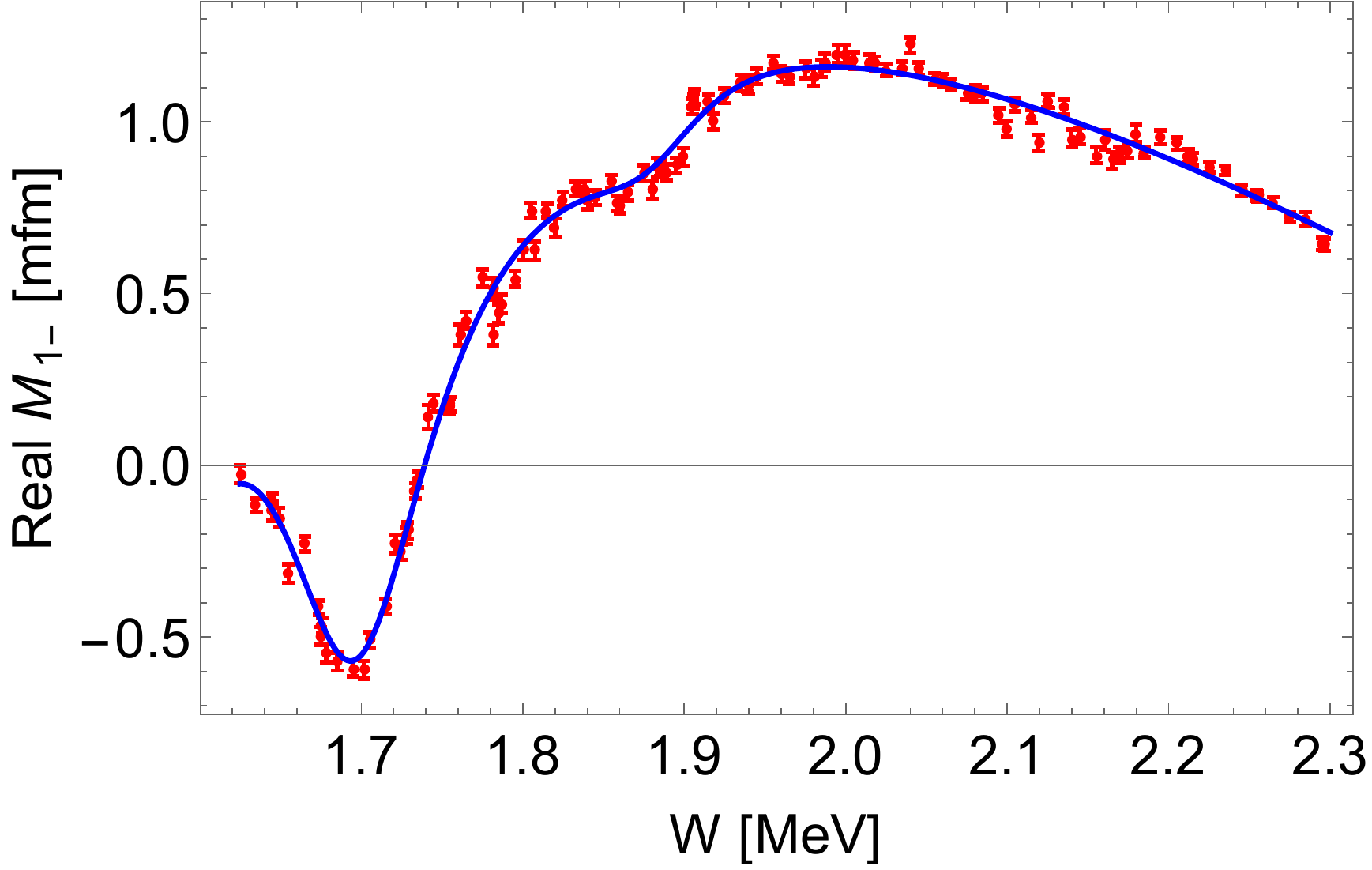} \hspace{0.5cm}
\includegraphics[width=0.37\textwidth]{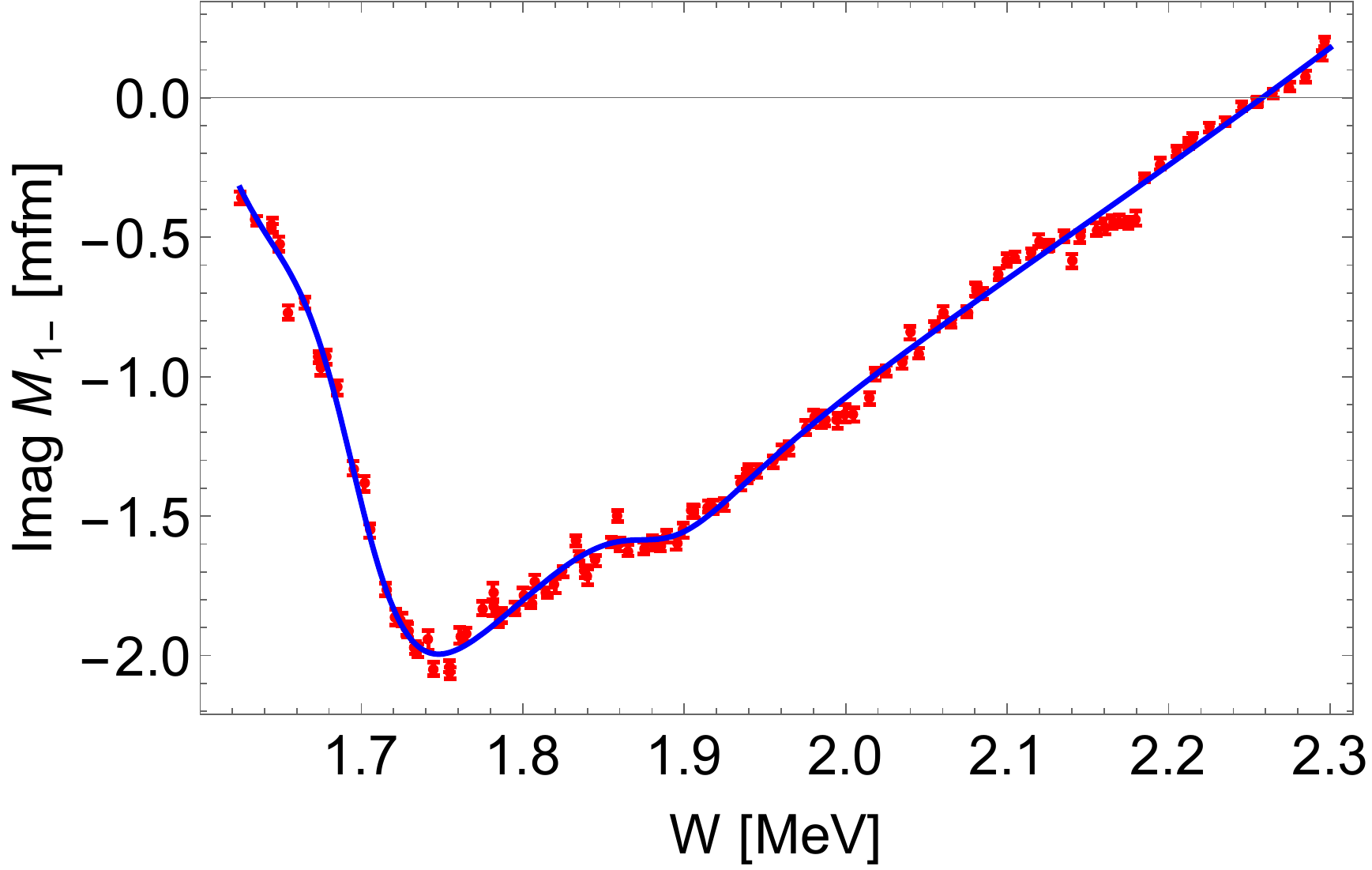} \hspace{0.5cm}
\includegraphics[width=0.37\textwidth]{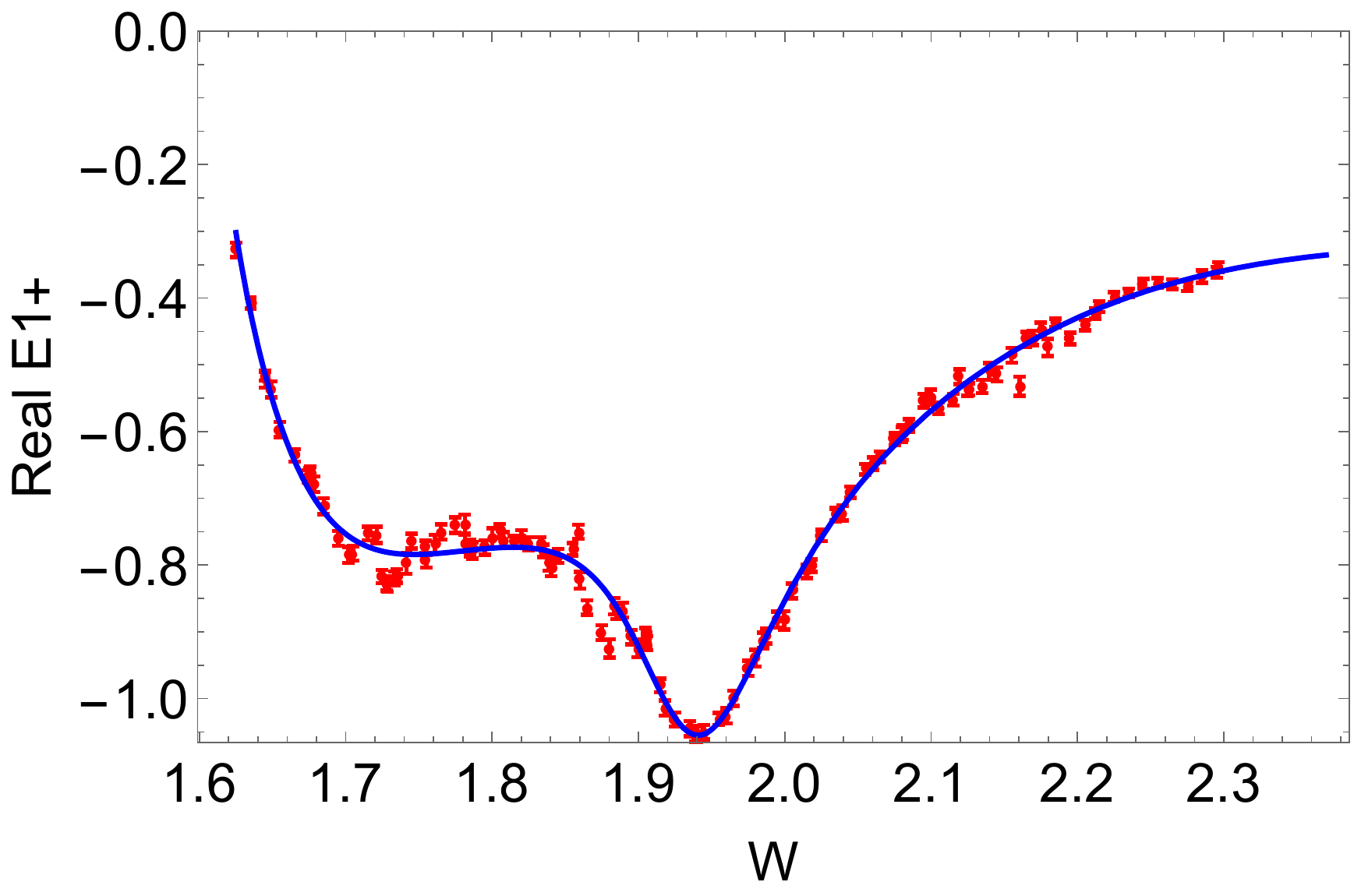} \hspace{0.5cm}
\includegraphics[width=0.37\textwidth]{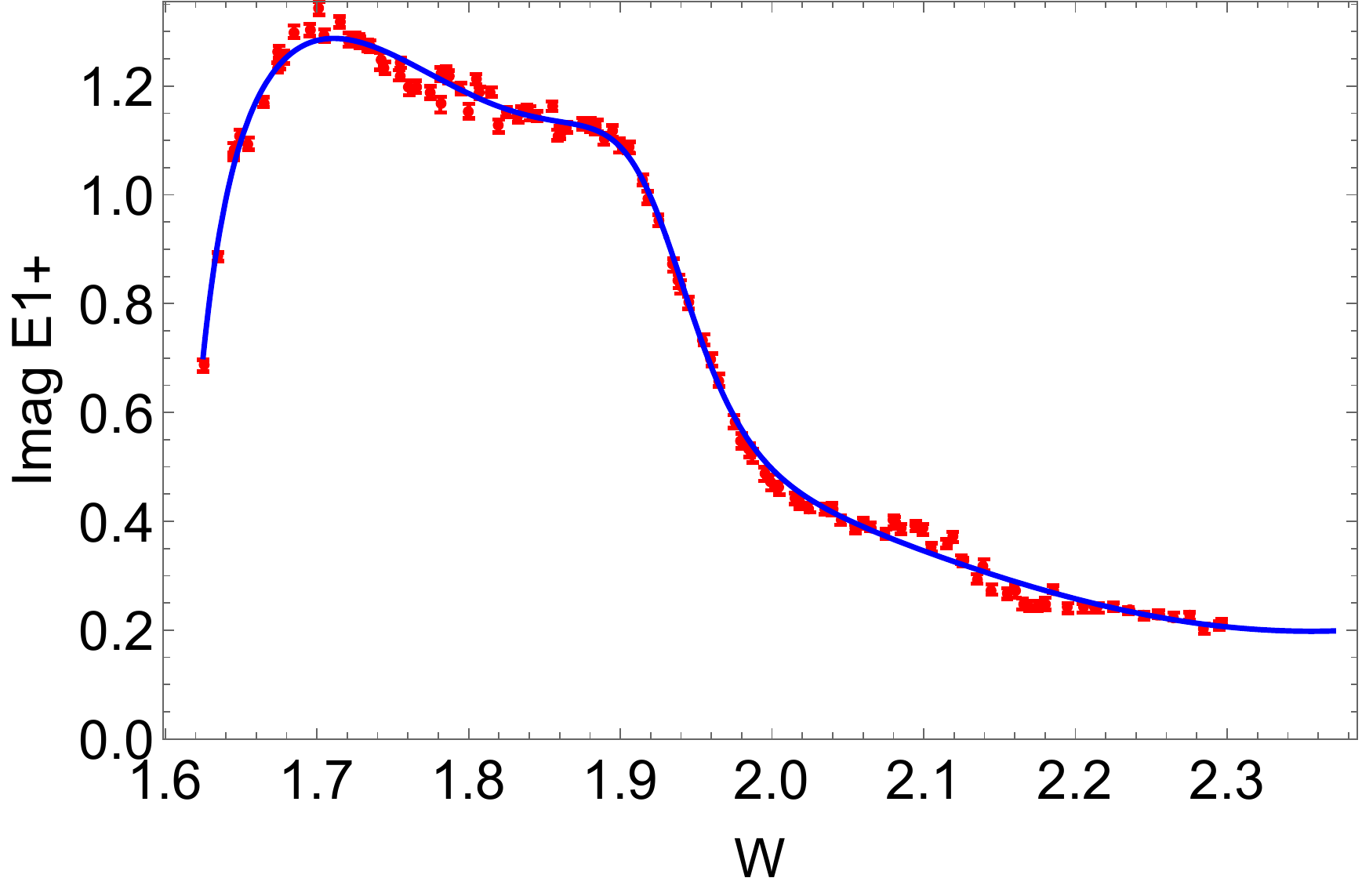} \hspace{0.5cm}
\includegraphics[width=0.37\textwidth]{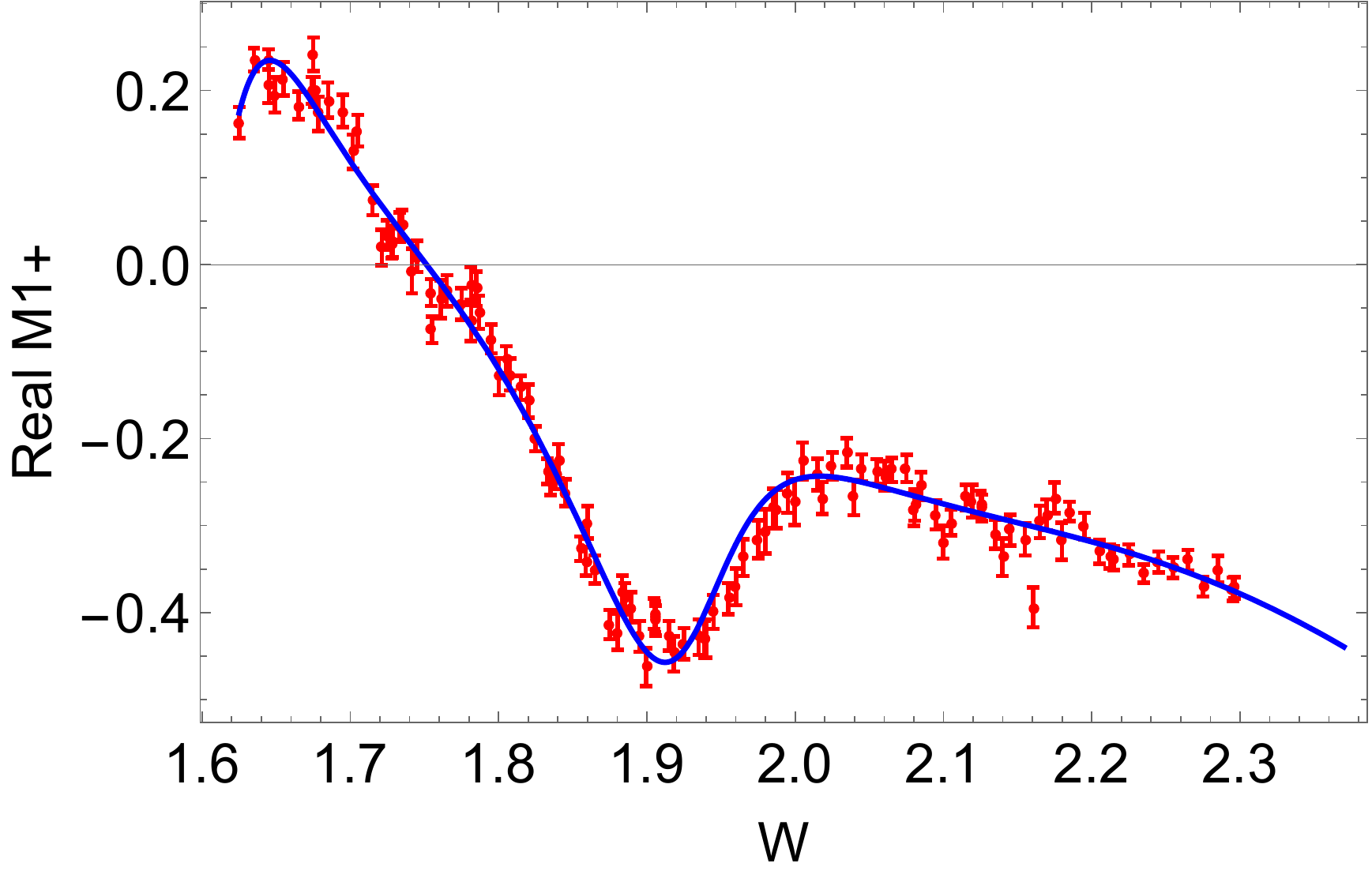} \hspace{0.5cm}
\includegraphics[width=0.37\textwidth]{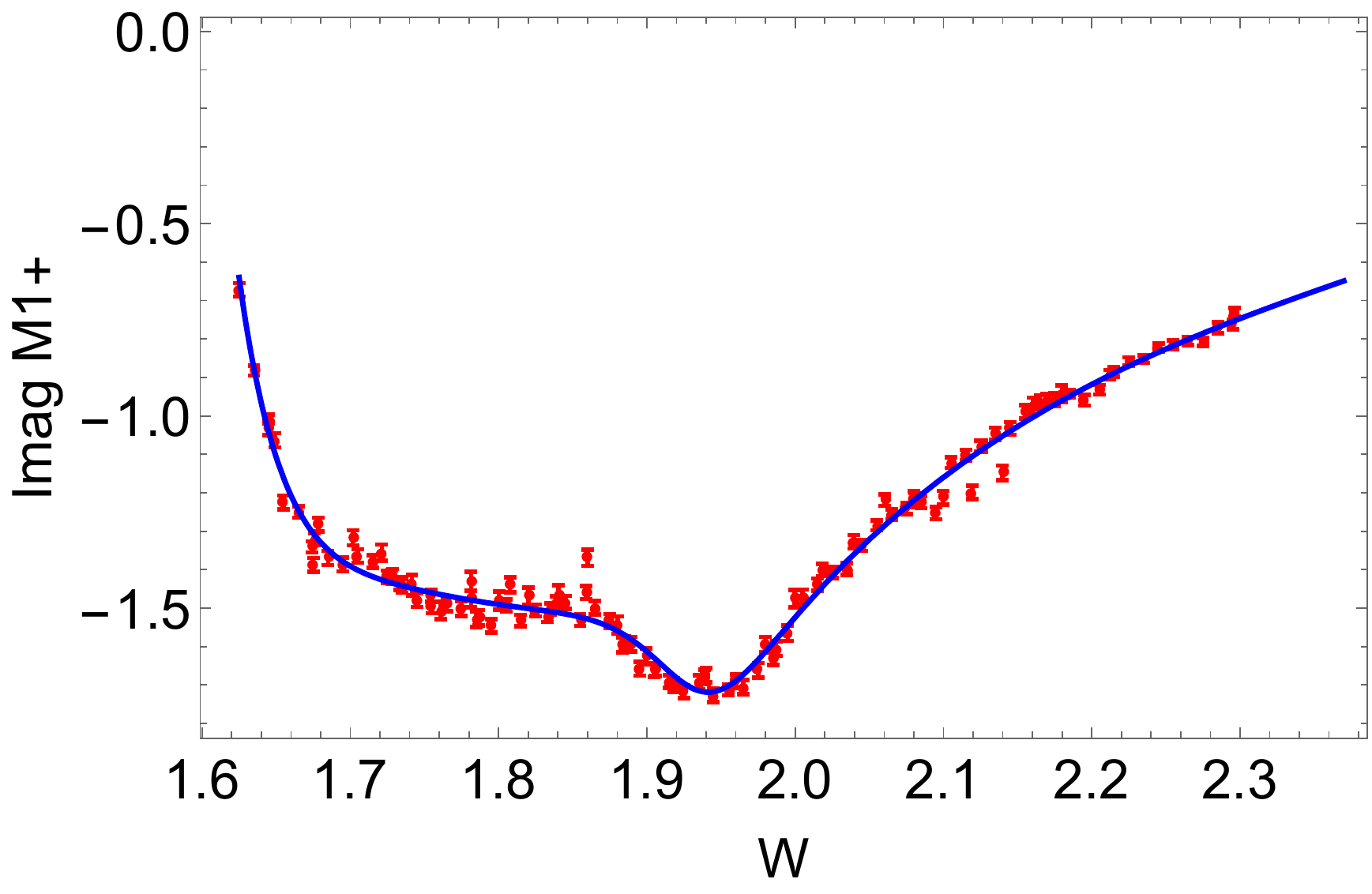}  \\

\caption{\label{L+Pfull1}(Color online) Multipoles obtained from full L+P PWA of SE AA/PWA solutions.
Red symbols are SE AA/PWA final results, and full blue line is the L+P fit.
Pole results are given in Table~\ref{tab:LPlusPIntermediateResultsI} } \ec
\end{figure}

\begin{figure}[h!]
\bc
\includegraphics[width=0.37\textwidth]{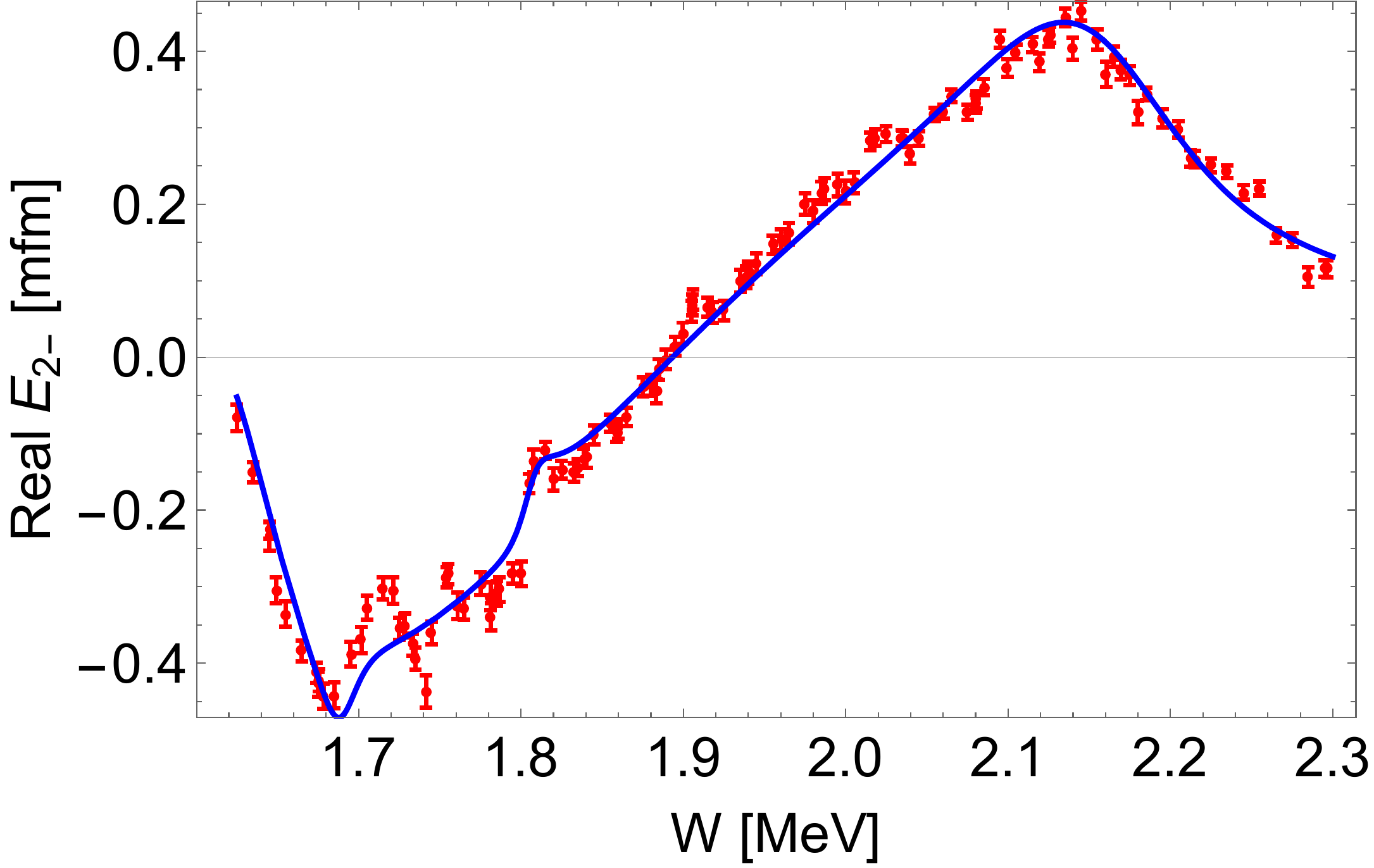} \hspace{0.5cm}
\includegraphics[width=0.37\textwidth]{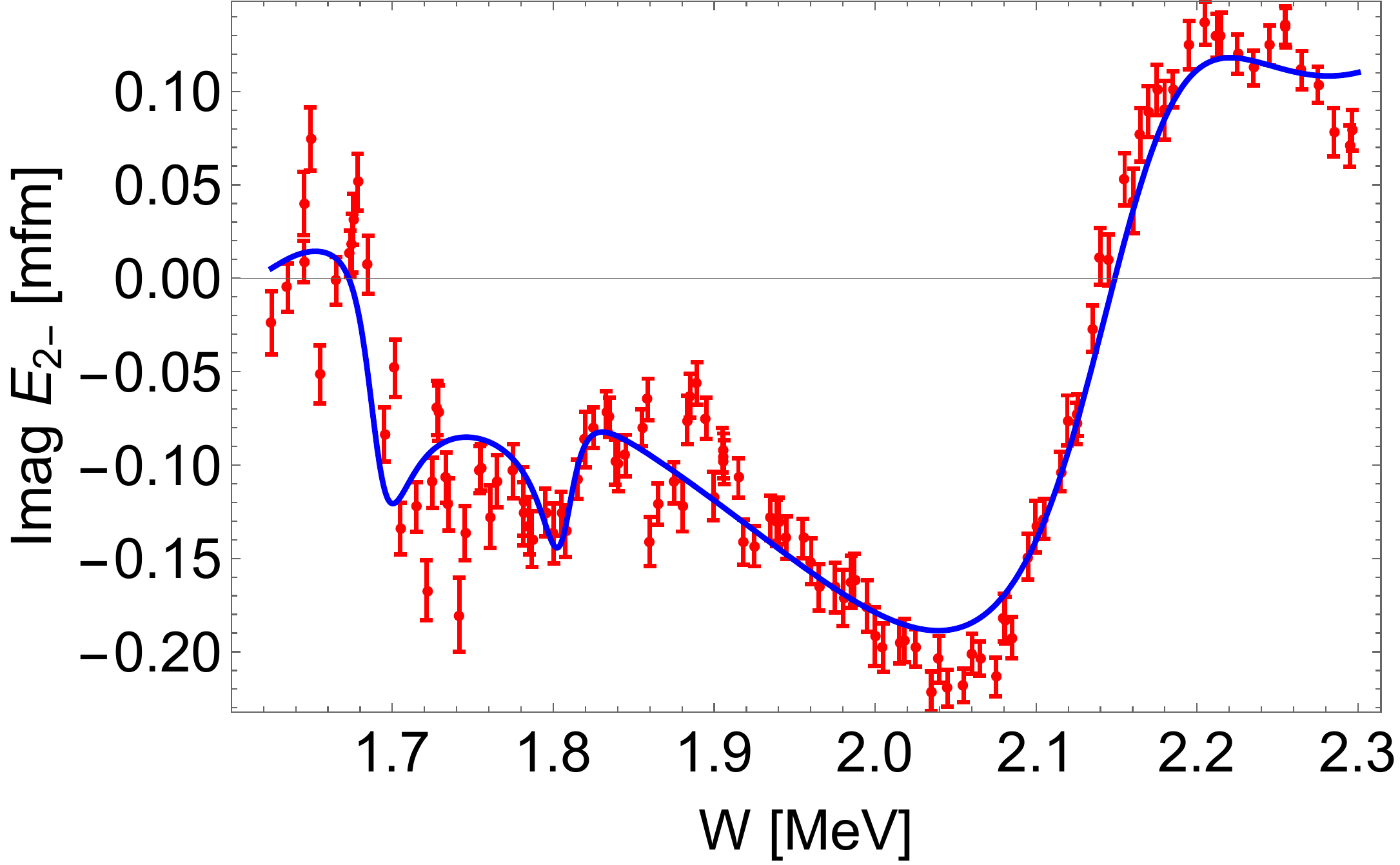}  \\
\includegraphics[width=0.37\textwidth]{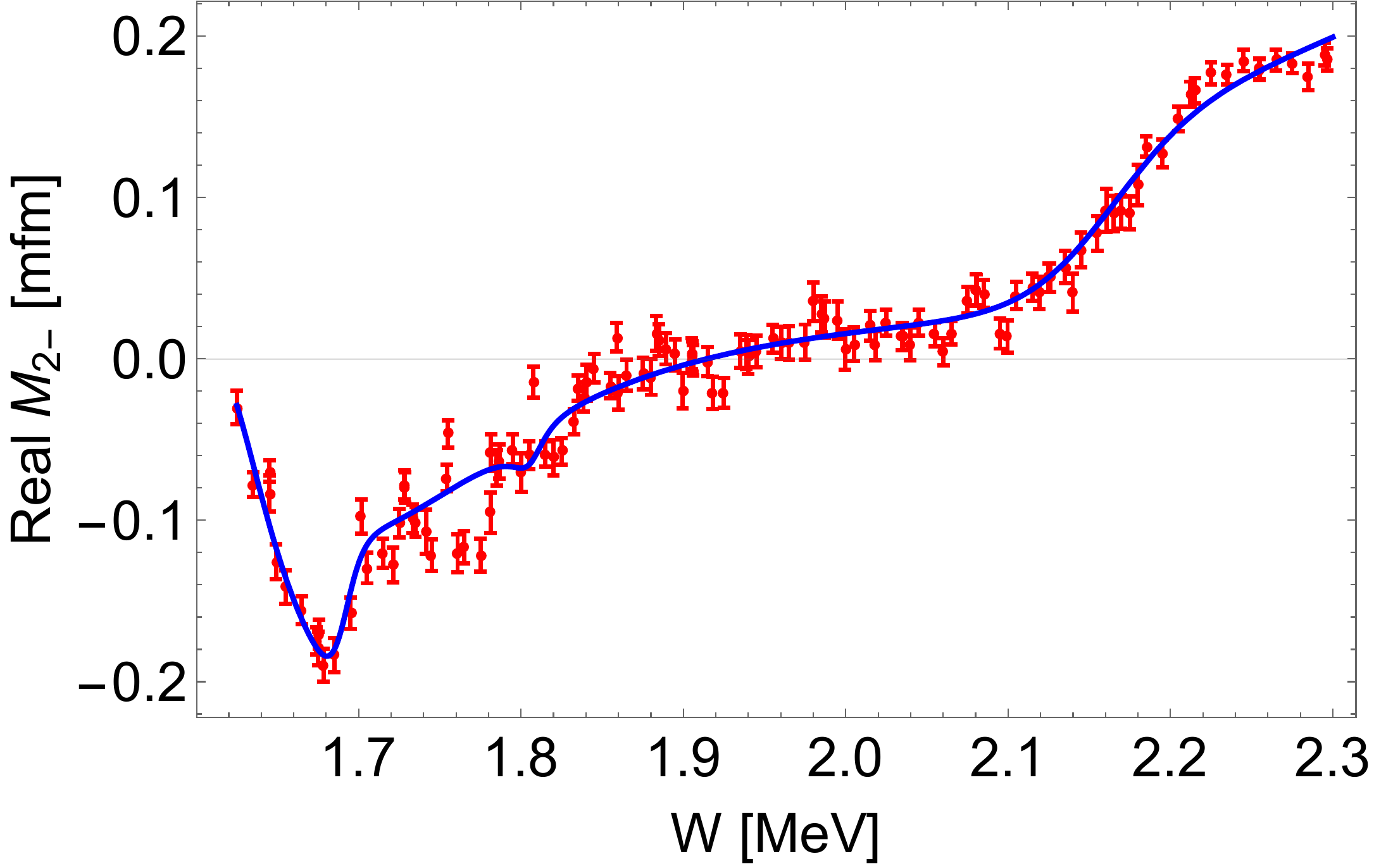} \hspace{0.5cm}
\includegraphics[width=0.37\textwidth]{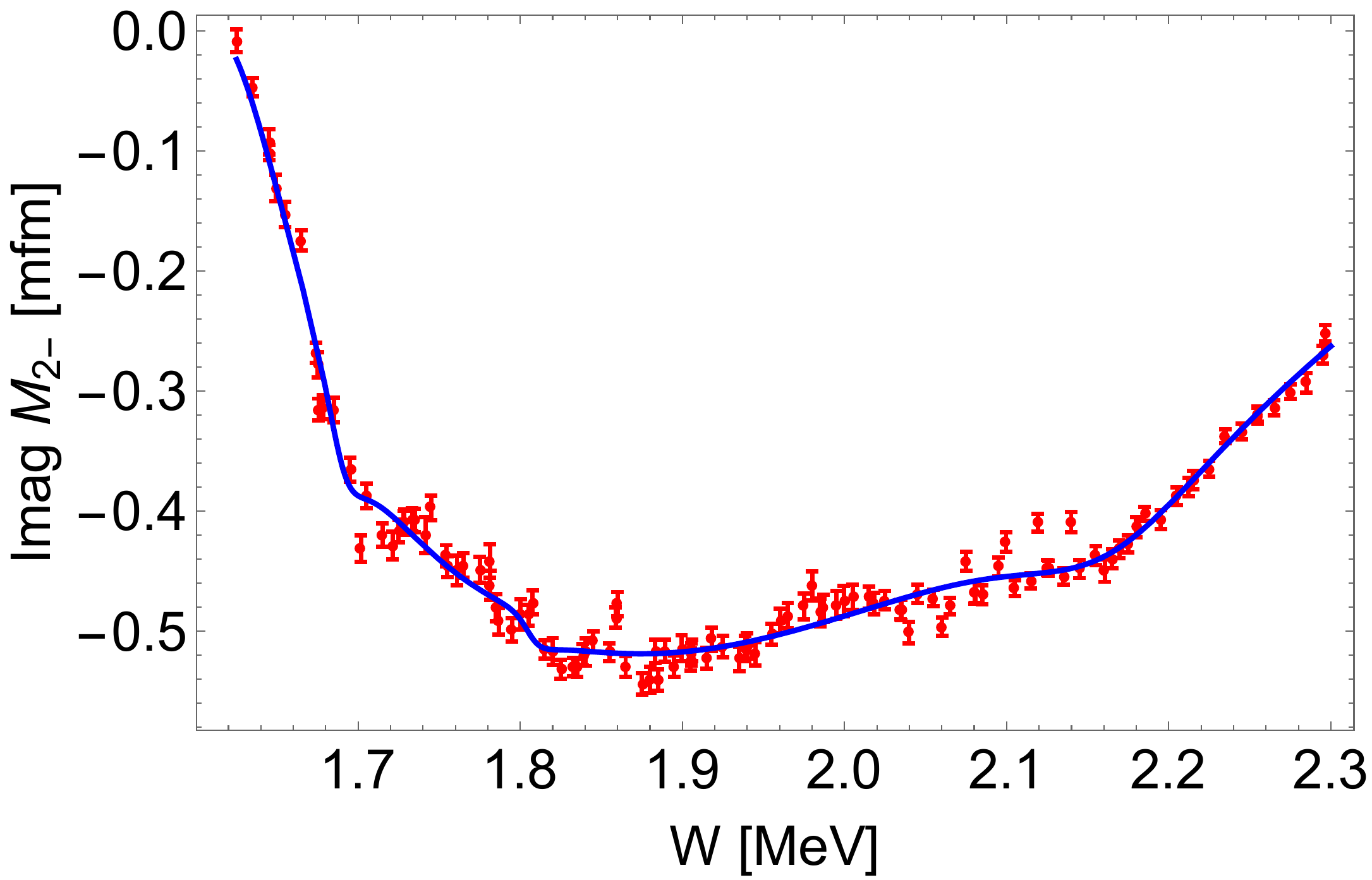}  \\
\includegraphics[width=0.37\textwidth]{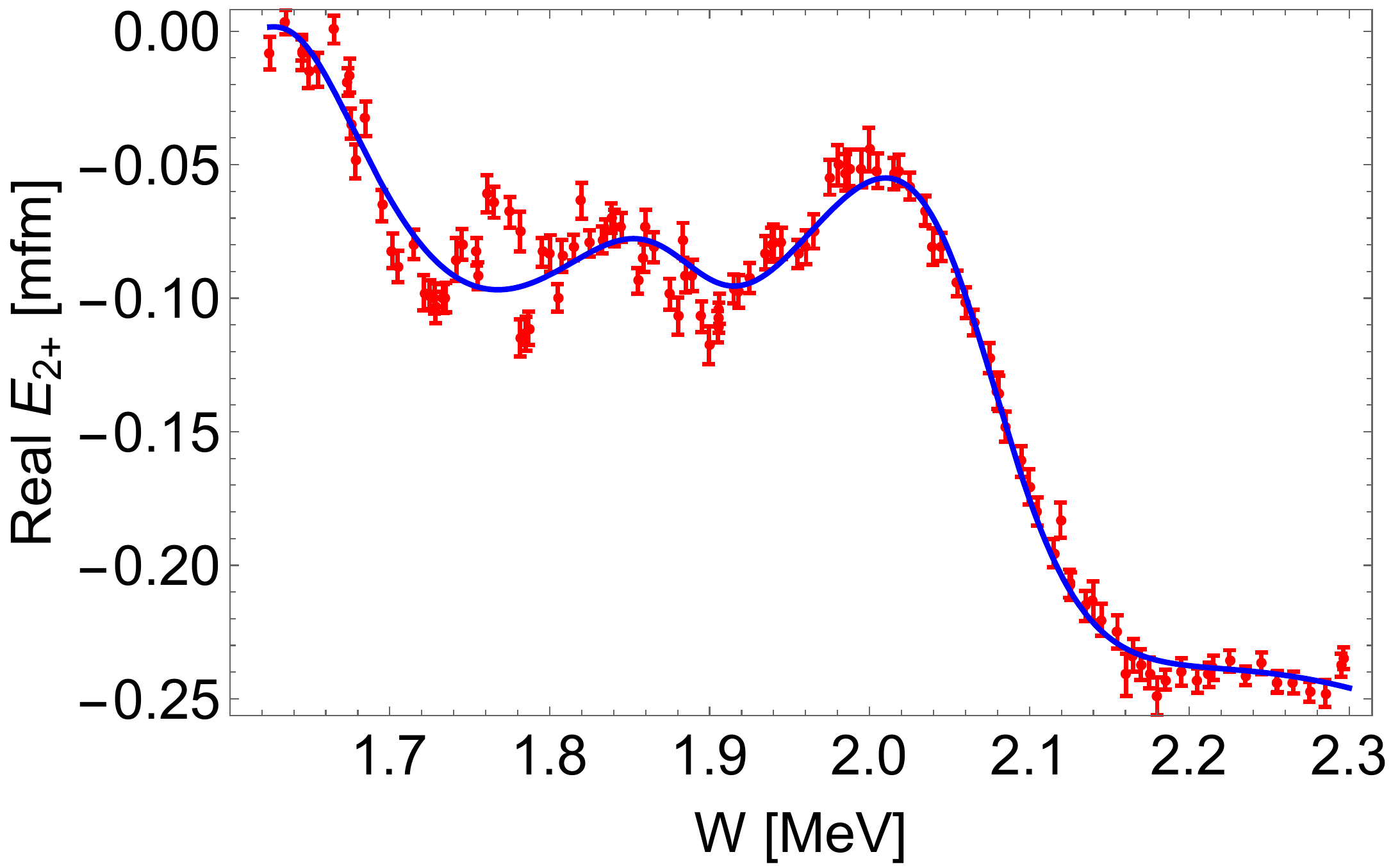} \hspace{0.5cm}
\includegraphics[width=0.37\textwidth]{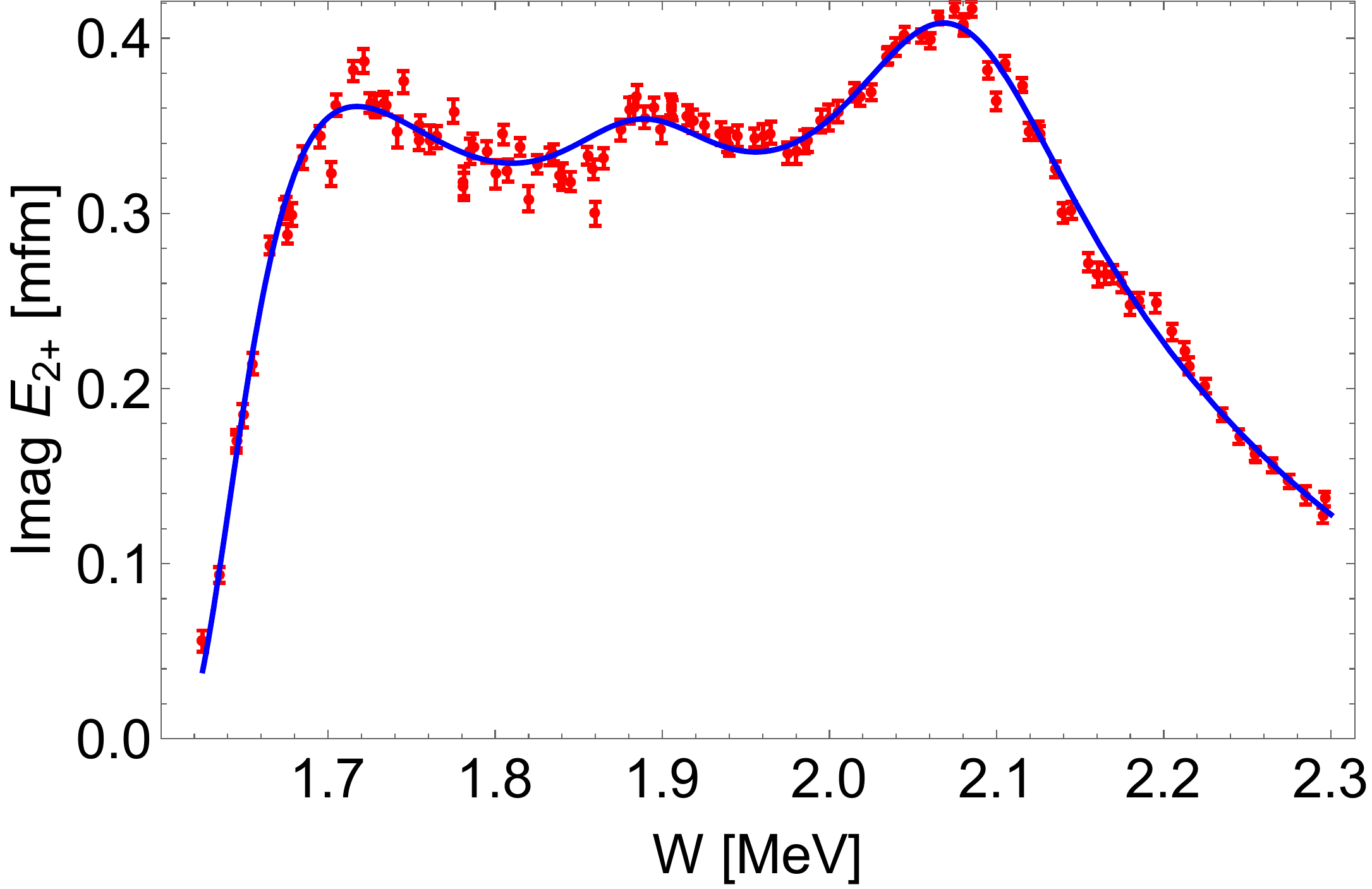}  \\
\includegraphics[width=0.37\textwidth]{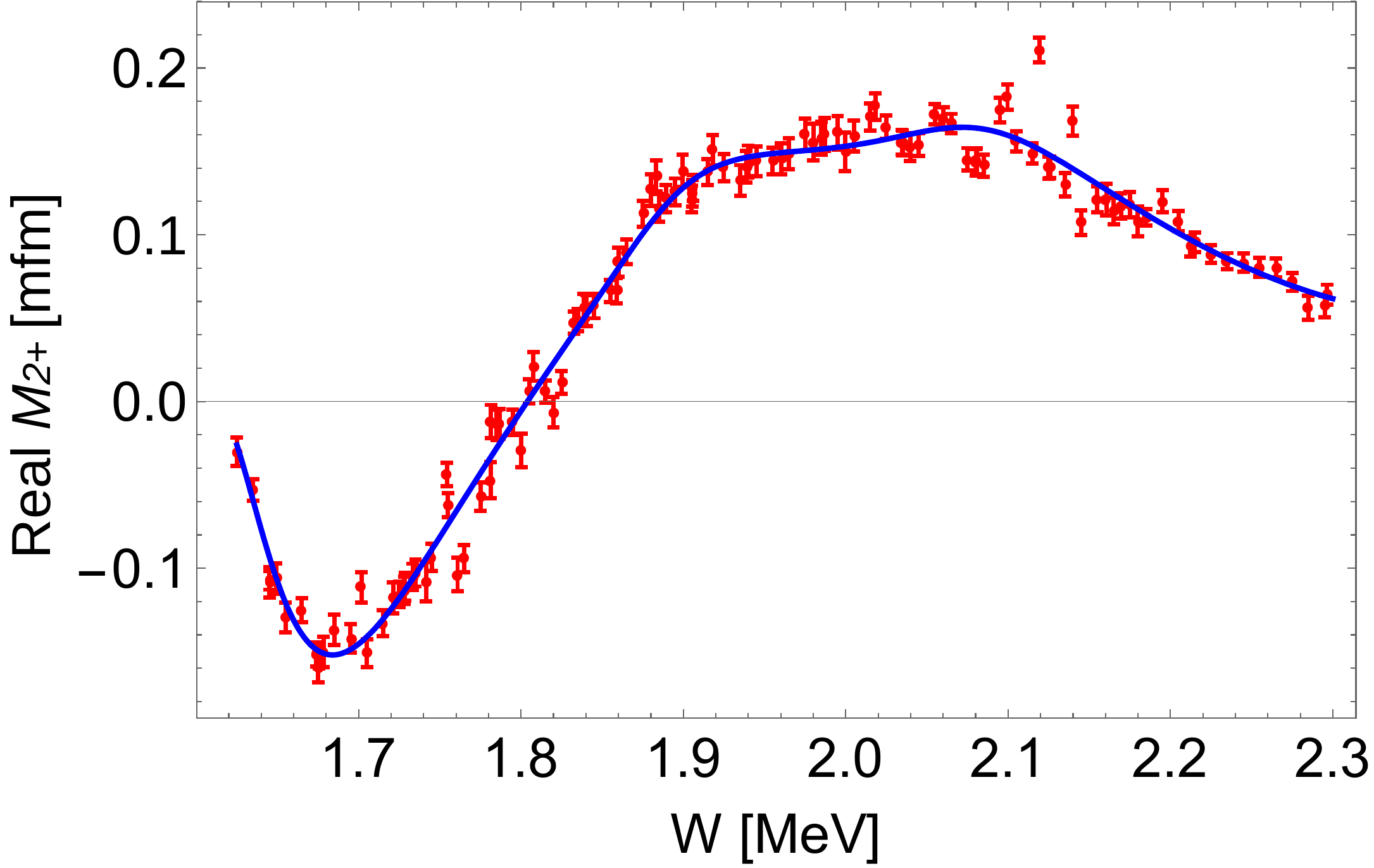} \hspace{0.5cm}
\includegraphics[width=0.37\textwidth]{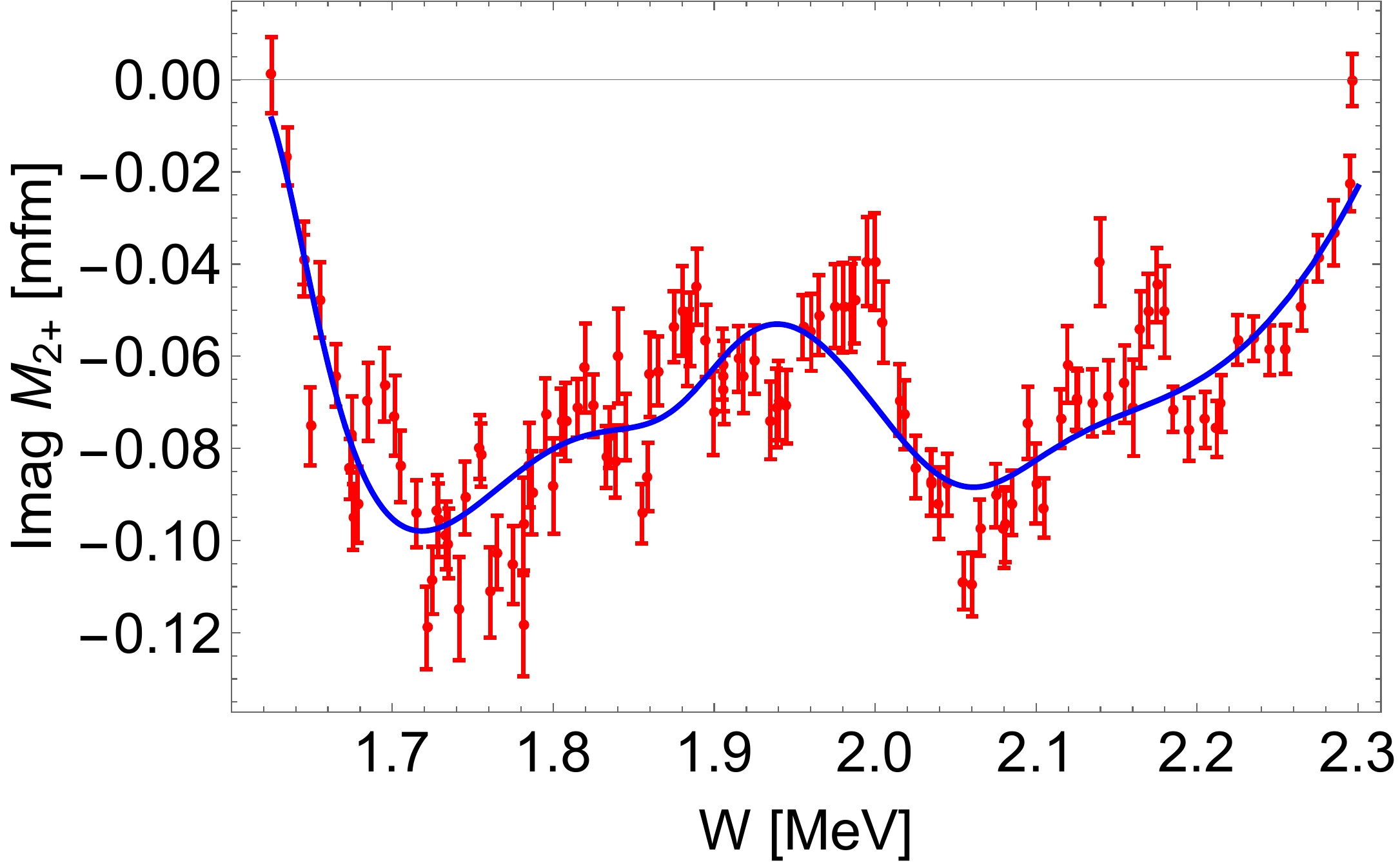}  \\

\caption{\label{L+Pfull2}(Color online) Multipoles obtained from full L+P PWA of SE AA/PWA solutions.
Red symbols are SE AA/PWA
results, and full blue line is the full L+P fit.
Pole results are given in Table~\ref{tab:LPlusPIntermediateResultsI} } \ec
\end{figure}

It is natural to explore the full analytic structure of the improved SE solutions displayed in Figs.~\ref{Multipoles:a} and \ref{Multipoles:b}. This can be done if the full L+P method of refs.~\cite{\LP,\LPapplication} is applied. In Table~\ref{tab:LPlusPIntermediateResultsI}, and in Figs.~\ref{L+Pfull1} and  \ref{L+Pfull2} we show the results.

We see that, for the lowest multipole $E_{0+}$, the pole positions and the general shape of multipole, comparing Fig.~\ref{Multipoles:a} and Fig.~\ref{L+Pfull1}, does not change significantly; it is well reproduced with the simplified L+P ED PWA. However, already for the $1-$ and  $1+$ multipoles, the shape of the function in Fig~\ref{L+Pfull1} is slightly different from the shape given in Fig.~\ref{Multipoles:a}. For multipoles $1+$, a second Pietarinen expansion has been added to compensate, and for multipole $1-$, a second resonance and a second Pietarinen expansion has been added. All results are collected in Table~\ref{tab:LPlusPIntermediateResultsI}, and in Fig.~\ref{L+Pfull1}.
Note that here, effective branch points are employed to represent all branch-cut contributions simultaneously (see Refs.~\cite{\LP,\LPapplication}).
\\ \\ \noindent

\begin{table}[h]
\centering
\begin{tabular}{ccccccccccc}
\hline
\hline
 $\ell \pm$ & $J^{P}$ &  & $\text{Re} M_1$ & $\text{Im} M_1$ & $\text{Re} M_2$ & $\text{Im} M_2$ & $\text{Re} M_3$ & $\text{Im} M_3$ &  $BP_{1}^{LP}$  & $BP_{2}^{LP}$    \\
  \hline
$0+$ & $\frac{1}{2}^{-}$ & \textbf{BG poles} & \textbf{1.658} & \textbf{0.051} & \textbf{1.906} & \textbf{0.05} &  &  &  & \\
 &  & Fitted BG amp. & 1.667 & 0.044 & 1.881 & 0.088 &  &  & 1.086  &  \\
 &  &L+P ED PWA  &  1.667 & 0.044  & 1.880 & 0.041 &  &  & 1.086  &  \\
  &  &Full L+P  & 1.688 & 0.073  & 1.869 & 0.043 &  &  & 1.086  &  \\
$1+$ &  $\frac{3}{2}^{+}$ & \textbf{BG poles} & \textbf{1.660} & \textbf{0.250} & \textbf{1.945} & \textbf{0.077} &  &  &  &  \\
 &  & Fitted BG amp. &  &  & 1.904 & 0.105 &  &  & 1.086  &  \\
 &  &L+P ED PWA &  &  & 1.933 & 0.067 &  &  & 1.086  &   \\
  &  &Full L+P  &  &  & 1.930 & 0.063 &  &  & 1.086  & 1.608   \\
$2+$ &  $\frac{5}{2}^{-}$ & \textbf{BG poles} & \textbf{1.654} & \textbf{0.075} & \textbf{2.040} & \textbf{0.195} & &    &  &  \\
 &  & Fitted BG amp. &  &  & 2.063 & 0.256 &  &  & 1.086  &  \\
 &  &L+P ED PWA &  &  & 2.098 & 0.093 &  &  & 1.086 &  \\
 &  &Full L+P & 1.671 & 0.318   & 2.070  & 0.093 & {\bf 1.888*} & {\bf 0.072*} & 1.086 & 1.562 \\

$3+$ &  $\frac{7}{2}^{+}$ & \textbf{BG poles} &  &  & \textbf{2.030} & \textbf{0.12} &  &  \\
 &  & Fitted BG amp. &  &  & 2.056 & 0.059 &  &   & 1.086 &  \\
 &  &L+P ED PWA &  &  &2.059 & 0.040 &  &  & 1.086  &   \\
$4+$ &  $\frac{9}{2}^{-}$ & \textbf{BG poles} &  &  & \textbf{2.195} & \textbf{0.235} &  &  \\
 &  & Fitted BG amp. &  &  &2.113 & 0.196 &  & & 1.086  &  \\
 &  &L+P ED PWA &  &  & 2.017 &0.180 &  & & 1.086 &   \\
$1-$ &  $\frac{1}{2}^{+}$ & \textbf{BG poles} & \textbf{1.697} & \textbf{0.042} & \textbf{1.875} & \textbf{0.0165} &  &  &  &  \\
 &  & Fitted BG amp. & 1.717 & 0.051 &  &  & &  & 1.086  &  \\
 &  &L+P ED PWA & 1.710 & 046 &  & & &  & 1.086  &  \\
 &  &Full L+P & 1.694 & 0.057 & 1.883  & 0.056 & &  & 1.086  & 1.461 \\
$2-$ &  $\frac{3}{2}^{-}$ & \textbf{BG poles} & \textbf{1.770} & \textbf{0.21} & \textbf{1.860} & \textbf{0.1} & \textbf{2.11} & \textbf{0.17} &  &   \\
 &  & Fitted BG amp. &  &  & & &  2.109  &  0.195  &  1.086 &  \\
 &  &L+P ED PWA & & &  &  & 2.012 & 0.11 &  1.086  &  \\
 &  &Full L+P  &1.691 & 0.016 & 1.806 & 0.013 & 2.158 & 0.088 &  1.086  & 0.94  \\
$3-$ &  $\frac{5}{2}^{+}$ & \textbf{BG poles} & \textbf{1.675} & \textbf{0.056} & \textbf{1.830} & \textbf{0.125} & \textbf{2.03} & \textbf{0.24} &  &  \\
 &  & Fitted BG amp. &  &  & &  & 2.013 & 0.135  &   \\
 &  &L+P ED PWA &  & & &  &  2.187 & 0.274 &    1.086 &  \\
$4-$ &  $\frac{7}{2}^{-}$ & \textbf{BG poles} &  &  & &  & \textbf{2.150} & \textbf{0.165} &   &   \\
 &  & Fitted BG amp. &  &  & 1.829 & 0.154 & & &  1.086 &   \\
 &  &L+P ED PWA &  & & 1.862 &0.322&  & & 1.086  &    \\
$5-$ & $\frac{9}{2}^{+}$ & \textbf{BG poles} & & &  &  & \textbf{2.150} & \textbf{0.22} &   &    \\
 &  & Fitted BG amp. &  &  & &  &1.969 & 0.105 &   1.086 &    \\
 &  &L+P ED PWA &  &  &  & & 2.163 & 0.273 & 1.086 &   \\
 \hline
\hline
\end{tabular}
  \caption{With Re$M_i$ and Im$M_i$, $i=1,2,3$ we denote  pole parameters for simplified L+P PWA and full L+P PWA in~GeV units.  $BP_{1}^{LP}$ and  $BP_{2}^{LP}$  denote the fixed $\pi N$ threshold, and the second, effective branch-points of Pietarinen expansions, also in GeV units.  Bolded numbers for each partial wave are Bonn-Gatchina values from the PDG (the value marked with * denotes one resonance which is demanded by the final L+P PWA fit of the 2+ multipole, and which is not given by the PDG~\cite{PDG}).  The second row is the result of fitting Bonn-Gatchina  transversity amplitudes with the simplified L+P PWA. The third row shows the result of simplified L+P ED PWA fit, and the fourth row where it exist, shows the results of full L+P fit on the final SE AA/PWA amplitudes.}
 \label{tab:LPlusPIntermediateResultsI}
\end{table}

\clearpage

\section{Summary and Conclusions}

For the first time, pole parameters have been used directly as fitting parameters in an ED L+P analysis of photoproduction data.
Previous L+P fits have been applied to sets of multipoles obtained in independent analyses of data.
As a proof-of-principle calculation, a simplified ED L+P PWA formalism was used, having a reduced number of non-pole parameters
and dropping weakly coupled poles. Phase information
from the multi-channel Bonn-Gatchina analysis was used to weakly mitigate the continuum ambiguity, plaguing single-channel fits. Here we have used the Bonn-Gatchina analysis to initialize the L+P parameters used in the fit to data.
To search for missing structure, as a second step, a constrained SE AA/PWA was performed with the result of an L+P ED PWA fit used as a constraint.  A comparison of multipoles, from the
ED and constrained SE fits, to those from Bonn-Gatchina, showed reasonable agreement for the largest multipoles but
significant deviations from the Bonn-Gatchina analysis in other cases. Closer agreement should not be expected.

To quantify the difference between our simplified L+P ED PWA and the constrained Step 2 discrete SE AA/PWA multipoles, which fit the data notably better, we have performed a full L+P analysis of the latter using formalism of \mbox{refs.~\cite{\LP,\LPapplication}} and compared the obtained poles. We conclude that for the dominant $E_{0+}$ multipole no corrections are needed, however all other multipoles are consistent with having at least one more pole, and required one extra Pietarinen expansion. Thus, some improvement of analytic structure of simplified L+P ED PWA is needed in future studies.

In the full L+P analysis we encountered one surprising result: in 2+ multipoles, the final fit required an extra resonance with parameters 1.888 + \i \, 0.072 GeV, not  by listed by the PDG. However, this could disappear once the above mentioned improvements to the L+P structure are implemented.

Some problems of principle remain to be fully addressed. The number of measurements required for a 'complete experiment' or a truncated
partial-wave analysis continues to be discussed. This has some impact on the ability to perform SE analyses, though arguments generally
must ignore the effect of experimental uncertainties. Here, by using a penalty function, our SE results are relatively smooth and tied
to an ED solution. However, some indications of a discontinuity where the number of observables changes from 8 to 4 may be visible in
the higher multipoles shown in Fig.~\ref{Multipoles:b}.
The main problem of SC methods is that a phase ambiguity~\cite{SE-PWA-uniqueness} remains, but is "hidden" in the choice of initial parameters of our Step 1 fit (fitting an analytically simplified L+P expansion of multipoles to the experimental data).
Being an overall phase, having no effect on observables, it is difficult to study. However, it  can change
the appearance of multipoles and therefore is an added problem in comparing the results of different groups.
In future, an upgrade of this proof-of-principle calculation will be implemented. As is commonly known, the presently used Mathematica software is notoriously inefficient for the minimization tasks, as the process involves inverting very large matrices, requiring a large amount of memory and CPU time.  Other,  faster minimization software will be used (MINUIT in FORTRAN90, for example), and applied with the new, better hardware. This will reduce CPU time, and will enable the introduction of more complex analytic forms (more poles, more Pietarinen expansions and more Pietarinen terms in a single expansion). Hence, the proposed iterative procedure will be fully implemented to formulate the final, full-scale model with the inclusion of error uncertainties and data consistency.

\clearpage

\begin{acknowledgments}

Deepest thanks go to our colleagues Lothar Tiator and Yannick Wunderlich who significantly contributed to an early version of this
work. Lothar Tiator found inconsistencies in early attempts and made suggestions for improvements; Yannick Wunderlich carefully read
early versions of the manuscript and made suggestions which improved its consistency and precision. We hope these interactions
continue and lead to future publications.
A.S. acknowledges the support from STRONG-2020 EU project, Grant agreement ID: 824093.
This work was supported in part by the U.S. Department of Energy, Office of Science, Office of Nuclear Physics,
under grant DE-SC001652.
\end{acknowledgments}

\appendix

\section{Formalism for $K \Lambda$ photoproduction} \label{AppendixA}
Partial wave decompositions are introduced through CGLN amplitudes:
\begin{align}
F_{1} \left( W, \theta \right) &= \sum \limits_{\ell = 0}^{\infty} \Big\{ \left[ \ell M_{\ell+} \left( W \right) + E_{\ell+} \left( W \right) \right] P_{\ell+1}^{'} \left( \cos \theta \right) \nonumber \\
 & \quad \quad \quad + \left[ \left( \ell+1 \right) M_{\ell-} \left( W \right) + E_{\ell-} \left( W \right) \right] P_{\ell-1}^{'} \left( \cos \theta \right) \Big\} \mathrm{,} \label{eq:MultExpF1} \\
F_{2} \left( W, \theta \right) &= \sum \limits_{\ell = 1}^{\infty} \left[ \left( \ell+1 \right) M_{\ell+} \left( W \right) + \ell M_{\ell-} \left( W \right) \right] P_{\ell}^{'} \left( \cos \theta \right) \mathrm{,} \label{eq:MultExpF2} \\
F_{3} \left( W, \theta \right) &= \sum \limits_{\ell = 1}^{\infty} \Big\{ \left[ E_{\ell+} \left( W \right) - M_{\ell+} \left( W \right) \right] P_{\ell+1}^{''} \left( \cos \theta \right) \nonumber \\
 & \quad \quad \quad + \left[ E_{\ell-} \left( W \right) + M_{\ell-} \left( W \right) \right] P_{\ell-1}^{''} \left( \cos \theta \right) \big\} \mathrm{,} \label{eq:MultExpF3} \\
F_{4} \left( W, \theta \right) &= \sum \limits_{\ell = 2}^{\infty} \left[ M_{\ell+} \left( W \right) - E_{\ell+} \left( W \right) - M_{\ell-} \left( W \right) - E_{\ell-} \left( W \right) \right] P_{\ell}^{''} \left( \cos \theta \right) \mathrm{.} \label{eq:MultExpF4}
\end{align}
Transversity amplitudes are defined as:
\begin{align}
 b_{1} \left( W, \theta\right) &= - b_{3} \left( W, \theta\right)
  - \frac{1}{\sqrt{2}} \sin \theta \left[ F_{3} \left( W, \theta\right) e^{- \imath \frac{\theta}{2}} + F_{4} \left( W, \theta\right) e^{ \imath \frac{\theta}{2}} \right] \mathrm{,} \label{eq:b1BasicForm} \\
 b_{2} \left( W, \theta\right) &= - b_{4} \left( W, \theta\right)
  +  \frac{1}{\sqrt{2}} \sin \theta \left[ F_{3} \left( W, \theta\right) e^{\imath \frac{\theta}{2}} + F_{4} \left( W, \theta\right) e^{- \imath \frac{\theta}{2}} \right] \mathrm{,} \label{eq:b2BasicForm} \\
 b_{3} \left( W, \theta\right) &= \frac{\imath}{\sqrt{2}} \left[ F_{1} \left( W, \theta\right) e^{- \imath \frac{\theta}{2}} -  F_{2} \left( W, \theta\right) e^{ \imath \frac{\theta}{2}} \right] \mathrm{,} \label{eq:b3BasicForm} \\
 b_{4} \left( W, \theta\right) &= \frac{\imath}{\sqrt{2}} \left[ F_{1} \left( W, \theta\right) e^{\imath \frac{\theta}{2}} -  F_{2} \left( W, \theta\right) e^{- \imath \frac{\theta}{2}} \right] \mathrm{.} \label{eq:b4BasicForm}
\end{align}

\newpage
All 16 polarization observables can be expressed in terms of transversity amplitudes:
\begin{table}[hb]
 \begin{tabular}{lr}
  \hline
 \hline
  Observable  &  Group  \\
  \hline
  $\sigma_{0} = \frac{1}{2} \left( \left| b_{1} \right|^{2} + \left| b_{2} \right|^{2} + \left| b_{3} \right|^{2} + \left| b_{4} \right|^{2} \right)$  &     \\
  $\hat{\Sigma} = \frac{1}{2} \left( - \left| b_{1} \right|^{2} - \left| b_{2} \right|^{2} + \left| b_{3} \right|^{2} + \left| b_{4} \right|^{2} \right)$  &   $\mathcal{S}$ \\
  $\hat{T} = \frac{1}{2} \left( \left| b_{1} \right|^{2} - \left| b_{2} \right|^{2} - \left| b_{3} \right|^{2} + \left| b_{4} \right|^{2} \right)$  &     \\
  $\hat{P} = \frac{1}{2} \left( - \left| b_{1} \right|^{2} + \left| b_{2} \right|^{2} - \left| b_{3} \right|^{2} + \left| b_{4} \right|^{2} \right)$  &     \\
  \hline
   $\hat{E}  = \mathrm{Re} \left[ - b_{3}^{\ast} b_{1} - b_{4}^{\ast} b_{2} \right]  = - \left| b_{1} \right| \left| b_{3} \right| \cos \phi_{13} - \left| b_{2} \right| \left| b_{4} \right| \cos \phi_{24}$  &  \\
   $\hat{F} = \mathrm{Im} \left[ b_{3}^{\ast} b_{1} - b_{4}^{\ast} b_{2} \right] = \left| b_{1} \right| \left| b_{3} \right| \sin \phi_{13} - \left| b_{2} \right| \left| b_{4} \right| \sin \phi_{24}  $ &  $\mathcal{BT} $ \\
   $\hat{G} = \mathrm{Im} \left[ - b_{3}^{\ast} b_{1} - b_{4}^{\ast} b_{2} \right] = - \left| b_{1} \right| \left| b_{3} \right| \sin \phi_{13} - \left| b_{2} \right| \left| b_{4} \right| \sin \phi_{24} $  &  \\
   $ \hat{H} = \mathrm{Re} \left[ b_{3}^{\ast} b_{1} - b_{4}^{\ast} b_{2} \right] = \left| b_{1} \right| \left| b_{3} \right| \cos \phi_{13} - \left| b_{2} \right| \left| b_{4} \right| \cos \phi_{24} $  &   \\
   \hline
   $\hat{C}_{x'}  = \mathrm{Im} \left[ - b_{4}^{\ast} b_{1} + b_{3}^{\ast} b_{2} \right]  = - \left| b_{1} \right| \left| b_{4} \right| \sin \phi_{14} + \left| b_{2} \right| \left| b_{3} \right| \sin \phi_{23}  $ &  \\
   $\hat{C}_{z'} = \mathrm{Re} \left[ - b_{4}^{\ast} b_{1} - b_{3}^{\ast} b_{2} \right] = - \left| b_{1} \right| \left| b_{4} \right| \cos \phi_{14} - \left| b_{2} \right| \left| b_{3} \right| \cos \phi_{23} $  &  $\mathcal{BR}$   \\
   $\hat{O}_{x'} = \mathrm{Re} \left[ - b_{4}^{\ast} b_{1} + b_{3}^{\ast} b_{2} \right] = - \left| b_{1} \right| \left| b_{4} \right| \cos \phi_{14} + \left| b_{2} \right| \left| b_{3} \right| \cos \phi_{23} $  &  \\
   $\hat{O}_{z'} = \mathrm{Im} \left[ b_{4}^{\ast} b_{1} + b_{3}^{\ast} b_{2} \right] = \left| b_{1} \right| \left| b_{4} \right| \sin \phi_{14} + \left| b_{2} \right| \left| b_{3} \right| \sin \phi_{23} $  &   \\
   \hline
   $\hat{L}_{x'} = \mathrm{Im} \left[ - b_{2}^{\ast} b_{1} - b_{4}^{\ast} b_{3} \right] = - \left| b_{1} \right| \left| b_{2} \right| \sin \phi_{12} - \left| b_{3} \right| \left| b_{4} \right| \sin \phi_{34}$  &   \\
   $\hat{L}_{z'}  = \mathrm{Re} \left[ - b_{2}^{\ast} b_{1} - b_{4}^{\ast} b_{3} \right]  = - \left| b_{1} \right| \left| b_{2} \right| \cos \phi_{12} - \left| b_{3} \right| \left| b_{4} \right| \cos \phi_{34}$  &  $\mathcal{TR}$  \\
   $\hat{T}_{x'} = \mathrm{Re} \left[ b_{2}^{\ast} b_{1} - b_{4}^{\ast} b_{3} \right] = \left| b_{1} \right| \left| b_{2} \right| \cos \phi_{12} - \left| b_{3} \right| \left| b_{4} \right| \cos \phi_{34}$  &    \\
   $\hat{T}_{z'} = \mathrm{Im} \left[ - b_{2}^{\ast} b_{1} + b_{4}^{\ast} b_{3} \right] = - \left| b_{1} \right| \left| b_{2} \right| \sin \phi_{12} + \left| b_{3} \right| \left| b_{4} \right| \sin \phi_{34}$ &   \\
   \hline
   \hline
 \end{tabular}
 \caption{The definitions of the $16$ polarization ob\-serva\-bles of pseudoscalar meson photoproduction
 are given here in terms of transversity amplitudes $b_{1}, \ldots, b_{4}$ (cf. Ref.~\cite{Chiang:1996em};
 sign conventions are consistent with~\cite{YannickPhD}). Expressions are given both in terms of real- and imaginary parts
 of bilinear products of amplitudes and in terms of moduli and relative phases of the amplitudes.
 Furthermore, the phase-space factor $\rho$ has been suppressed in the given expressions (i.e. we have set $\rho = 1$).
 The four different groups of four observables each are indicated as well.}
 \label{tab:PhotoproductionObservables}
\end{table}

\clearpage



\begin{thebibliography}{99}


\bibitem{Kamano2013} H. Kamano, S. X. Nakamura, T. -S. H. Lee, T. Sato, Phys. Rev. C \textbf{88},  035209 (2013).

\bibitem{Kamano:2016bgm}
H.~Kamano, S.~X.~Nakamura, T.~S.~H.~Lee and T.~Sato,
Phys. Rev. C \textbf{94}, no.1, 015201 (2016).

\bibitem{Kamano:2019gtm}
H.~Kamano, T.~S.~H.~Lee, S.~X.~Nakamura and T.~Sato,
[arXiv:1909.11935 [nucl-th]].

\bibitem{Anisovich:2011fc}
A.~V.~Anisovich, R.~Beck, E.~Klempt, V.~A.~Nikonov, A.~V.~Sarantsev and U.~Thoma,
Eur. Phys. J. A \textbf{48}, 15 (2012)


\bibitem{Anisovich:2016vzt}
A.~V.~Anisovich, R.~Beck, M.~D\"oring, M.~Gottschall, J.~Hartmann, V.~Kashevarov, E.~Klempt, U.~G.~Mei\ss{}ner, V.~Nikonov and M.~Ostrick, \textit{et al.}
Eur. Phys. J. A \textbf{52}, no.9, 284 (2016).

\bibitem{Ronchen:2012eg}
D.~R\"onchen, M.~D\"oring, F.~Huang, H.~Haberzettl, J.~Haidenbauer, C.~Hanhart, S.~Krewald, U.~G.~Mei\ss{}ner and K.~Nakayama,
Eur. Phys. J. A \textbf{49}, 44 (2013).

\bibitem{Ronchen:2018ury}
D.~R\"onchen, M.~D\"oring and U.~G.~Mei\ss{}ner,
Eur. Phys. J. A \textbf{54}, no.6, 110 (2018).


\bibitem{Hunt:2018tvt}
B.~C.~Hunt and D.~M.~Manley,
Phys. Rev. C \textbf{99}, no.5, 055203 (2019).

\bibitem{Hunt:2018mrt}
B.~C.~Hunt and D.~M.~Manley,
Phys. Rev. C \textbf{99}, no.5, 055204 (2019).

\bibitem{Hunt:2018wqz}
B.~C.~Hunt and D.~M.~Manley,
Phys. Rev. C \textbf{99}, no.5, 055205 (2019).

\bibitem{Workman:2012hx}
R.~L.~Workman, R.~A.~Arndt, W.~J.~Briscoe, M.~W.~Paris and I.~I.~Strakovsky,
Phys. Rev. C \textbf{86}, 035202 (2012).

\bibitem{Ciulli}S. Ciulli and J. Fischer in Nucl. Phys. \textbf{24}, 465 (1961).

\bibitem{CiulliFisher}I. Ciulli, S. Ciulli, and J. Fisher, Nuovo Cimento \textbf{23}, 1129 (1962).

\bibitem{Pietarinen} E. Pietarinen, Nuovo Cimento Soc. Ital. Fis. \textbf{12A}, 522 (1972).

\bibitem{Pietarinen1} E. Pietarinen, Nucl. Phys. \textbf{B107}, 21 (1976).


\bibitem{Svarc2013} A. \v{S}varc, M. Had\v{z}imehmedovi\'{c}, H. Osmanovi\'{c}, J. Stahov, L. Tiator,
and R.L. Workman, Phys,  Rev.  \textbf{C 88}, 035206 (2013).

\bibitem{Svarc2016} A. \v{S}varc, M. Had\v{z}imehmedovi\'{c}, H. Osmanovi\'{c}, J. Stahov, L. Tiator, R.L. Workman, Phys. Lett. \textbf{B755}, 452 (2016).

\bibitem{Svarc2014}  A. \v{S}varc, M. Had\v{z}imehmedovi\'{c}, R. Omerovi\'{c}, H. Osmanovi\'{c}, and  J. Stahov, Phys,  Rev.  \textbf{C 89}, 045205 (2014).

\bibitem{Svarc2014a} A. \v{S}varc, M. Had\v{z}imehmedovi\'{c}, H. Osmanovi\'{c},  J. Stahov, L. Tiator, and R.L. Workman,  \textbf{C 89}, 065208 (2014).

\bibitem{Svarc2015}  A. \v{S}varc, M. Had\v{z}imehmedovi\'{c}, H. Osmanovi\'{c},  J. Stahov, and R.L. Workman,  \textbf{C 91}, 015205 (2015).

 \bibitem{Svarc2016a} L. Tiator, M. D\"{o}ring, R. L. Workman, M. Had\v{z}imehmedovi\'{c}, H. Osmanovi\'{c}, R. Omerovi\'{c},  J. Stahov, and A. \v{S}varc,  \textbf{C 94}, 65204 (2016).

\bibitem{SE-PWA-uniqueness} A.~\v{S}varc, Y. Wunderlich, H. Osmanovi\'{c}, M. Had\v{z}imehmedovi\'{c}, R. Omerovi\'{c}, J. Stahov, V. Kashevarov, K. Nikonov, M. Ostrich, L. Tiator, R. Workman, Phys.\ Rev.\ C\ {\bf 97}, 054611 (2018).

\bibitem{Svarc2018} A.~\v{S}varc, Y. Wunderlich, H. Osmanovi\'{c}, M. Had\v{z}imehmedovi\'{c}, R. Omerovi\'{c}, J. Stahov, V. Kashevarov, K. Nikonov, M. Ostrich, L. Tiator, R. Workman, Few Body Systems  {\bf 59}, 96 (2018).

\bibitem{Svarc2020} A. \v{S}varc, Y. Wunderlich , and L. Tiator, Phys. Rev. \textbf{C 102}, 064609 (2020).

\bibitem{Svarc2022} A. \v{S}varc, Y. Wunderlich , and L. Tiator, Phys. Rev. \textbf{C 105}, 024614 (2022).

\bibitem{Hoehler} G. H\"{o}hler, Pion Nucleon Scattering, Part 2, Landolt-Bornstein: Elastic and Charge Exchange Scattering of Elementary Particles, Vol. 9b (Springer-Verlag, Berlin, 1983).


\bibitem{PDG} P.A. Zyla et al. (Particle Data Group), Prog. Theor. Exp. Phys. 2020, 083C01 (2020) and 2021 update, mini-review on N and $\Delta$ resonances.

\bibitem{Chiang:1996em}
  W.~T.~Chiang and F.~Tabakin,
  Phys.\ Rev.\ C {\bf 55}, 2054 (1997).

\bibitem{YannickPhD}
Y.~Wunderlich, "The complete experiment problem of pseudoscalar meson photoproduction in a truncated partial wave analysis", PhD thesis, University of Bonn (2019) [arXiv:2008.00514 [nucl-th]].

\bibitem{Phase-ambiguity} A.\v{S}varc, Phys. Rev. \textbf{C 104}, 014605 (2021).

\bibitem{Anisovich2017} A.V. Anisovich, V. Burkert, M. Had\v{z}imehmedovi\'{c}, D.G. Ireland, E. Klempt, V.A. Nikonov, R. Omerovi\'{c}, H. Osmanovi\'{c},  A.V. Sarantsev, J. Stahov, A. \v{S}varc, and U. Thoma,  Phys. Rev. Lett. \textbf{119}, 062004 (2017).

\bibitem{Anisovich2017a} A.V. Anisovich, V. Burkert, M. Had\v{z}imehmedovi\'{c}, D.G. Ireland, E. Klempt, V.A. Nikonov, R. Omerovi\'{c}, H. Osmanovi\'{c},  A.V. Sarantsev, J. Stahov, A. \v{S}varc, and U. Thoma, Eur. Phys. J. A \textbf{53}: 242 (2017).

\bibitem{BGpoles} Numbers for Bonn-Gatchina model of Refs.~\cite{\BGED} are taken from Ref.~\cite{PDG}.

\bibitem{Osmanovic2018}  H. Osmanovi\'{c}, M. Had\v{z}imehmedovi\'{c}, R. Omerovi\'{c}, J. Stahov, V. Kashevarov,
K. Nikonov, M. Ostrick, L. Tiator,  and A. \v{S}varc, Phys. Rev.  \textbf{C 97}, 015207 (2018).

\bibitem{Osmanovic2019}
H. Osmanovi\'{c}, M. Had\v{z}imehmedovi\'{c}, R. Omerovi\'{c}, J. Stahov,  M. Gorchtein,  V. Kashevarov, K. Nikonov, M. Ostrick,
L. Tiator, and A. \v{S}varc, Phys. Rev. \textbf{C 100}, 055203 (2019).


\bibitem{Osmanovic2021}
H. Osmanovi\'{c}, M. Had\v{z}imehmedovi\'{c}, R. Omerovi\'{c}, J. Stahov, V. Kashevarov, M. Ostrick, L. Tiator, and A. \v{S}varc,
Phys. Rev. \textbf{C 104}, 034605 (2021).


\bibitem{BG-web} https://pwa.hiskp.uni-bonn.de/.

\bibitem{GWU-web} https://gwdac.phys.gwu.edu/.

\bibitem{Bradford} R.~Bradford et al., Phys. Rev.\textbf{C 73}, 035202 (2006).

\bibitem{McCracken} M.E.~McCracken et al., Phys. Rev.\textbf{C 81}, 025201 (2010).

\bibitem{Lleres} A.~Lleres et al., Eur. Phys. J. \textbf{A 31}, 79 (2007).

\bibitem{Paterson} C.A. Paterson et al., Phys. Rev. \textbf{C 93}, 065201 (2016).

\bibitem{Nakayama:2018yzw}
  K.~Nakayama, Phys.\ Rev.\ C {\bf 100}, 035208 (2019).

\bibitem{Tiator2012} L. Tiator. Towards a model-independent partial wave analysis for pseudoscalar meson photoproduction. AIP Conf. Proc., 1432:162-167, 2012.

\bibitem{Wunderlich2014} Y. Wunderlich, R. Beck, and L. Tiator, Phys.\ Rev.\ C {\bf 89}, 055203 (2014).

\bibitem{CPU} Tests were carried out using a Lenovo Legion laptop with AMD Ryzen 7 4800H with Radeon Graphics processor,  2.90 GHz and 16 Gb RAM
with Mathematica 11.0 code. Shorter run times could be achieved using Fortran or C but Mathematica is a more versatile language for model testing.

\end{thebibliography}
\end{document}